%% file: paper-arxiv.tex
\documentclass[letter,fleqn]{cas-sc}	
\AtBeginDocument{%
	}

	\usepackage[numbers]{natbib}
	\usepackage{myMac-els}	
	\usepackage{float} 
\input{lmac}

	\usepackage[ruled]{algorithm2e}
	\def\tsc#1{\csdef{#1}{\textsc{\lowercase{#1}}\xspace}}
	\tsc{WGM}
	\tsc{QE}
	\tsc{EP}
	\tsc{PMS}
	\tsc{BEC}
	\tsc{DE}

\begin{document}
	\let\WriteBookmarks\relax
	\def\floatpagepagefraction{1}
	\def\textpagefraction{.001}
	\shorttitle{A Novel Approach to Initial Value Problems}
	\shortauthors{Zhang and Yap}

\title[mode=title]{
	A Novel Approach to the Initial Value Problem
	with a Complete Validated Algorithm
	}
\tnotemark[1]

	\tnotetext[1]{This document contain the results of the
	research funded by NSF Grant \#CCF-2212462.}

\author[1]{Bingwei Zhang}[
	type=editor,
	auid=000,
	bioid=1,
	orcid=0009-0002-2619-9807
	]
	\fnmark[1]
	\ead{bz2517@nyu.edu}
	
	\affiliation[1]{organization=
	{The Courant Institute of Mathematical Sciences},
		addressline={Department of Computer Science},
		city={New York, NY 10012},
		country={USA}}
	
\author[1]{Chee Yap}[
	type=editor,
	auid=000,
	bioid=2,
	orcid=0000-0003-2952-3545
	]
	\cormark[1]
	\ead{yap@cs.nyu.edu}
	
	\cortext[cor1]{Corresponding author}
	
	\ignore{
	\nonumnote{This work demonstrates $a_b$ the formation Y\_1 of a
	complete validated algorithm for solving initial value problems
	through novel computational approaches.}
	}

\begin{abstract} 
	We consider the first order autonomous differential equation
	(ODE) $\bfx'=\bff(\bfx)$ where $\bff: \RR^n\to\RR^n$ is locally
	Lipschitz.
	For $\bfx_0\in\RR^n$ and $h>0$, the initial value problem (IVP)
	for $(\bff,\bfx_0,h)$ is to determine if there is a unique
	solution, i.e., a function $\bfx:[0,h]\to\RR^n$ that satisfies the
	ODE with $\bfx(0)=\bfx_0$. Write $\bfx =\ivp_\bff(\bfx_0,h)$ for
	this unique solution.
	
	We pose a corresponding computational problem,
	called the \dt{End Enclosure Problem}:
	given $(\bff,B_0,h,\veps_0)$ where $B_0\ib\RR^n$ is a box
	and $\veps_0>0$, to compute a pair of non-empty boxes $(\ulB_0,B_1)$
	such that $\ulB_0\ib B_0$, width of $B_1$ is $<\veps_0$,
	and for all $\bfx_0\in \ulB_0$, $\bfx=\ivp_\bff(\bfx_0,h)$
	exists and $\bfx(h)\in B_1$.
	We provide a complete validated algorithm for this problem.
	Under the assumption (promise) that for all $\bfx_0\in B_0$,
	$\ivp_\bff(\bfx_0,h)$ exists, we prove the halting of our algorithm.  
	This is the first halting algorithm for IVP problems
	in such a general setting.
	
	We also introduce novel techniques for subroutines
	such as StepA and StepB, and a scaffold datastructure 
	to support our End Enclosure algorithm.
	Among the techniques are new ways refine
	full- and end-enclosures based on
	a \dt{radical transform} combined with logarithm norms.
	Our preliminary implementation demonstrates the viability of the
complete algorithm.  Comparisons with existing validated IVP software
also help identify the principal computational costs associated with
guaranteeing termination and a prescribed end-enclosure accuracy.
\end{abstract}

\ignore{
\begin{abstract}
	The Initial Value Problem (IVP) is concerned with finding
	solutions to a system of autonomous
	ordinary differential equations (ODE)
		$\bfx' = \bff(\bfx)$
	with given initial condition $\bfx(0)\in B_0$
	for some box $B_0\ib \RR^n$.
	Here $\bff:\RR^n\to\RR^n$
	and $\bfx:[0,1]\to\RR^n$
	where $\bff$ and $\bfx$ are $C^1$-continuous.
	Let $\texttt{IVP}_\bff(B_0)$ denote the set of all
	such solutions $\bfx$.
	
	In this paper, we introduce a novel
	way to exploit the theory of \textbf{logarithmic norms}:
	we introduce the concept of a \textbf{radical transform}
	$\pi:\RR^n\to\RR^n$
	to convert the $(\bfx,\bff)$-system 
	into another
	system $\bfy' = \bfg(\bfy)$ so that
	the $(\bfy,\bfg)$-space has negative
	logarithmic norm in any desired small enough neighborhood.
	
	Based on such radical transform steps,
	we construct a complete validated algorithm for the
	following \textbf{End-Enclosure Problem}:
	\begin{center}
		INPUT:\hspace*{5mm} $(\bff, B_0,\veps_0)$,\\
		OUTPUT:\hspace{2mm} $(\ulB_0,B_1)$,\hspace*{3mm}
	\end{center}
	where $B_0\ib \RR^n$ is a box, $\veps_0>0$,
	and 
	$(\ulB_0,B_1)$,
	such that
	$\ulB_0\ib B_0$, 
	the diameter of $B_1$ is at most $\veps_0$,
	and $B_1$ is an end-enclosure for $\ivp(\ulB_0)$,
	i.e., for all
	$\bfx\in \ivp(\underline{B}_0)$,
	$\bfx(1)\in B_1$.
	This is a ``promise'' algorithm in the
	sense that the input contains a promise that for each
	$\bfp_0\in B_0$,
	there is a unique solution
	$\bfx\in \ivp(B_0)$ with $\bfx(0)=\bfp_0$.
	By setting $B_0=\set{\bfp_0}$, our algorithm has the unique
	ability to approximate the endpoint $\bfx(1)$
	to any desired $\veps_0$.
	Our algorithm is guaranteed to halt and contains no hidden
	hyperparamters.
	A preliminary implementation of our algorithm shows promise.
\end{abstract}
}%

\ignore{
	\begin{graphicalabstract}
	\includegraphics{figs/cas-grabs.pdf}
	\end{graphicalabstract}
	
	\begin{highlights}
	\item Research highlights item 1
	\item Research highlights item 2
	\item Research highlights item 3
	\end{highlights}
	}%


\maketitle
\input{inc/1intro}
\input{inc/2tools}

\input{inc/3overview}
\input{inc/4stepAB}
\input{inc/5comparison}
\input{inc/6refine}
\input{inc/7mainAlgo}

\input{inc/8conclusion}

\input{inc/appendix}




\input{paper-arxiv.bbl}

\end{document}

%% file: lmac.tex
\newcommand{\issacArxiv}[2][]{#2}	

\newcommand{\bw}[2][]{\cored{Bingwei says: #2}\comagenta{New: #1}}
\newcommand{\bwX}[2][]{#1}
\newcommand{\chee}[2][]{\coblue{Chee says: #2}\cocyan{New: #1}}
\newcommand{\cheeX}[2][]{#1}

\newcommand{\ccheeX}[2][]{#1}

\newcommand{\Tube}{\mbox{Tube}} 
\newcommand{\chull}{\mbox{CHull}} 
\newcommand{\wmax}{w_{\max}} 
\newcommand{\wmin}{w_{\min}} 

\newcommand{\nnorm}[1][\cdot]%
	{\left\lVert #1 \right\rVert_{\max}}

\def\lognorm{{logNorm}}

\def\ivp{\mbox{{\ttt{IVP}}}}		
\def\stepA{{\mbox{\tt StepA}}}
\def\stepB{{\mbox{\tt StepB}}}
\def\stepb{{\mbox{\tt Step B}}} 

\def\standardIVP{\mbox{{\tt standard\_IVP}}}

\def\endEncIVP{\mbox{{\tt EndEncl\_IVP}}} 
\def\endEncAlgo{\mbox{{\tt EndEnc}}} 
\def\ourAlgo{{\endEncAlgo}}
\def\capdCr{\mbox{\tt CAPD}}
\def\vnodelp{\mbox{\tt VNODE}}
\def\Ours{\mbox{\tt Ours}}

\def\Time{\mbox{\rm Time}}

\def\Refine{\mbox{\ttt{Refine}}}

\def\Extend{\mbox{\ttt{Extend}}}

\def\Standard{\mbox{\ttt{OurSimple}}}
\def\StandardT{\mbox{\ttt{OurSimpleT}}}
\def\OurNoT{\mbox{\ttt{OurNoT}}}
\def\OurNoSubr{\mbox{\ttt{OurNoEuler}}}

\def\bigStep{\mbox{\ttt{bigStep}}}
\def\smallStep{\mbox{\ttt{smallSteps}}}
\def\miniStep{\mbox{\ttt{miniSteps}}}

\def\AvoidsZero{{\ttt{AvoidsZero}}}
\def\Transform{{\ttt{Transform}}}

\def\TransformBound{{\ttt{TransformBound}}}
\def\SubrSeven{{\ttt{EulerTube}}}
\def\eulertube{{\ttt{EulerTube}}}
\def\Bisect{{\ttt{Bisect}}}

\def\ellFlag{{\ttt{$\ell$Flag}}}

\def\stage{{\calS}} 
\def\ministep{{\bfG}}

\newcommand{\std}{\text{std}}
\newcommand{\xform}{\text{xform}}

\newcommand{\euler}{^{\text{euler}}}


%% file: inc/1intro.tex
\sect{Introduction}
	We consider the following system of
	first order ordinary differential equations (ODEs)
	\beql{bfx'}
		\bfx' = \bff(\bfx)
	\eeql
	where $\bfx=(x_1\dd x_n)\in C^1([0,h]\to \RR^n)$
	are functions of time and $\bfx'=(x_1'\dd x_n')$
	indicate differentiation with respect to time,
	and $\bff=(f_1\dd f_n):\RR^n\to\RR^n$.
	Since this is an autonomous ODE, we may assume the initial time
	$t=0$.  Up to time scaling, we often assume that the end time
	is $h=1$.  This assumption is just for simplicity but
	our results and implementation allow any value of $h>0$.

\begin{figure}
		\centering
	        \includegraphics[width=0.4\linewidth]{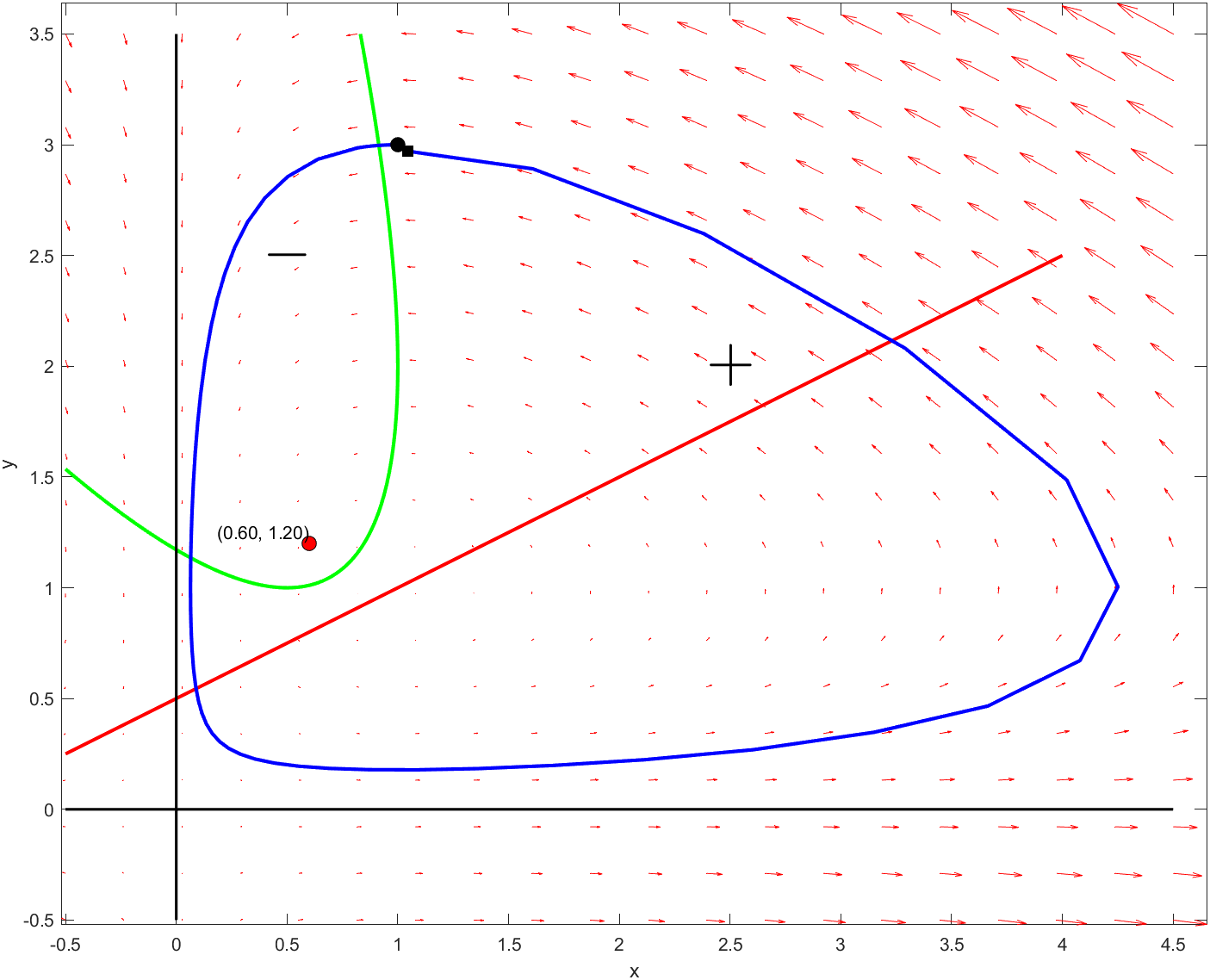}
		\caption{ Volterra system (Eg1).  The negative zone of
			the system is the region above the green parabola.}
		\label{fig:Volterra-21-13}
		\end{figure}

	Given $\bfp_0\in\RR^n$ and $h>0$,
	the \dt{initial value problem} (IVP) for $(\bfp_0,h)$ is the
	mathematical problem of finding a \dt{solution},
	i.e.,
	a continuous function $\bfx: [0,h]\to\RR^n$
	that satisfies \refeq{bfx'}, 
	subject to $\bfx(0)=\bfp_0$.
	Let $\ivp_\bff(\bfp_0,h)$ denote the set of all such solutions.
    Since $\bff$ is usually fixed or understood,
	we normally omit $\bff$ in our notations.
	We say that $(\bfp_0,h)$ is \dt{valid} if
	the solution exists and is unique, i.e.,
	$\ivp(\bfp_0,h)=\set{\bfx_0}$ is a singleton. 
	In this case, we write $\bfx_0= \ivp(\bfp_0,h)$.
	It is convenient to write
			$\bfx(t;\bfp_0)$ for $\bfx_0(t)$.
	See \refFig{Volterra-21-13} for the solution to the
	Volterra system (Eg1 in \refTab{problems}).
	The IVP problem has numerous applications such as
	modeling physical, chemical and biological systems.



	The mathematical IVP gives rise to a variety
	of algorithmic problems since we generally
	cannot represent a solution $\bfx_0= \ivp(\bfp_0,h)$ explicitly.
	We are interested in \dt{validated algorithms}
	\cite{moore:bk} meaning that all
	approximations must be explicitly bounded (e.g.,
	numbers are enclosed in intervals). In this setting,
	we introduce the simplest {\em algorithmic} IVP problem,
	that of computing an enclosure for $\bfx(h;\bfp_0)$.
	\ignore{
		Another possible algorithmic IVP problem is to compute
		a polygonal $\delta$-approximation to the trajectory
		that $\bfx(t;\bfp_0)$.
	}
	In real world applications, only
	approximate values of $\bfp_0$ are truly meaningful
	because of modeling uncertainties.
	So we replace $\bfp_0$ by a \dt{region} $B_0\ib\RR^n$: 
	$B_0$ is a non-empty set like a box or ball. 
	Let $\ivp(B_0,h)\as \bigcup_{\bfp_0 \in B_0}\ivp(\bfp_0,h)$.  
	Call $B_1\ib \RR^n$ an \dt{end-enclosure} for $\ivp(B_0,h)$
	if $B_1$ contains the set
		$\set{\bfx(h) : \bfx\in \ivp(B_0,h)}$.
	\savespace{
		Corliss \cite[Section 3]{corliss:survey-ode-intvl:89}
		confirms that this is the interval viewpoint.
		}
	So our formal algorithmic problem is
	the following \dt{End Enclosure Problem}:

	\renewcommand{\alt}[2]{#2} 
	\renewcommand{\alt}[2]{#1} 
	\alt{ 
		\beql{endEncProb}
			\hspace*{30mm}
			\includegraphics[width=0.50\columnwidth]{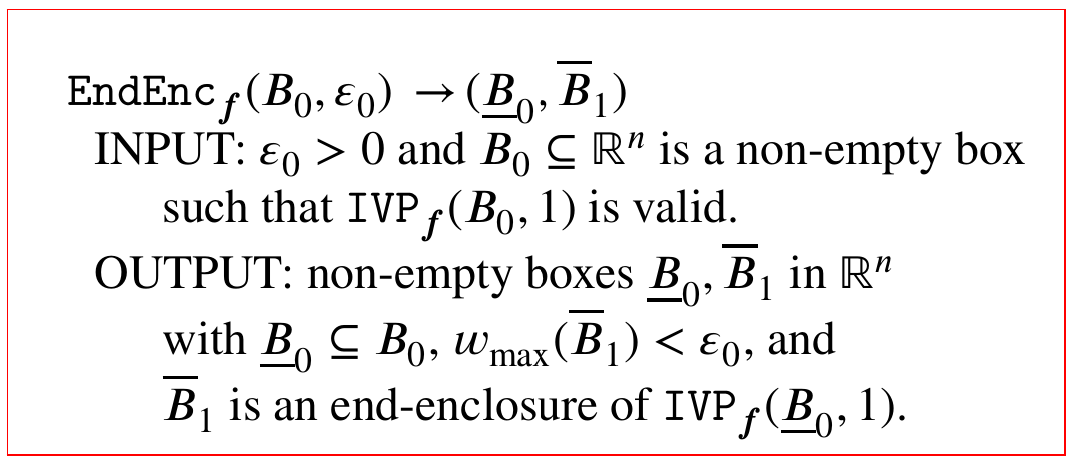}
		\eeql
	}{
		\Ldent\progb{
			\lline[-2] $\endEncAlgo_\bff(B_0,\veps_0)$
						$\ssa (\ulB_0,\olB_1)$
			\lline[0] INPUT: $\veps_0>0$ and
						$B_0\ib \RR^n$ is a non-empty box
			\lline[5] such that $\ivp_\bff(B_0,1)$ is valid.
			\lline[0] OUTPUT: non-empty boxes
					$\ulB_0, \olB_1$ in $\RR^n$
			\lline[5] with $\ulB_0\ib B_0$, $\wmax(\olB_1)<\veps_0$, and
			\lline[5] $\olB_1$ is an end-enclosure of
						$\ivp_\bff(\ulB_0,1)$.
		}
	}
	
	This is basically the \dt{Reachability Problem} in
		the non-linear control systems and verification literature
		(e.g.,	\cite{shen+2:tight-reach:21}).
	Our formulation of \refeq{endEncProb}
	has a natural but rare requirement that
	the end-enclosure $\olB_1$ satisfies $\wmax(\olB_1)<\veps_0$.
	In order to achieve this,
	we must allow input $B_0$ to shrink to some $\ulB_0$.
	But if we are promised that $\ivp(B_0,1)$ has an end-enclosure
	$\olB_1$ with $\wmax(\olB_1)<2\veps_0$, then it is possible
	to turn off shrinking and let $\ulB_0=B_0$.  
	In the common formulation of the IVP problem,
	$B_0$ is a singleton, $B_0=\set{\bfp_0}$.
	In this case, our algorithm will output $\ulB_0=B_0$.

	\ignore{
		If shrinking $B_0$ is objectionable, we could modify our
	algorithm to another version of End Enclosure problem:
	Given $(B_0,\veps)$ where $\ivp(B_0,1)$ is valid,
	compute a box $\olB_1$ such that
		$$B_1 \ib
		 Box(\set{\bfx(h): \bfx\in IVP(B_0,h)})\pm Box(\pm \veps).$$
	}%
	\ignore{
		Idea of Bingwei: since $\ivp(B_0,1)$ is valid,
		there is an admissible triple $(B_0,1,F)$.
		Let $\olmu \as \mu_2(J_\bff(F))$.
		If $r_0<\veps/2e^{\olmu}$ then our algorithm
		will succeed... Is that true?
	}
	\bwX{The reason we require \(\wmax(\olB_1) <
	\varepsilon\) is that it allows us to prove the termination of the
	algorithm without introducing an additional parameter \(h_{\min}\).
	This condition naturally leads to the need to scale \(B_0\). If
	\(\varepsilon\) is sufficiently large, then we have \(\ulB_0 = B_0\).
	}

\ssect{What is Achieved}
	Our formulation of the end-enclosure problem
	in \refeq{endEncProb} is new.  We will present a
	complete validated algorithm
	for this problem.  Our algorithm is
	\dt{complete} in the sense that if the input is valid,
	then [C0] the algorithm halts, and
	[C1] if the algorithm halts,
	the output $(\ulB_0,\olB_1)$ is correct.
	Algorithms that only satisfy [C1] are said\footnote{
		Completeness and partial correctness are standard
		terms in theoretical computer science.
	} to be \dt{partially correct}.
	To our knowledge, current validated IVP algorithms 
	are only partially correct since halting is not proved.

	The input to
	$\endEncIVP_\bff(B_0,\veps)$ 
	assumes the validity of $(B_0,1)$. 
	All algorithms have requirements
	on their inputs, but they are typically syntax requirements
	which are easily checked.  But validity of $(B_0,1)$
	is a semantic requirement which is non-trivial to check.
	Problems with semantic conditions on the input are called
	\dt{promise problems}
	\cite{goldreich:promise-problems:06}.
	Many numerical algorithms are actually
	solutions of promise problems. Checking if the promise
	holds is a separate decision problem.  To our knowledge,
	deciding validity of $(B_0,1)$ is an open problem although 
	some version of this question is
	undecidable in the analytic complexity framework
	\cite{graca+2:max-intvl-ivp:09, ko:real:bk}.


	Hans Stetter \cite{stetter:validatedODE:90}
	summarized the state-of-the-art over 30 years ago as follows:
	{\em To date, no programs that could be truly called
	`scientific software' have been produced. 
	AWA is state-of-the-art, and can be used by a sufficiently
	expert user -- it requires selection of step-size, order and
	suitable choice of inclusion set represention.}  
	Corliss \cite[Section 10]{corliss:survey-ode-intvl:89}
	made similar remarks.
		We believe our algorithm meets Stetter's and Corliss'
		criteria.  The extraneous inputs such as step-size,
		order, etc, noted by Stetter are usually called
		\dt{hyperparameters}.  Our algorithm\footnote{
			Hyperparameters are useful when used correctly.
			Thus, our implementation has some hyperparameters
			that may be used to improve performance, but
			they are optional and have
			no effect on completeness.
		}
		does not require any hyperparameters.
		Our preliminary implementation shows the
		viability of our algorithm, and
		its ability to do certain computations where
		current IVP software fails.
\ssect{In the Shadow of Lohner's AWA}
	In their comprehensive 1999 review,
	Nedialkov et al.~\cite{nedialkov+2:validated-ode:99}
	surveyed a family of validated IVP algorithms 
	that may be\footnote{
		This $(AB)^+$ motif is shared with homotopy path
		algorithms where $A$ and $B$ are usually
		called predictor and corrector (e.g.,
		\cite{xu-burr-yap:homotopy:18}).
	}
	called \dt{A/B-algorithms} because each computation
	amounts to a sequence of steps of the form
			$\underbrace{ABAB\cdots AB}_{2m}$ $=(AB)^m$
	for some $m\ge 1$,
	where $A$ and $B$ refer to two
	subroutines which\footnote{
		Nedialkov et al.~ called them Algorithms I and II.
	} we call \stepA\ and \stepB.
	It appears that all validated algorithms
	follow this motif, including
	Berz and Makino \cite{makino-berz:taylorModels:17}
	who emphasized their \stepB\ based on Taylor models.
	Ever since Moore
		\cite{nickel:wrapping-effect:86}
	pointed out the \dt{wrapping effect},
	experts have regarded the mitigation of
	this effect as essential. The solution
	based on iterated QR transformation by Lohner
		\cite{lohner:thesis}
	is regarded as the best technique to do this.
	It was implemented in the software
	called AWA\footnote{
		AWA abbreviates ``Anfangswertaufgabe'', the German term for IVP.
	} 
	and recently updated by Bunger
		\cite{bunger:taylorODE:20}
	in a INTLAB/MATLAB implementation. 
	The complexity and numerical issues
	of such iterated transformations have not been studied
	but appear formidable.
	See Revol \cite{revol:affine-iter-wrapping:22} for
	an analysis of the special case of iterating
	a fixed linear transformation. 
	\ignore{ Natalie Revol
			\cite{revol:affine-iter-wrapping:22}
		studies methods of controlling errors in
		interated linear transformations of the form
			$x_{n+1} \ass Ax_n + b$ for $n\ge 1$.
		She compares Lohner's QR approach 
		versus SVD approach.
		Revol's paper is in our refs folder.
	}%
	In principle, Lohner's transformation could be
	incorporated into our algorithm.  By not doing this,
	we illustrate the extent to which other techniques could
	be used to produce viable validated algorithms.
	
	In this paper, we introduce
	[N1] new methods to achieve variants of \stepA\ and \stepB,
	and [N2] data structures and subroutines to support
	more complex motifs than $(AB)^m$ above.
	Our algorithm is a synthesis of [N1]+[N2].
	The methods under [N1] will
	refine full- and end-enclosures by exploiting
	\lognorm\ and radical transforms
	(see next).
	Under [N2], we design subroutines
	and the scaffold data-structure to support new algorithmic
		motifs such as $(AB^+)^m$, i.e.,
	$A$ followed by one or more $B$'s.
	Moreover, $B^+$ is periodically replaced
	by calling a special ``\SubrSeven'' to refine end-enclosures.
	This will be a key to our termination proof.
\ssect{How to exploit Logarithmic Norm with Radical Transform}
	A \dt{logNorm bound} of $B_1\ib \RR^n$ is any upper bound on
		\beql{lognormbound}
		\mu_2(J_\bff(B_1))\as \sup \set{
					\mu_2(J_\bff(\bfp)) : {\bfp\in B_1}}
		\eeql
	where $J_\bff$ is the Jacobian of $\bff$
	and $\mu_2$ is a \lognorm\ function
	(\refSSec{lognorm}).
	Unlike standard operator norms,
	\lognorm s can be negative.
	We call $B_1$ a \dt{contraction zone} if it has a
	negative \lognorm\ bound.
	We exploit the fact that
		$$\|\bfx(t;\bfp_0)-\bfx(t;\bfp_1)\|
			\le \|\bfp_0-\bfp_1\| e^{t\olmu}$$
	(\refThm{ne} in \refSSec{lognorm} where $\olmu$ is
	a logNorm bound).
	In the Volterra example in \refFig{Volterra-21-13}, it can be
	shown that the exact contraction zone is the
	region above the green parabola.  In tracing a solution
	$\bfx(t;\bfp_0)$ for $t\in [0,h]$ through a contraction zone,
	we can compute a end-enclosure $B$
	for $\ivp(B_0,h,B_1)$ with $\wmax(B)<\wmax(B_0)$
	(i.e., the end-enclosure is ``shrinking'').
	\ignore{
	This is superior to direct Taylor bounds such as
				$\bfx(t; B_0)= B_0+ \bff(B_0)t + \cdots$
		which cannot shrink.  But
	\bw{direct-method may shrink. }
	\chee{thanks for pointing out that the "direct"
		method can shrink, and, in the limit,
		it is related to the C1-Method of CAPD,
		which can lead to lognorms.
	}%
	}
	Previous authors have exploited \lognorm s in the IVP problem
	(e.g., Zgliczynski \cite{zgliczynski:lohner:02},
	Neumaier \cite{neumaier:theoryI:94}).
	We will exploit it in new way via a transform:
	for any box $B_1\ib\RR^n$,
	we introduce a ``radical map'' $\pi:\RR^n\to\RR^n$
	(\refSec{xform}) with $\bfy=\pi(\bfx)$.
	Essentially, this transform is
			\beql{essentially}
				\bfy=(x_1^{-d_1}\dd x_n^{-d_n}) \qquad
					\text{(for some $d_1\dd d_n\ne 0$)}
			\eeql
	where $\bfx=(x_1\dd x_n)$.
	The system $\bfx'=\bff(\bfx)$ transforms to
	another system $\bfy'=\bfg(\bfy)$ in which the \lognorm\
	of $\pi(B_1)$ has certain properties (e.g., $\pi(B_1)$ is
	a contraction zone in the $(\bfy,\bfg)$-space).
	By computing end-enclosures 
	in the $(\bfy,\bfg)$-space, we infer a corresponding end-enclosure
	in the $(\bfx,\bff)$-space.  Our analysis of the
	1-dimensional case (\refSSec{comparison}),
	suggests that the best bounds are obtained when the \lognorm\ of
	$\pi(B_1)$ is close to $0$.
	In our current code, computing $\pi(B_1)$ is expensive,
	and so we avoid doing a transform if $\mu_2(J_\bff(B_1))$ is
	already negative.

\ssect{Brief Literature Review} 
	The validated IVP literature appeared almost from the start of
	interval analysis, pioneered
		by Moore, Eijgenraam, Rihm and others
	\cite{eijgenraam:ivp:81,moore:autoAnalysis:65,moore:bk,
		rihm:interval-ivp:94}.
	Corliss \cite{corliss:survey-ode-intvl:89} surveys this early
		period.
	Approaches based on Taylor expansion is dominant as they benefit
		from techniques such automatic differentiation
		and data structures such as the \dt{Taylor model}.  The latter,
		developed and popularized by Makino and Berz
		\cite{ bunger:taylorODE:20,
			bunger:taylor-precond:21,makino-berz:taylorModels:17},
		has proven to be very effective. 
	A major activity is the development of techniques
		to control the ``wrapping effect''.
		Here Lohner's approach \cite{adams:enclosure:87,lohner:thesis}
		has been most influential. 
		Another advancement is the $C^r$-Lohner method
		developed by Zgliczyński et al.
		\cite{zgliczynski:lohner:02,walawska-wilczak:ode:16}.
		This approach involves 
		solving auxiliary IVP systems to estimate
		higher order terms in the Taylor expansion.
		The field of validated methods, including IVP,
		underwent great development in the decades of 1980-2000.
		Nedialkov et al provide an excellent survey of
		the various subroutines of validated IVP
		\cite{jackson+nedialkov:advances-ode:02, nedialkov:thesis:99,
		nedialkov-jackson-pryce:HOI:01, nedialkov+2:validated-ode:99}.
	
	In Nonlinear Control Theory
		(e.g., \cite{fan+3:simulation-reach:18,scott-barton:bounds:13}),
		the End-Enclosure Problem is studied under various
		\dt{Reachability} problems.
	In complexity theory, Ker-i Ko \cite{ko:real:bk} has shown
		that IVP is PSPACE-complete.
		This result makes the very strong assumption
		that the search space is the unit square ($n=1$).
		Bournez et al \cite{bournez+2:ode-unbounded:11}
		avoided this restriction by assuming that $\bff$
		has analytic extension  to $\CC^d$.
	In this paper, we	
	adopt the ``naive'' but standard view of 
	real computation throughout in order to avoid 
	discussing arbitrary precision computation as in \cite{ko:real:bk};
	however all our methods only depend on interval bounds
	and full rigor can be achieved
	(cf.~the AIE pathway in \cite{xu-yap:real-zero-system:19}).
		
	The concept of \dt{logarithmic norm}\footnote{
			This concept goes by other names, including
				{logarithmic derivative},
				{matrix measure} and
				{Lozinski\u{i} measure}.
		}
		(or \dt{\lognorm} for short) was independently introduced
		by Germund Dahlquist and Sergei M.~Lozinski\u{i} in 1958
		\cite{soderlind:log-norm-history}.
		The key motivation was to improve bound errors in IVP.
		Neumaier \cite{neumaier:theoryI:94}
		is one of the first to use \lognorm s in validated IVP.
		The earliest survey is
		T.~Str\"om (1975) \cite{strom:log-norm:75}.
		\savespace{
			M.~Vidyasagar (1978) \cite{vidyasagar:matrix-measure:78}
			introduced a modification of matrix measure which 
			has most of the properties
			needed in the applications. 
		}
		The survey of Gustaf S\"{o}derlind
		\cite{soderlind:log-norm-history}
		extends the classical theory of \lognorm s to the general
		setting of {functional analytic} via Banach spaces.
	
	
	One of the barriers to the validated IVP literature
	is cumbersome notations and lack of precise
	input/output criteria for algorithms.
		For instance, in the $A/B$ algorithms, it is not stated
		if a target time $h>0$ is given (if given, how 
		it is used other than to terminate).
		Algorithm 5.3.1 in \cite{nedialkov+2:validated-ode:99}
		is a form of \stepA\ has a $\veps>0$ argument
		but how it constrains the output is unclear; so it is
		unclear how to use this argument in the A/B algorithm.
	We provide a streamlined notation, largely by focusing
	on autonomous ODEs, and by introducing
	high-level data structures such as the scaffold.
	Besides non-halting and unclear input/output
	specification, another issue is the use of
	``failure modes''
	(e.g., \cite[p.458, Figure 1]{nedialkov-jackson-pryce:HOI:01}).
	Typical failure modes have no clear semantics
	but appear as an artifact of the implementation.
	
\ssect{Paper Overview}
	The remainder of the paper is organized as follows: 
	\dt{Section 2} introduces some key concepts and computational tools. 
	\dt{Section 3} gives an overview of our algorithm.
	\dt{Section 4} describes our \stepA\ and \stepB\ subroutines.
	\dt{Section 5} describes our transform approach to obtain
		tighter enclosures.
	\dt{Section 6} describes the \Extend\ and \Refine\ subroutines.
	\dt{Section 7} presents our main algorithm and some global
			experiments.
	We conclude in \dt{Section 8}.
	\issacArxiv[
		\dt{Appendix A} gives some critical proofs. Due
		to space limitation, the remaining proofs
		in found in arXiv \cite{zhang-yap:ivp:25arxiv}.
	]{
		\dt{Appendix A} gives all the proofs.
		\dt{Appendix B} provide details of the affine
		transform $\olpi$.
	}

	\ignore{
	In Sections 8 and 9, we present the theoretical details of our
		refinement method, with Section 8 addressing the case where the
		logarithmic norm is negative and Section 9 covering the case where
		the logarithmic norm is non-negative. 
	Finally, in Section 10, we provide the algorithmic details of our
		method and prove that it is guaranteed to terminate.

	Appendix???
	}

%% file: inc/2tools.tex
\section{Basic Tools}
\ssectL[Notations]{Notations and Key Concepts}
\cheeX{ Jan'26: Decided to eliminate the column/row vector convention!
}
	We use bold fonts such as $\bfx=(x_1\dd x_n)$
	or $\bfp\in\RR^n$ for vectors.
	Vector--matrix or matrix--matrix products are indicated by
	$\Bigcdot$ (e.g., $A \Bigcdot \bfp$ or $A\Bigcdot B$).
	Let $\intbox\RR^n$ denote the set of \dt{boxes} in
	$\RR^n$ where each box $B$ is a Cartesian product
	$B=\prod_{i=1}^n I_i$ and $I_i=[a_i,b_i]$, $a_i\le b_i$.
	For any $S \subseteq \RR^n$, let $Box(S)$ and $Ball(S)$
	denote, respectively, the smallest box and the smallest
	Euclidean ball containing $S$.
	The $Box$ and $Ball$ operators can take numerical parameters:
	if $\bfr = (r_1 \dd r_n) \ge \0$ and $\bfp \in \RR^n$, then
	\[
	Box(\bfr) \as \prod_{i=1}^n [-r_i, r_i],
	\qquad
	Box_\bfp(\bfr) \as \bfp + Box(\bfr).
	\]
	We simply write $Box_\bfp(r)$ if $r_i = r$ for all $i$.
	For the box $B = Box_\bfp(\bfr)$, its \dt{midpoint} and \dt{width}
	are $\bfm(B) \as \bfp$ and $\bfw(B) \as 2\bfr$, respectively.
	Similarly, $Ball_\bfp(r)$ is the ball centered at $\bfp$
	of radius $r$; simply write ``$Ball(r)$''
	if $\bfp$ is the origin $\0$.
	The \dt{width} and \dt{midpoint} of
	an interval $I = [a,b]$ is 
	$w(I) \as b - a$ and $m(I) \as (a + b)/2$.
	The \dt{width} and \dt{mid point} of a box
	$B = \prod_{i=1}^n I_i$ is
			$\bfw(B) \as (w_1\dd w_n)$ and
			$\bfm(B) \as (m_1\dd m_n)$,
	where $w_i=w(I_i)$ and $m_i=m(I_i)$.
	The \dt{diameter} of $B$ is 
		$\|\bfw(B)\|_2=\sqrt{\sum_{i=1}^n w_i^2}$.
	It follows that $Ball(B)=Ball_{\bfm(B)}(r(B))$.
	where $r(B)$ be half the diameter of $B$.
	Also define the \dt{max-width} and \dt{min-width} of $B$ as
		$ \wmax(B) \as \max_{i=1}^n w(I_i)$,
		$\wmin(B) \as \min_{i=1}^n w(I_i)$.
	We use the Euclidean norm on $\RR^n$, writing 
	$\|\bfp\|=\|\bfp\|_2$.
	If $S\ib\RR^n$, then let $\|S\|\as \sup\set{\|\bfp\|: \bfp\in S}$.
	For any function $f: X \to Y$, we
	re-use `$f$' to denote its \dt{natural set extension},
	$f:2^X\to 2^Y$ where $2^X$ is the power set of $X$ and
	$f(S)=\set{f(x): x\in S}$ for all $S\ib X$.

	\ignore{
	Let $C^k([0, h] \to \RR^n)$ be
	the set of $C^k$-continuous functions ($k \geq 0$)
	from $[0, h]$ to $\RR^n$.
	Assume that the set of solutions
	$\ivp(B_0, h)$ is a subset of $C^\infty([0, h] \to \RR^n)$.

	Now, consider $\bfx \in C^2([0, h] \to \RR^n)$. If
	$\bfx \in \ivp(B_0, h)$,
	it satisfies the set of solutions to \refeq{bfx'},
	where any solution $\bfx$ to $\ivp(B_0, 1)$ belongs to
	$C^2([0, 1] \to \RR^n)$.
	In particular, $\bfx(t)$ is defined and bounded for all
	$t \in [0, 1]$. 
	However, $\ivp(B_0, 1)$ might not always have a solution.
	Specifically, if $\bfx(0) \in B_0$ leads to $\bfx(t)$ becoming
	undefined for some $t \in (0, 1]$, the system lacks a valid
	solution.

	E.g., $n=1$ then \refeq{bfx'} 
	 $x'=x^2$ has solution $x(t)= \efrac{x_0\inv -t}$, $x(0)=x_0$
	 to $\ivp(x_0,1)$.  If $B_0\ib [1,\infty)$ then $\ivp(B_0,1)$
	 has no solutions since $x(t)=\infty$ for
	 when $t=1/x_0 \in [0,1]$.  To avoid this issue, we assume
	 the promise problem in which 
	 $\ivp(\bfp_0, 1)$ is assumed to have
	 a solution for all $\bfp_0 \in B_0$.
	}%

	The \dt{image} of a function $f: A \to B$ is
		$\image(f) \as \set{f(a) : a \in A}.$  
	The \dt{image} of $\ivp(B_0, h)$ is the union
		$\bigcup_{\bfx \in \ivp(B_0, h)} \image(\bfx)$.
		%
	A \dt{full-enclosure} of $\ivp(B_0, h)$ is a
	set $B_1\ib\RR^n$ that contains
		$\image(\ivp(B_0, h))$.
	We call $(B_0,h,B_1)$ an \dt{admissible triple} if
	$(B_0,h)$ is valid and $B_1$ is a full-closure of $\ivp(B_0,h)$.
	In this case, we can also write $\ivp(B_0,h,B_1)$ for $\ivp(B_0,h)$,
	and call $(h,B_1)$ an \dt{admissible pair} for $B_0$.
	Finally, $\ulB_1\ib B_1$ is
	an \dt{end-enclosure} for $\ivp(B_0, h, B_1)$
	if for all solution $\bfx \in \ivp(B_0, h, B_1)$, 
	we have $\bfx(h) \in \ulB_1$.  Call $(B_0,h,B_1,\ulB_1)$
	an \dt{admissible quadruple} (or quad).

	\ignore{
	If $P$ is a problem or an algorithm, its header has
	the format 
		``$P(...inputs)\ssa (outputs...)$''.
	This is illustrated in the introduction by
	$\endEncIVP(B_0,\delta)\ssa (\ulB_0,\olB_1)$.
	Moreover, the exact relation between $(inputs..., outputs...)$
	will be explicitly stated.
	}
	
	\ccheeX{Please bring some of the notations of pap2 here.
	We want the notations of the 2 papers to be as consistent as possible.
		E.g., $Box(\bfeps)$ and $Box_\bfp(\bfeps)$.
	}
	\ignore{
	If \(\ivp(B_0, h)\) is valid, then under the assumption
	\(\bff \not\equiv \mathbf{0}\),
	we have the following: for any
	\(\bfx_0 \in B_0\), if \(\bfx(t)\)
	is a solution with \(\bfx(0) = \bfx_0\), then
	for all \(t \in [0, h)\), it holds that \(\bff(\bfx(t)) \neq
	\mathbf{0}\).
	}%
	
	\savespace{
	  Another example for $n=2$ is the system:  
	  \[
	  	\mmat{x'\\ y'} = \mmat{f_1\\ f_2}
			=\mmat{ x + \frac{\sqrt{e}}{\sqrt{e} - 1}\\
	  			x - y + \frac{\sqrt{e} + 1}{2\sqrt{e} - 2}}  
	  \]  
	  with the initial condition $[x, y] = [1, 1]$. One can check that
	  $f_1=f_2=0$ when $t=0.5$.
	}

	\savespace{	
		It is not strictly necessary to assume $\ivp(B_0, 1)$ is
		valid. It is enough
		to know a point $\bfp_0$ in the interior of $B_0$
		such that $\ivp(\bfp_0,1)$ is valid.
		Our current algorithm can be defined to shrink
		towards $\bfp_0$, and halting can be assured.
	}
\ssectL[implicit]{Implicit use of Interval Computation}
	For any $\bfg:\RR^n\to\RR$, a \dt{box form} of $\bfg$
	is any function $\bfG:\intbox\RR^n\to\intbox\RR$
	which is (1) conservative, and (2) convergent.
	This means
	(1) $\bfg(B)\ib\bfG(B)$ for all $B\in\intbox\RR^n$,
	and (2)
	$\bfg(\bfp)= \lim_{i\to\infty} \bfG(B_i)$ for any
	infinite sequence $B_1, B_2, B_3, \ldots$ that converges to 
	a point $\bfp\in\RR^n$.  In some proofs,
	it may appear that we need the
	additional condition, (3) that $\bfG$ is
	\dt{isotone}. This means $B\ib B'$ implies $\bfG(B)\ib\bfG(B')$.
	In practice, isotony can often be avoided.

	We normally denote a box form $G$ of $\bfg$ by $\intbox\bfg$
	(if necessary, adding subscripts or superscripts to distinguish
	various box forms of $\bfg$).  See
	\cite{hormann-kania-yap:range:21,hormann-yap-zhang:hermite:23}.
	In this paper, we will often ``compute'' exact bounds
	such as ``$\bff(E_0)$'' (e.g., in \stepA, \refSSec{stepA}).
	But in implementation,
	we really compute a box form $\intbox\bff(E_0)$.  In the
	interest of clarity, we do not explicitly
	write $\intbox\bff$ since the mathematical function
	$\bff(E_0)$ is clearer.  
	\ignore{
	We adopt the ``naive'' but standard view of 
	real computation throughout in order to avoid 
	discussing arbitrary precision computation;
	but see discussions of the AIE pathway to full rigor
	in \cite{xu-yap:real-zero-system:19}.
	}
\ssect{Normalized Taylor Coefficients}
	For any solution $\bfx$ to the
	ODE \refeq{bfx'}, its $i$th \dt{normalized Taylor coefficient}
	is recursively defined as follows:
		\beql{normalizedTaylorCoef}
			\bff\supn[i](\bfx) =
					\clauses{ \bfx & \rmif\ i=0,\\
						\efrac{i} \Big( J_{\bff\supn[i-1]} \Bigcdot
						\bff\Big)(\bfx) & \rmif\ i\ge 1}
		\eeql
	where $J_\bfg$ denotes the Jacobian of any function
	$\bfg=\bfg(\bfx)\in C^1(\RR^n\to\RR^n)$
	in the variable $\bfx=(x_1\dd x_n)$.
	For instance,
		$\bff\supn[1] = \bff$
	and
		$\bff\supn[2](\bfx) = \half (J_\bff\cdot \bff)(\bfx).$
	It follows that the order $k\ge 1$ Taylor expansion of $\bfx$
	at the point $t=t_0$ is
		\beql{taylorExpand} 
			\bfx(t_0+h) = \Big\{
				\sum_{i=0}^{k-1} h^i \bff\supn[i](\bfx(t_0)) \Big\}
					+ h^k \bff\supn[k](\bfx(\xi))
		\eeql
	where $0\le \xi-t_0 \le h$. 
	If $\bfx(\xi)$ lies in a box $B\in\intbox\RR^n$,
	then interval form is
		\beql{taylor}
			\bfx(t_0+h) \in \Big\{
				\sum_{i=0}^{k-1} h^i \bff\supn[i](\bfx(t_0)) \Big\}
					+ h^k \bff\supn[k](B) \eeql
	It is well-known that such Taylor coefficients
	can be automatically generated, and they can be evaluated
	at interval values using automatic
	differentiation. 

	\savespace{
	\dt{Running Example}:
		Let $\bff = \mmat{x^2+1\\ -y^2+7x}$.
		Then
			{\small \beqarrys
			\bff\supn[1] &=& \bff(\bfx)\\
			\bff\supn[2] &=& \half (J_\bff\cdot \bff)(x,y)\\
				&=& \half \mmat{2x(x^2+1)\\ 7(x^2+1-2yx)+2y^3}\\
			\bff\supn[3] &=& \frac{1}{3}(J_\bff\supn[2]\cdot \bff)(x,y)\\
			&=& \frac{1}{3} \mmat{(3x^2+1)(x^2+1)\\
					-3y^4+7(x^3-x^2y+4xy^2-7x^2+x-y)}.
			\eeqarrys
			}
	}
\ssect{Banach Space $X_h$} 
	We give a brief summary of the basic theory.
	If $X,Y$ are topological spaces, let $C^k(X\to Y)$ ($k\ge 0$)
	denote the set of $C^k$-continuous functions
	from $X$ to $Y$.
	{\em We fix $\bff\in C^k(\RR^n\to\RR^n)$ throughout the paper,
	and thus $k\ge 1$ is a global constant.}
		It follows that $\ivp_\bff(B_0,h) \ib C^k([0, h] \to \RR^n)$.
	Let $X_h \as C^k([0,h]\to \RR^n)$.
		Then $X_h$ is a real linear space where
		$c\in\RR$ and $\bfx,\bfy\in X_h$ implies
		$c\bfy\in X_h$ and
		$\bfx\pm \bfy\in X_h$.
		Let $\0\in X_h$ denote the additive identity
		in $X_h$: $\bfx\pm \0=\bfx$.
		Next $X_h$ becomes a normed space with norm 
			$\|\bfx\|_{\max} \as
			\max_{t\in [0,h]} \|\bfx(t)\|_{2}$
		where $\|\cdot\|_{2}$ is the $2$-norm.
		If $S\ib X_h$, we let
			$\|S\|_{\max}\as \sup_{\bfx\in S}\|\bfx\|_{\max}$.
		We turn $X_h$ into a complete metric space $(X_h,d)$
		with metric $d(x,y)= \|x-y\|_{\max}$.
		Thus, $X_h$ is a Banach space.
		To prove existence and uniqueness of solutions in $X_h$,
		we need a compact subset $Y_h\ib X_h$.  Let
	$Y_h=Y_h(\bfp_0,r)\as \set{\bfx\in C^1([0,h]\to
			Ball_{\bfp_0}(r)): \bfx(0)=\bfp_0}$.
		Then $Y_h$ is also a complete metric space induced by $X_h$.
		The \dt{Picard operator} on $T_\bff: X_h\to X_h$, is given by
		$T_\bff[\bfx](t) = \bfx(0)+\int_0^t \bff(\bfx(s))ds$
		for $\bfx\in X_h$ and $t\in [0,h]$.  We have 2 basic results:

		\begin{description}
		\item (M) If $h<r/M$ where
		$M=\sup\set{\|\bff(\bfp)\|_2: \bfp\in Ball_{\bfp_0}(r)}$,
		then $T_\bff[Y_h]\ib Y_h$, i.e., $T_\bff$ is an endomorphism
		of $Y_h$.
	\bwX{	\item (M) If $h<r/M$ where
			$M=\sup\set{\|\bff(\bfp)\|_2: \bfp\in Ball_{\bfp_0}(r)}$,
			then $T[Y_h]\ib Y_h$, i.e., $T$ is an endomorphism
			of $Y_h$.}
		\item (L) If we also have $h<L\inv$ where
		$L$ is a Lipshitz constant for $\bff$ on the ball
		$Ball_{\bfp_0}(r)$, then $T_\bff$
		is a contraction map on $Y_h$, i.e.,
		$\|T_\bff[\bfx]-T_\bff[\bfy]\|_{\max}<\|\bfx-\bfy\|_{\max}$.
		\end{description}

		It follows from (M) and (L) that there is unique
		$\bfy^*\in Y_h$ such that $T_\bff[\bfy^*]=\bfy^*$, i.e.,
		$\bfy^*$ is the unique fixed point of $T_\bff$.
		All this Banach space theory finally leads to the
		central result of IVP:

	\bthmT[pl]{Picard-Lindel\"of Theorem}
		Let $B=Ball_{\bfp_0}(r)\ib \RR^n$ be the ball
		centered at $\bfp_0$ and $r>0$.
		Then there exists $h>0$ such
			$\ivp(\bfp_0,h,B)$ is valid.
		In fact, it is sufficient to choose
			$h\le \min\set{r/M, 1/L}$.
	\bwX{In fact, it is sufficient to choose $h\le \min\set{r/M, 1/L}$.}
	\ethmT

	For our algorithmic development,
	we need  ``constructive'' tools to ensure the conditions
	of the Picard-Lindel\"of theorem.
	The following is the key tool:
	
	\bthmT[admiss]{Admissible Triple} \
		Let $\bff\in C^k(\RR^n\to\RR^n)$, $k\ge 1$ and $h>0$.
		Let $F\ib \RR^n$ be a compact convex set.
		If $E_0$ is contained in the interior of $F_1$ and
		satisfies the inclusion
		\beql{tay}
			\sum_{i=0}^{k-1}
		 	[0,h]^i \bff^{[i]}(E_0)+[0,h]^k \bff^{[k]}(F_1)
				\subseteq F_1,
		 \eeql
		then $(E_0,h,F_1)$ is an admissible triple.
	\ethmT
	\ccheeX{Bw, note that I changed $k\ge 2$ to $k\ge 1$.
		Check that this is still correct? 
	}
	Note that our result is very similar to the
	High-Order Enclosure Method (HOE) in
	\cite[Theorem 4.1]{nedialkov-jackson-pryce:HOI:01}
	who quoted 
	\cite[Theorem 3]{corliss-rihm:hi-order-enclosure:96}.
	We give a self-contained proof in the appendix.
\ssectL[lognorm]{Logarithmic norms and Euler-tube Method}
	Let $\|A\|_p$ be the induced $p$-norm of a $n\times n$ matrix $A$
	with complex entries.
	Then the \dt{logarithmic $p$-norm} of $A$ is defined as
		\[\mu_p(A) \as \lim_{h\to 0+}\frac{\|I+hA\|_p-1}{h}.\]
	We shall focus on $p=2$, and call $\mu_2$ the \lognorm.
	E.g., if $n=1$ then $A=a\in\CC$ and $\mu_p(A)=\Re(a)$.
	%
	%
	%
	%
	%
	%
	%
	We have these bounds for logarithmic $p$-norms:

	\bleml[lognorm] \ 

	\benum[(a)]
	\item
		$\mu_p(A+B)\le \mu_p(A)+\mu_p(B)$
	\item
		$\mu_p(A)\le \|A\|_p$
	\item
		$\mu_2(A)=\max_{j=1\dd k}(\frac{1}{2}(\lambda_{j} (A+A^T)))$
			where $\lambda_1(A)\dd \lambda_k(A)$
			is the set of eigenvalues of $A$.
	\item
		Let $A$ be an $n\times n$ matrix  and let
		$\max_{i=1}^n (\Re(\lambda_i))$ where
		$\lambda_i's$ are the eigenvalues of $A$.
		Then
		\bitem
			\item $\max_{i=1}^n (\Re(\lambda_i))\le
					\mu(A)$ holds for any \lognorm.
			\item For any $\veps \ge 0$, there exists an
				invertible matrix $P$ such that 
			\[\max_i(Re(\lambda_i))\le \mu_{2,P}(A)\le
				\max_i(Re(\lambda_i))+\veps.\]
			where $\mu_{2,P}(A)\as \mu_2(P^{-1}AP).$
		\eitem
	\eenum
	\eleml
	For parts(a-c) see
		\cite{desoer-haneda:measure:72},
	and part(d), see
		Pao \cite{pao:log-der-matrix:73}.
	In some estimates, we use these standard bounds:
	\beql{matrixnorm}
		\grouping[l l c l ]{
		\|A\|_2 &=&		\max_i(\sqrt{\lambda_{i}(A^*A)})\\
		\|AB\|_2 &\le&	\|A\|_2\|B\|_2
		}
	\eeql
	\savespace{
		We will use these standard bounds on norms.
		\bleml[matrixnorm] \
	
		\benum[(a)]
		\item
			$\|A\|_2=\max_i(\sqrt{\lambda_{i}(A^*A)})$
		\item $\|AB\|_2\le \|A\|_2\|B\|_2$
		\eenum
		\eleml
	}

	%
	
	We have the following result from Neumaier
	\citep[Corollary 4.5, p.~331]{neumaier:theoryI:94}
	(also \cite[Theorem I.10.6]{hairer+2:ode1:bk}):

\ccheeX{Nov'25: FIXED!  We made mistakes
		in citing this theorem of Neumaier!!!
		(look at the arxiv version to see what went wrong).
		Bingwei, PLEASE, YOU MUST CHECK THESE THINGS.
	}

	\bthmT[ne]{Neumaier} \

		Let $\bfx\in \ivp_\bff(\bfp_0,h)$
		and $\xi(t)\in C^1([0,h]\to \mathbb{R}^n)$
		be any ``approximate solution''.

		Let\footnote{
			For our purposes, matrix $P$ in this theorem can be the
			identity matrix.
		}
		$P$ be an invertible matrix.
		Assume the constants $\veps,\delta,\olmu$ satisfy

		\begin{enumerate}
			\item 
				$\veps \ge
					\|P\inv\Bigcdot(\xi'(t) - \bff(\bfx(t)))\|_2$
				for all $t \in [0, h]$ 
			\item
				$\delta \ge 
					\|P\inv\Bigcdot(\xi(0)-\bfx(0))\|_2 $
			\item 
				$\olmu \ge
					\mu_2  \big(P\inv\Bigcdot J_{\bff}(s\bfx(t) +
					(1-s)\xi(t))\Bigcdot P\big)$
				for all $s \in [0,1]$ and $t \in [0, h]$
		\end{enumerate} 
		Then for all $t \in [0, h]$,
		\beql{xibfx}
			\|P\inv\Bigcdot(\xi(t) - \bfx(t))\|_2 \le
				\begin{cases}
			\delta e^{\olmu t} + \frac{\veps}{\olmu}(e^{\olmu t} - 1),
					& \olmu \ne 0, \\
			\delta + \veps t, & \olmu = 0.
				\end{cases}
		\eeql
	\ethmT

	\bcorl[cor-1] \ 
		Let $\bfx_i\in \ivp(Ball_{\bfp_0}(r),h,F)$ for $i=1,2$
	and $\olmu\ge \mu_2(J_\bff(F))$.
	\benum[(a)]
	\item
	For all $t\in [0,h]$,
	\beql{bfx12appendix}
	\|\bfx_1(t)-\bfx_2(t)\|_2
	\le \|\bfx_1(0)-\bfx_2(0)\|_2 e^{\olmu t}.\eeql 
	\item
	If $\bfx_1(0)=\bfp_0$ then
	an end-enclosure for $\ivp(Ball_{\bfp_0}(r),h,F)$ is
	$$Ball_{\bfx_1(h)}(re^{\olmu h}).$$
	\eenum
	\ecorl

\begin{figure}
		\centering
		\includegraphics[width=0.4\linewidth]{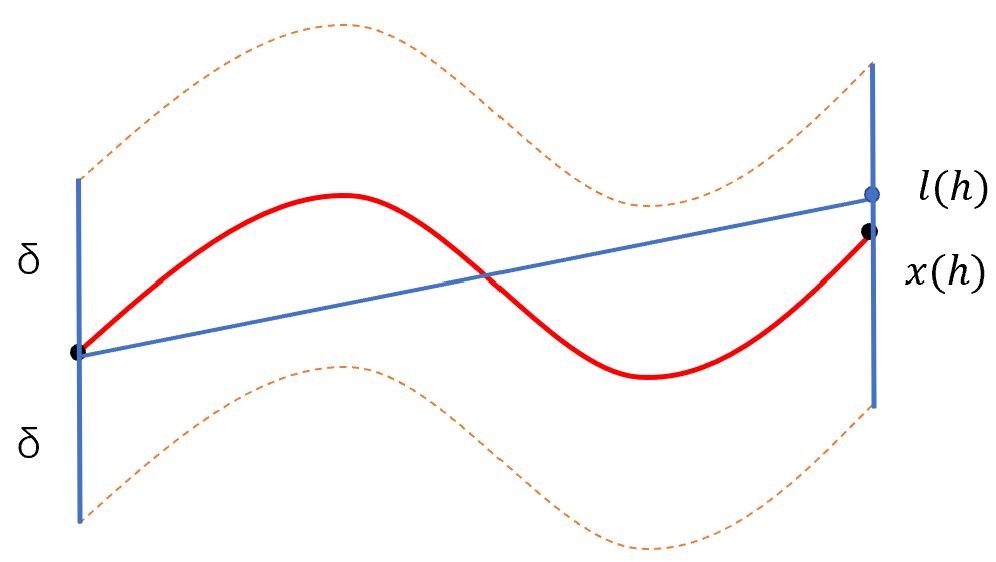}
		\caption{The dashed lines in the figure form a \(\delta\)-tube
		around the red solid curve representing \(\bfx(t)\). The segment
		\(l(t)\) is a line segment inside this \(\delta\)-tube.}
		\label{fig:tube}
	\end{figure}

	\dt{Euler-tube Method:}
	For any $\bfx \in \ivp(E_0, h)$ and
	$\delta>0$, the \dt{$\delta$-tube} of $\bfx$ is
	the set
		$$\Tube_\delta(\bfx) \as
				\set{(t,\bfp): \|\bfp-\bfx(t)\|_2 \le \delta, 0\le t\le h}
				\quad \big(\ib [0,h]\times\RR^n\big)$$
	We say that a function $\ell: [0,h]\to\RR^n$ belongs to the
	$\delta$-tube of $\bfx$ is for all $t\in[0,h]$,
	$(t,\ell(t)) \in \Tube_\delta(\bfx)$,see graph \ref{fig:tube} for
	illustration.
	
	\blemT[eulerStep]{Euler Tube Method}\ 

		Let $(B_0,H,B_1)$ be admissible triple,
	$\olmu\ge \mu_2(J_{\bff}(B_1))$ and
	$ \olM\ge \|\bff^{[2]}(B_1)\|$.
	
	Consider the polygonal
	path $\bfQ_{h_1}=(\bfq_0,\bfq_1\dd \bfq_m)$ 
	from the Euler method with $m$ steps of uniform step-size $h_1$
	given by 
		\beql{h1}
			h_1= h\euler(H,\olM,\olmu,\delta) \as
			\begin{cases}
		\min\set{H, \frac{2\olmu\delta}
			{\olM \cdot (e^{\olmu H}-1)}}
		&\rmif\ \olmu\ge 0\\
		\min\set{H, \frac{2\olmu\delta}
			{\olM \cdot (e^{\olmu H}-1)-\olmu^2\delta}}
		&\rmif\ \olmu<0.
	\end{cases}
	\eeql

	If each $\bfq_i\in B_1$ ($i=0\dd m$) then the path $\bfQ_{h_1}$
	lies inside the $\delta$-tube of $\bfx(t;\bfq_0)$.
	\\ In other words, for all $t\in [0,H]$, we have
		\beql{eulerBd}
		\|\bfQ_{h_1}(t)-\bfx(t;\bfq_0)\|\le \delta.
		\eeql
	\elemT

	This lemma gives us a technique to
	refine end- and full-enclosures
	(see \refLem{delta-distance} below).

%% file: inc/3overview.tex
\sect{Overview of our Algorithm}
	We will develop an algorithm for the End-Enclosure
	Problem \refeq{endEncProb}, by elaborating
	on the classic Euler method or corrector-predictor
	framework for homotopy path 
	(e.g., \cite{sommese+2:intro-nag:10,xu-burr-yap:homotopy:18}).
	The basic motif is to 
	repeatedly call two subroutines\footnote{
		Nedialkov et al.~\cite{nedialkov+2:validated-ode:99}
		call them Algorithms I and II.
	}
	which we call \stepA\ and \stepB, respectively:
	\savespace{They correspond roughly
	to predictor/corrector steps of homotopy path methods.
	}
	
	\beql{stepA-}
		\hspace*{30mm}
			\includegraphics[width=0.45\columnwidth]{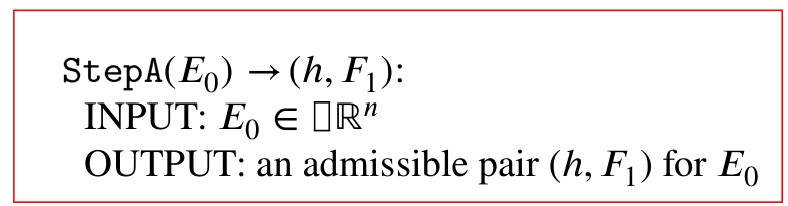}
	\eeql
	\beql{stepB}
		\hspace*{30mm}
			\includegraphics[width=0.45\columnwidth]{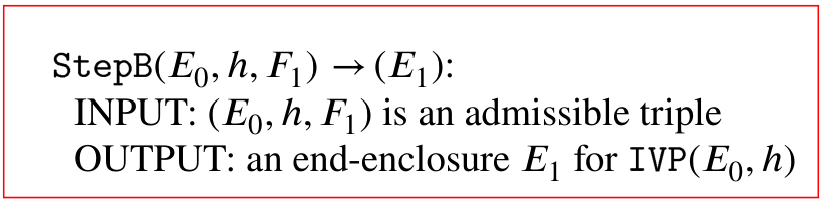}
	\eeql
	
	\ignore{
	\Ldent\progb{	
		\lline[-2] $\stepA(E_0)\ssa (h,F_1)$:
		\lline[0] INPUT: $E_0\in\intbox\RR^n$
		\lline[0] OUTPUT:  an admissible pair $(h,F_1)$
							for $E_0$
	}
	\Ldent\progb{	
		\lline[-2] $\stepB(E_0,h,F_1)\ssa (E_1)$:
		\lline[0] INPUT: $(E_0,h,F_1)$ is an admissible triple
		\lline[0] OUTPUT:  an end-enclosure $E_1$
							for $\ivp(E_0,h)$
	}
	}%
\DeclareRobustCommand{\loongrightarrow}{%
  	\DOTSB \relbar\joinrel \relbar\joinrel
  		\relbar\joinrel \relbar\joinrel
		\rightarrow }
\DeclareRobustCommand{\looongrightarrow}{%
  \DOTSB \relbar\joinrel \relbar\joinrel
  		 \relbar\joinrel \relbar\joinrel
  		 \relbar\joinrel \relbar\joinrel \rightarrow }
	
	Thus we see this progression
		\beql{stepAstepB}
			E_0 \overset{\stepA}{\looongrightarrow}
			(E_0,h_0,F_1) \overset{\stepB}{\looongrightarrow}
			(E_0,h_0,F_1,E_1) 
		\eeql
	where $\stepA$ and $\stepB$ successively
	transforms $E_0$ to an admissible triple and quad.
	By iterating \refeq{stepAstepB} with $E_1$ we can get to
	the next quad $(E_1, h_1, F_2,E_2)$, and so on.  {\em This is
	the basis of most validated IVP algorithms.}
	We encode this as:

	\renewcommand{\alt}[2]{#2} 
	\renewcommand{\alt}[2]{#1} 
	\alt{ 
		\beql{simpleivp}
			\hspace*{30mm}
			\includegraphics[width=0.45\columnwidth]{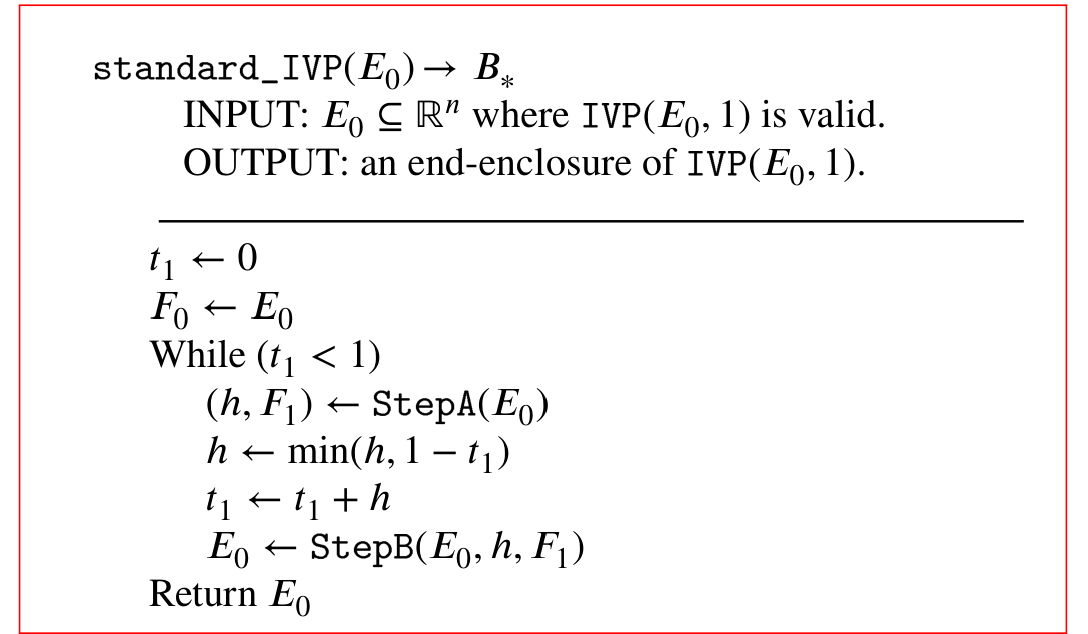}
		\eeql
	}{ 
		\Ldent\progb{
		\lline[0] \standardIVP($E_0$)$\ssa B_*$
		\lline[8]   INPUT: $E_0\ib\RR^n$
									where $\ivp(E_0,1)$ is valid.
		\lline[8]   OUTPUT: an end-enclosure
						of $\ivp(E_0,1)$.
		\lline[5] \myhlineX[3]{0.6}
		\lline[5] $t_1\ass 0$
		\lline[5] $F_0\ass E_0$
		\lline[5] While $(t_1<1)$
		\lline[10] 	$(h,F_1)\ass \stepA(E_0)$
		\lline[10] 	$h\ass \min(h, 1-t_1)$
		\lline[10] 	$t_1\ass t_1+h$
		\lline[10] 	$E_0 \ass \stepB(E_0, h,F_1)$
		\lline[5] Return $E_0$
		}
	}
	
	Note that the iteration of \refeq{stepAstepB}
	above is not guaranteed to halt (i.e., to reach $t=1$).
	Towards ensuring halting, we will strengthen $\stepA$
	with this concept:
	let $\bfveps=(\veps_1\dd \veps_n)\ge 0$.
	Call $(h,F_1)$
	an \dt{$\bfveps$-admissible pair} for $E_0$ if
	$(h,F_1)$ is an admissible pair for $E_0$ and
		\beql{bfveps-admiss}
			\veps_i \ge \sup_{\bfp \in F_1} | (\bff^{[k]}(\bfp))_i |,
				\qquad (i=1\dd n)
		\eeql
	where $(\bfx)_i$ is the $i$th component of a vector $\bfx$.
	As usual, $(E_0,h,F_1)$ is a $\bfveps$-admissible
	triple iff $(h,F_1)$ is a $\bfveps$-admissible pair for $E_0$.
	When $\veps_i=\veps$ for all $i$, then we simply
	say $\veps$-admissible pair/triple.
	See \refThm{admiss} for the context of this definition.
	We modify $\stepA$ to produce $\bfveps$-admissible pairs:

		\cheeX{Nov'25: Update StepA to reflect the vector $\bfveps$.
	Propagate to rest of paper!}

	\renewcommand{\alt}[2]{#2} 
	\renewcommand{\alt}[2]{#1} 
	\alt{ 
	\beql{stepA}
		\hspace*{30mm}
		\includegraphics[width=0.45\columnwidth]{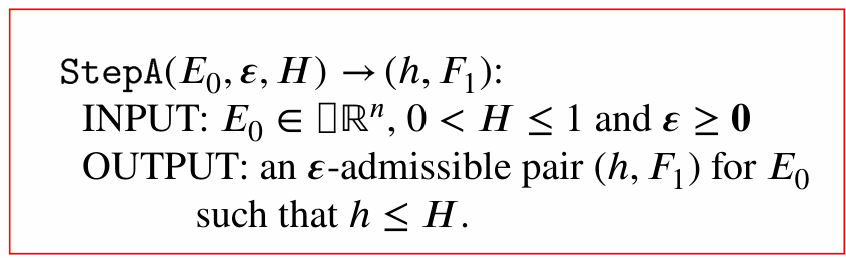}
	\eeql
	}{
	\Ldent\progb{	
		\lline[-2] $\stepA(E_0,\bfveps,H)\ssa (h,F_1)$:
		\lline[0] INPUT: $E_0\in\intbox\RR^n$,
					$0<H\le 1$ and $\bfveps=(\veps_1\dd \veps_n)\ge \0$
		\lline[0] OUTPUT:  an $\bfveps$-admissible pair $(h,F_1)$
							for $E_0$
		\lline[10]		such that $h\le H$.
	}
	}

\begin{figure}
		\begin{center}
	        \includegraphics[width=0.6\columnwidth]{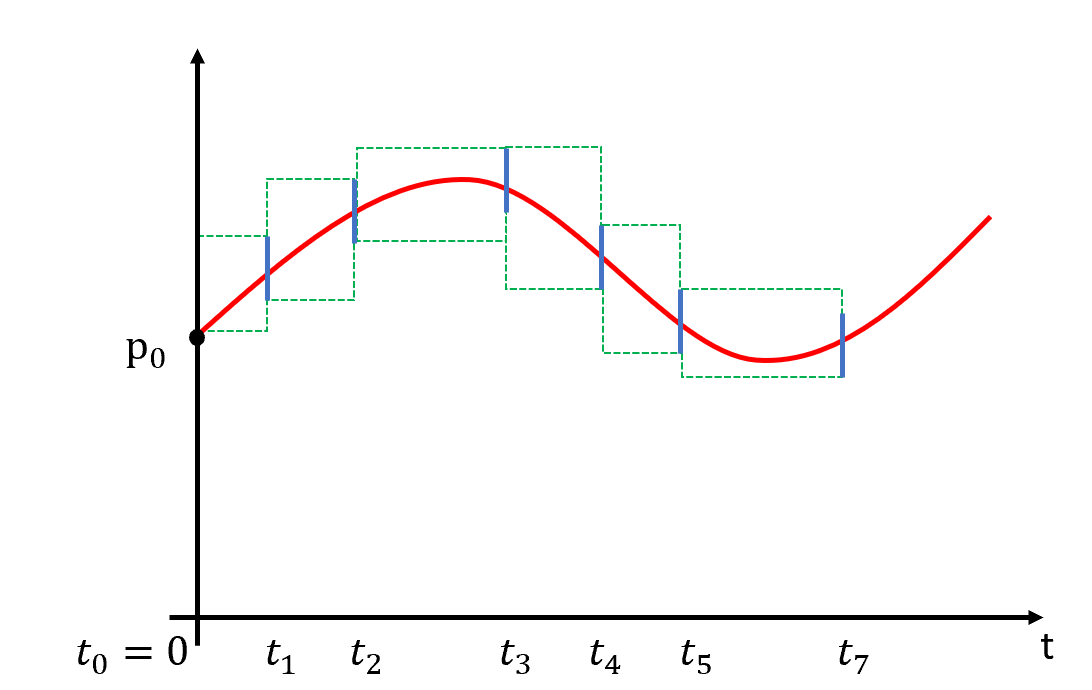}
		\caption{ A 7-step scaffold. The horizontal
			axis represents time, and the vertical axis represents
			$\RR^n$. The red curve corresponds to $\bfx(t)$, the blue
			line segments represent end-enclosures, and the green boxes,
			represent full-enclosures.}
		\label{fig:explain-D}
		\end{center}
		\end{figure}

\ssectL[scaffold]{Scaffold Data Structure}
	To go beyond the simple AB-iteration of \refeq{simpleivp}, we
	introduce a data structure called a ``scaffold'' to encode the
	intermediate information needed for this computation.
	\refFig{explain-D} shows such a scaffold.

	A \dt{scaffold} with $m\ge 1$ \dt{stages} is a quad
		$\stage =(\bft,\bfE,\bfF,\ministep)$
	where
		$\bft= (0\le t_0 < t_1 < \cdots < t_m)$,
		$\bfE = (E_0 \dd E_m)$, $\bfF = (F_1 \dd F_m)$ and
		$\ministep = (G_1\dd G_m)$
		such that for all $i=1\dd m$,
		\beql{Ei-1}
			(E_{i-1}, t_i-t_{i-1}, F_i,E_i)
			\text{ is an admissible quad}
			\eeql
		and $G_i$ is the $i$th \dt{refinement substructure}
		whose discussion is deferred to \refSec{extend-refine}.
		But basically, the time span $[t_{i-1},t_i]$ will be
		refined into many mini-steps of uniform step sizes
		and $G_i$ encodes this data.
	The quad in \refeq{Ei-1}
	is called the $i$th \dt{quad} of $\stage$, also denoted
		$$(E_{i-1}(\stage), \Delta t_i(\stage), F_i(\stage),
				E_i(\stage)).$$
	The \dt{end-} and \dt{full-enclosure} of $\stage$ is $E_m$ and $F_m$
	(respectively).
	The \dt{time sequence} and \dt{time-span} of $\stage$
	refer to $\bft$ and $[t_0,t_m]$, respectively.

	\dt{Observation}:
		For each $i=0\dd m$,
		$\bfE[i]$ is an end enclosure of $\ivp(\bfE[0],\bft[i])$.
	
	This observation is an immediate consequence of the
	admissibility of \refeq{Ei-1}.
	
	
	Let $\stage$ and $\stage'$ be scaffolds
	with $m$ and $m'$ stages respectively:
	\bitem
	\item 
	Call $\stage'$ an \dt{extension} of $\stage$ if $\stage$ is 
	a prefix of $\stage'$. If $m'=m+1$, then $\stage'$ is
	a \dt{simple extension}.
	\item
	Call $\stage'$ a \dt{refinement} of $\stage$ if
	they have the same time sequence $\bft(\stage)=\bft(\stage')$
	(so $m=m'$) and for each $i=1\dd m$, we have
	$E_i(\stage') \ib E_i(\stage)$, and
	$F_i(\stage') \ib F_i(\stage)$.

	\eitem
	
	\dt{How to use a scaffold $\stage$:}
	We call 2 subroutines on $\stage$,
	to {\em extend} and to {\em refine} it.
	The extension/refinement subroutines are analogous to
	to \stepA/\stepB.
	We invoke the $\Extend$ subroutine as follows:
			$$\stage.\Extend(\cdots).$$
	Here we view $\stage$ as an object in the sense
	of Object-Oriented Programming Languages (OOPL).
	The dot-invocation (``object.subroutine'') allows
	the subroutine to modify the object.  Here is the header
	for the $\Extend$ subroutine:

	\renewcommand{\alt}[2]{#2} 
	\renewcommand{\alt}[2]{#1} 
	\alt{
		\beql{extend}
		\hspace*{15mm} 
			\includegraphics[width=0.70\columnwidth]{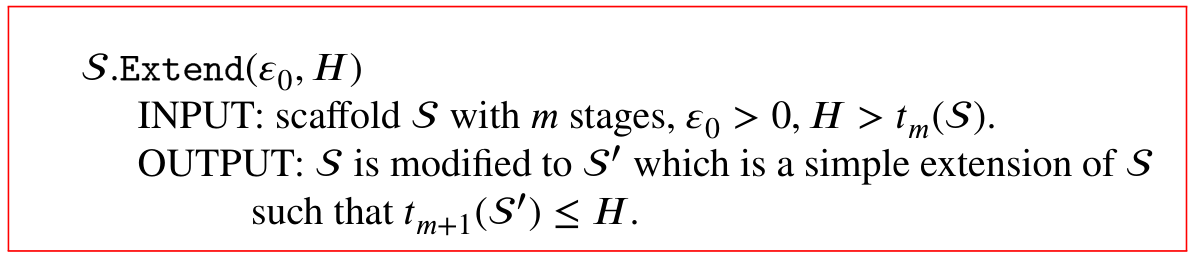}
		\eeql
	}{
		\Ldent\progb{
			\lline[0] 
			$\stage.\Extend( \veps_0,H)$ 
			\lline[5] INPUT:  scaffold $\stage$ with $m$ stages,
					$\veps_0>0, H>t_m(\stage).$
			\lline[5] OUTPUT:  $\stage$ is modified to $\stage'$
				which is a simple extension of $\stage$
			\lline[15]
				such that $t_{m+1}(\stage')\le H$.
		}
	}

	Basically, $\Extend$ could be implemented by calling $\stepA$
	followed by $\stepB$ to produce a new admissible quad.
	In previous approaches (exemplified by \refeq{simpleivp}),
	one basically keep calling $\Extend$ until
	the end-time of $\stage$ reaches the desired target $H$.
	But to ensure halting, we now insist that $\stage$ be
	sufficiently ``refined'' before it can be extended again.
	Sufficient means that the end-enclosure of the last stage
	of $\stage$ satisfies
	$\wmax(E_m(\stage))\le\veps_0$ where $\veps_0$ is input 
	in \refeq{endEncProb}).
	Here then is the header of the \Refine\ subroutine:

	\ignore{
		[MOVE mechanism]
	Here is the mechanism for making $\stage$
	$\veps_0$-bounded.  
	We maintain a value $\delta_i>0$ (for $i=1\dd m$)
	in the $i$th stage of $\stage$.
	Call $\delta_i$ the \dt{delta-value} of the $i$th stage.
	For instance, right
	after calling $\stage.\Extend$, the delta-value
	of the last stage is $\veps_0$ (the desired 
	bound on the final end-enclosure in \refeq{endEncProb}).
	}%

	\renewcommand{\alt}[2]{#2} 
	\renewcommand{\alt}[2]{#1} 
	\alt{
		\beql{refine}
		\hspace*{20mm}
			\includegraphics[width=0.45\columnwidth]{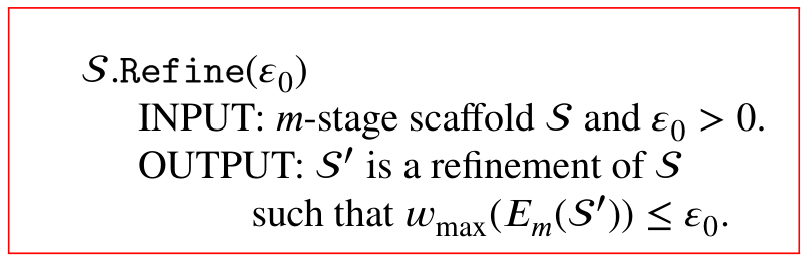}
		\eeql
	}{ 
		\Ldent\progb{
		\lline[0] $\stage.\Refine( \veps_0) $
		\lline[5] INPUT:  $m$-stage scaffold $\stage$ and
				$\veps_0>0$.
		\lline[5] OUTPUT: $\stage'$ is a refinement of $\stage$ 
		\lline[15]
				 such that $\wmax(E_{m}(\stage'))\le \veps_0.$}
	}

\ignore{
	Moreover, once stage $i$ becomes delta-bounded, we will instantly
	update half its delta-value: $\delta_i\ass \tfrac{1}{2}\delta_i$.
	For efficiency reasons,
	we do not require that each stage is delta-bounded,
	but we will prove that if $\Refine$ is called
	infinitely often, then each stage will be
	delta-bounded infinitely often. 
}%
	
	Within the \Refine\ procedure, we will refine the $i$th 
	time-span $[t_{i-1}, t_i]$ (for $i=1,2,\ldots$)
	using "light-weight" techniques
	to improve the full- and end-enclosures.  Specifically, the
	interval is uniformly subdivided into mini-step size $h_i$, and
	refinement is performed on such mini-intervals.  Thus, we can
	refine the width of enclosures without modifying original
	time spans of the stages, allowing for more efficient and targeted
	refinement.  See \refSec{extend-refine} below.

\ssectL[keyIdea]{The Radical Transformation Method}
	We now introduce new techniques
	to compute enclosures more efficiently based
	on \lognorm\ estimates via \refThm{ne}.
	Consider an admissible triple $(E_0, h_1, F_1)$.
	Let $\olmu$ be a logarithmic norm bound for $(\bff, F_1)$ (see
	\refeq{lognormbound}).  We briefly review the two methods.

	The first method is based on the Euler-tube method:
	When the Euler step size
	is less than $h\euler(H, \olM, \olmu, \delta)$ (see
	\refLem{eulerStep}), we can invoke \refCor{cor-1}
	to derive improved full- and end-enclosures for $(E_0,h_1,F_1)$.
	This solution lies inside an explicit
	$\delta$-tube, a strong property we will need.
	
	The second method is
	based on the \dt{radical transformation} of the original
	system to reduce the logarithmic norm.
	We distinguish two cases:
	
	\dt{Easy Case: $\olmu \le 0$.}
	In this case, $F_1$ is a contraction zone. By \refThm{ne}, we have
	$\wmax(E_1) \le \wmax(E_0)$.
	Therefore, we directly estimate the full- and end-enclosures using
	the logarithmic norm without further transformation.
	
	\dt{Hard Case: $\olmu > 0$.}
	The key idea here is to construct an invertible transformation
		$
			\pi: \RR^n \to \RR^n.
		$
	Let $\bfy = (y_1, \ldots, y_n) \as \pi(\bfx)$ and consider the
	transformed differential system:
	\beql{bfy'}
		\bfy' = \bfg(\bfy), \qquad
		\bfg(\bfy) \as J_{\pi}(\pi^{-1}(\bfy)) \cdot \bff(\pi^{-1}(\bfy)).
	\eeql
	This is considered with the admissible triple
	$(\pi(E_0), h, \pi(F_1))$.
	
	We define the transformation as a composition:
	\beql{circ}
	\pi = \whpi \circ \olpi,
	\eeql
	where $\olpi$ is an affine map (see Appendix B), and
		$
		\whpi(\bfx) = (x_1^{-d_1}, \ldots, x_n^{-d_n})
		$
	for some exponent vector $\bfd = (d_1, \ldots, d_n)$ to be determined.
	The map $\whpi$ is invertible provided $d_i \ne 0$ for all $i$.
	Due to the component-wise inversion, we refer to $\pi$ as the
	\dt{radical transform}.
	
	Assuming $\ivp(B_0, h, B_1)$ is valid (\cored{Section 2.1}),
	and that $B_1$ is sufficiently small, we can show that $\pi(B_1)$ is
	a contraction zone for $(\pi(B_1), \bfg)$.
	This brings the problem back to the easy case. After computing a
	shrunken enclosures in the transformed space, we pull it back to
	obtain an enclosures for the original IVP.
	For consistency, in the easy case, we define $(\pi, \bfg)$
	as $(\id, \bff)$.



%% file: inc/4stepAB.tex

\sectL[stepAB]{Steps A and B}
	Nedialkov et al
	\cite{nedialkov-jackson-pryce:HOI:01,nedialkov+2:validated-ode:99}
	provide a careful study of various algorithms
	for \stepA\ and \stepB.
	In the following, we provide new forms of \stepA\ and \stepB.  
\ssect{Step A}\label{ssec:stepA}
	We now present an algorithm for $\stepA(E_0,H,\veps)$.
	Its input/output specification has been given in \refeq{stepA}.
	We can regard its main goal
	as computing the largest possible $h>0$ (subject to $h\le H$)
	such that $(E_0,h,F_1)$ is $\veps$-admissible for some $F_1$.
	When calling $\stepA$, we are at some time $t_1\in [0,1)$,
	and so the largest $h$ we need is $H=1-t_1$.   We therefore
	pass this value $H$ to our subroutine.  Our approach should
	be compared to
	\cite[p.458, Figure 1]{nedialkov-jackson-pryce:HOI:01},
	who uses a complicated heuristic formula for $H$
	based the previous step.

	\bleml[ad]\ 

	Let $ H > 0 $, $\bfveps = (\veps_1\dd \veps_n)$, and 
	$ E_0 \ib \RR^n $.  
	Also let $\bfM=(M_1\dd M_n)$ with
		$$M_i \as \sup_{\bfp \in \olB}
			\Big|\big(\bff^{[k]}(\bfp)\big)_i\Big|,
			\quad (i=1\dd n)$$
	where $(\bfx)_i$ is the $i$th coordinate of $\bfx\in\RR^n$ and
		$$\olB \as \sum_{i=0}^{k-1} [0, H]^i \bff^{[i]}(E_0)
			+ Box(\bfveps).$$
	If
	\beql{hF1}
		h = \min \Big\{ H, \min_{i=1}^n
		\Big(\tfrac{\veps_i}{M_i}\Big)^{1/k} \Big\}
		\quad\text{and}\quad
		F_1 \as \sum_{i=0}^{k-1} [0, h]^i \bff^{[i]}(E_0)
			+ Box(\bfveps).
	\eeql  
	then $(E_0,h, F_1)$ is an admissible triple.
	\eleml

	Using \refLem{ad}, we can define $\stepA(E_0,H,\bfveps)$
	as computing $(h,F_1)$ as given by \refeq{hF1}.
	Call this the \dt{non-adaptive $\stepA$}, denoted $\stepA_0$.
	This non-adaptive $h$ may be too pessimistic.
	Instead, we will compute $h$ adaptively:

	\Ldent\progb{
		\lline[-2] \stepA($E_0,H, \bfveps$)$\ssa (h,F_1)$
		\lline[5] INPUT: $E_0\in \intbox\RR^n$, $H>0$,
					$\bfveps=(\veps_1\dd \veps_n)$
		\lline[5] OUTPUT: $0<h\le H$, $F_1\in \intbox\RR^n$ such
					that $(int(E_0),h,F_1)$ is $\bfveps$-admissible.
		\lline[5] \myhlineX[3]{0.6}
		\lline[5]  $h\ass 0$
		\lline[5]	While $(H>h)$
		\lline[10] 	$F_1 \ass Box\big(\sum_{i=0}^{k-1} [0, H]^i
					\bff^{[i]}(E_0)\big) + Box(\bfveps)$
		\lline[10] 	$\bfM \ass \bfw(Box(\bff^{[k]}(F_1)))$
		\lline[10]	$h \ass \min\set{H, \min_{i=1}^n
						\Big(\tfrac{\veps_i}{M_i}\Big)^{1/k}}$
					\qquad (where $\bfM=(M_1\dd M_n)$)
		\lline[10] 	\hspace{-4.0mm}
			\Commentx{\cored{Invariant:} $(E_0,h,F_1)$
						is $\bfveps$-admissible}
		\lline[10] 	$H\ass H/2$
		\lline[5]	Return $(h,F_1)$
	}

	This is a bisection search for the ``largest'' $h$
	that satisfies the \refLem{ad}, i.e.,
	such that $(E_0,h,F_1)$ is $\bfveps$-admissible.
	Within each iteration of the while-loop,
	the computed values of $h$ and $F_1$ form
	an $\bfveps$-admissible triple $(E_0,h,F_1)$
	(by \refLem{ad}).
	We halve $H$ before repeating
	the while-loop -- this ensures that the next value of $h$
	increases.  We exit when $H\le h$.
	Our experiments will show that our approach is highly effective.

	\blemT[stepA]{Correctness of \stepA}
		$\stepA(E_0,H, \bfveps)\ssa (h,F_1)$ is correct:\\
		I.e., the algorithm halts
		and outputs an $\bfveps$-admissible triple
				$(E_0,h,F_1)$.
	\elemT

	Implementation Notes:
	the code for \stepA\ is written for readability, but
	easily modified to gain some efficiency:
	\bitem
	\item
		The summation $\sum_{i=0}^{k-1}$ in \stepA\
		should be evaluated with Horner's rule
		(see \cite[p.~458]{nedialkov-jackson-pryce:HOI:01}).
	\item
		We should pre-compute the values
		$D_i\ass \bff^{[i]}(E_0)$ (for $i=0\dd k-1$) outside
		the while-loop.  Inside the while-loop, we update $F_1$
		using the formula
			$Box(\sum_{i=0}^{k-1} [0,H^i]D_i)+ Box(\bfveps)$.
	\eitem
\ssect{Step B}\label{ssec:stepB}
	For \stepb, there are several methods such as
	the ``Direct Method''
	\cite{nedialkov:thesis:99,nedialkov+2:validated-ode:99},
	Lohner's method \cite{lohner:thesis},
	and $C^1$-Lohner method \cite{wilczak-zgliczynski:lohner:11}.
	The Direct Method is basically the
	mean value form of Taylor expansion to order $k$.
	Given $(E_0,h,F_1)$, the Direct method 
	given by Nedialkov
	\cite{nedialkov:thesis:99,nedialkov+2:validated-ode:99},
	returns
		\beqarrayl{directa}
		E_1	&=& \underbrace{
				\sum_{i=0}^{k-1}
					h^i\bff^{[i]}(\bfm(E_0))}_{\text{Point}}
			~+~ \underbrace{
				\Big(\sum_{i=0}^{k-1} h^iJ_{\bff^{[i]}}(E_0)\Big)
					\Bigcdot\big(E_0-\bfm(E_0)
				\big)}_{\text{Centered Remainder}}
			~+~ \underbrace{
				 h^k\bff^{[k]}(F_1)}_{\text{Full Remainder}}\\
			&=& \underbrace{
				\sum_{i=0}^{k-1}
				h^i\bff^{[i]}(\bfm(E_0))
				 + h^k\bff^{[k]}(\bfm(F_1))}_{\text{Point($q_0$)}}
			~+~ \underbrace{
				\Big(\sum_{i=0}^{k-1}h^iJ_{\bff^{[i]}}(E_0)\Big)
					\Bigcdot\big(E_0-\bfm(E_0) \big)
				 + h^k\bff^{[k]}(F_1-\bfm(F_1))
				 }_{\text{Centered Remainder}}
			\label{eq:direct}
		\eeqarrayl
	where the final expression \refeq{direct}
	is obtained from \refeq{directa} by
	writing the (Full Remainder)
	$h^k\bff^{[k]}(F_1)$ as the sum
	$h^k\bff^{[k]}(\bfm(F_1)) +
		h^k\bff^{[k]}(F_1-\bfm(F_1))$
	is the sum of a point
	$\bfq_0$ and a ``centered remainder set''.
	Note that $\bfq_0$ is basically the order $k$
	Taylor expansion of the midpoint $\bfm(E_0)$.
	Both the Lohner and the $C^1$-Lohner
	methods are refinements of the Direct method. The Lohner method aims
	to reduce the wrapping effect in the Remainder Set.
	The $C^1$-Lohner method goes
	further by considering the limit when $k\to\infty$: then 
	$V\as \left( \sum_{i=0}^{\infty} h^i J_{\bff^{[i]}} \right)$
	satisfies another ODE: $V'=J_f\cdot V$.
	By solving this ODE, the method
	effectively computes a tighter end-enclosure.

	For reference, the Direct Method based on
	\refeq{direct} will be called ``$\stepB_0$'', i.e.,
	$\stepB_0(E_0,h,F_1)\ssa E_1$.
	What we will call ``$\stepB$'' is a simple
	modification of $\stepB_0$ to
	include a logarithmic norm estimate from \refCor{cor-1}:
	%

		\Ldent\progb{
		\lline[0] \stepB($E_0,h,F_1$)$\ssa E_2$
		\lline[5]   INPUT: A admissible triple $(E_0,h,F_1)$
		\lline[5]   OUTPUT: $E_1$ is an end-enclosure for
						$(E_0,h,F_1)$.
		\lline[5] \myhlineX[3]{0.6}
		\lline[5] $\bfq_0\ass \sum_{i=0}^{k-1} (h^i\bff^{[i]}(\bfm(E_0))$.
		\lline[5] $E_1 \ass  \bfq_0 + h^k\bff^{[k]}(F_1)
					+ \Big(\sum_{i=0}^{k-1}h^iJ_{\bff^{[i]}}(E_0)\Big)
				 		\Bigcdot\big(E_0-\bfm(E_0)\big)$.
		\lline[5] $\olmu\ass\mu_2(J_{\bff}(F_1))$
		\lline[5] $r_0\ass \half \|\bfw(E_0)\|_2 e^{\olmu h}$
		\lline[5] $B\ass Box_{\bfq_0}(r_0) 
						+ Box(h^k\bff^{[k]}(F_1))$
		\lline[5] Return $E_2\ass E_1\cap B$
	}
	
	\blemT[stepB]{Correctness of \stepB}
		$\stepB(E_0,h,F_1)\ssa E_2$ is correct,
		i.e., $E_2$ is an end-enclosure for $(E_0,h,F_1)$
	\elemT
\ssect{Refinements of Full and End Enclosures using Euler}
	\label{ssec:ne}  
	So far, the full- and end-enclosures
	using Steps A and B are based on Taylor's formula. 
	We now show how to refine such enclosures using Euler steps.


	\blemT[delta-distance]
		{1-Step Euler Enclosures with \lognorm}\ \\ 
	Consider an admissible triple $(E_0,H,F_1)$ where
		$E_0\as Ball_{\bfp_0}(r_0)$. \\
	Let
			$\olmu = \mu_2(J_\bff(F_1))$,
			$\olM = \|\bff\supn[2](F_1)\|$,
			and $\delta>0$.\\
	If $\bfq_0=\bfp_0+h_1\bff(\bfp_0)$ is obtained from
		$\bfp_0$ by an Euler step of size $h_1$ where
		$h_1\le  h\euler(H,\olM,\olmu,\delta)$ (see~\refeq{h1}),
	then:
	\benum[(a)]
	\item  \dt{($\delta$-Tube)}\\
		The linear function
		$\ell(t)\as (1-t/h_1)\bfp_0+(t/h_1)\bfq_0$
		lies in the $\delta$-tube of $\bfx_0=\ivp(\bfp_0,H)$.
	\item \dt{(End-Enclosure)}\\
		Then
		$Ball_{\bfq_0}(r_0 e^{\olmu h_1}+\delta)$
		is an end-enclosure for $\ivp(E_0,h_1)$.
	\item \dt{(Full-Enclosure)}\\
		The convex hull
			$\chull(Ball_{\bfp_0}(r'), Ball_{\bfq_0}(r'))$
				where
			$r'=\delta+\max(r_0 e^{\olmu h_1},r_0)$
		is a full-enclosure for $\ivp(E_0,h_1)$.
	\eenum
	\elemT

	\dt{The refinement strategy for a stage:}
	To exploit \refLem{delta-distance}, we refine the $i$th stage by subdividing
	the time interval $[t_{i-1},t_i]$ into $2^{\ell_i}$ uniform Euler steps, where
	$(t_i-t_{i-1})/2^{\ell_i}$ is the $i$th \dt{mini-step size}.
	Recall that if this mini-step size is smaller than
	$h_1 = h\euler(H,\olM,\olmu,\delta)$ (see \refeq{h1}),
	then the Euler trajectory remains inside the $\delta$-tube
	around the exact solution $\bfx(t;\bfq_0)$.
	Here the parameters $H,\olM,\olmu$ are taken from the data of the $i$th stage.
	In this case, we can compute a $\delta$-bound end-enclosure.
	We denote by $\delta=\delta(G_i)$ the \dt{Euler-tube target} associated with
	the $i$th stage, where $G_i$ is the corresponding refinement substructure.
	For the final stage of an $m$-stage scaffold, we set
	$\delta(G_m)=\veps_0$, the input tolerance of \Refine.
	
	We call this the \SubrSeven\ Subroutine.
	Empirically, this procedure becomes inefficient when $\olM$ and $\olmu$
	are large. We therefore introduce an alternative adaptive method,
	called Bisection, to reduce $\olM$ and $\olmu$.

	\bitem
	\item Bisection Method:
	we subdivide the interval $[0,H]$ into $2^\ell$
	mini-steps of size
		$h_\ell\as H/2^{\ell}$ (for $\ell=1,2,\ldots$). 
	At each \dt{level} $\ell$, we can compute full-
	and end-enclosures 
			$(\bfF_i[j],\bfE_i[j])$
	of the $j$th mini-step ($j=1\dd 2^\ell$) using
	the following formulas:
		\beql{taylor-full} 
			\bfF_i[j] \ass \sum_{p=0}^{k-1}
				[0,h_\ell]^p \bff^{[p]}(\bfE_i[j-1]
			+ [0,h_\ell]^k \bff^{[k]}(\cored{F_1}),
		\eeql  
	and 
		\beql{stepBmini}
			\bfE_i[j] \ass \stepB(\bfE_i[j-1], h_\ell,
					\bfF_i[j],\mu_2(J_\bff(\bfF_i[j]))).
		\eeql
	\item
		When $\ell$ is sufficiently
	large, i.e., $h_\ell\le h_i$, then we
	can call the \SubrSeven\ subroutine above.
	Our experiments show, this subroutine is more accurate.
	\eitem

	\savespace{
	\bpf
	By \refLem{endenclosure}, we have
		$E_1 =  Ball_{\bfq_0}(r_0e^{\olmu h}+\delta)$ is the
	end-enclosure for $\ivp(E_0, h)$.  
	
	Next, we prove $(ii)$.  
	We show that for any $T \in [0, h]$,  the end-enclosure of 
	$\ivp(E_0,T)$ is a subset of $Box(Ball_{\bfp_0}(r' +\delta),
	Ball_{\bfq_0}(r' +\delta))$.
	Note that by \refLem{endenclosure}, we have
	$E_1 =  Ball_{\ell(T)}(r_0e^{\olmu T}+\delta)$ is the
	end-enclosure for $\ivp(E_0, T)$.
	
	Let $\ell(T)_i$ denote the $i$-th component of $\ell(T)$
	and $r(T)\as r_0e^{\olmu T}+\delta$. Then, we only need to
	prove that for any $i=1\dd n$, the interval
		$\ell(T)_i \pm r(T)$ satisfies  
	\[
	\ell(T)_i \pm r(T) \subseteq Box((\bfp_0)_i\pm (r'+\delta),
	(\bfq_0)_i\pm (r'+\delta)),
	\]  
	where $(\bfp_0)_i $ and $(\bfq_0)_i $ are the $i$-th components of
	$\bfp_0$ and $\bfq_0$,
	respectively.  
	
	Since $\ell(T)$ is a line segment, it follows that  
	\[
		\min((\bfq_0)_i, (\bfp_0)_i)
			\leq \ell(T)_i \leq \max((\bfq_0)_i, (\bfp_0)_i).
	\]  
	Additionally, we have $r(T) \leq r' + \delta$.  
	
	Combining these observations, we conclude that 
	\[
	\ell(T)_i \pm r(T) \subseteq Box((\bfp_0)_i\pm (r'+\delta),
	(\bfq_0)_i\pm (r'+\delta)).
	\] 

	\epf
	}
	
	\savespace{
	\begin{figure}[h]
		\centering
		\includegraphics[width=0.7\linewidth]{figs/delta-distance}
	
		\caption{The red curve $\bfx_c(t)$ represents the solution to  
			$\ivp(\bfp_0, h)$. The blue curve $l(t)$ lies within the
			$\delta$-tube of $\bfx_c$. According to
			\refLem{delta-distance}, the ball
			$B = Ball_{\ell(h)}(r_0e^{\olmu
			h} + \delta)$ provides an end-enclosure for
			$\ivp(Ball_{\bfp_0}(r_0), h)$, as shown in the left graph. In
			the right graph, the green box,
			$Box(Ball_{\bfp_0}(r' + \delta),
			Ball_{\bfq_0}(r' + \delta))$, projected onto the
			vertical axis, represents the full-enclosure.}  

		\label{fig:delta-distance}
	\end{figure}
	}
	
	\savespace{
	Next, we show that the Euler method with a step size
	$0<h_1=h_1(\delta)\le h$ will be able to compute a series of  points
	$P=(\bfp_0\dd \bfp_{N+1})$ such that for any $i=1\dd N+1$ 
	the line segment $\ol{\bfp_{i-1}\bfp_i}$ lies in the $\delta$-tube of
	$\bfx_c$.
	
	The classic Euler method for $\ivp(\bfp_0, h)$
	produces a polygonal path of the form
			$$[(0,\bfp_0), (t_1,\bfp_1) \dd (t_N,\bfp_N), (h,\bfp_{N+1})]$$
	of length $N+1$.  In case the time step is uniformly $h_1$,
	then $t_i = i h_1$ for $i=0\dd N\as \floor{h/h_1}$ and
	we can represent the path\footnote{
		Allowing the degenerate case where $t_N=h$.
	} as
			$P=[\bfp_0, \bfp_1 \dd \bfp_N, \bfp_{N+1}]$.

	We now consider a variant of this classical subroutine
	to compute a polygonal approximation $P$ to 
			\beql{bfxp0}
				\bfx =\ivp_\bff(\bfp_0,h,F_1)\eeql
	so that $P$ lies in the $\delta$-tube of $\bfx$,
	for any user chosen $\delta>0$. 
	
	\Ldent\progb{
		\lline[0]  ${\sc Euler}_\bff(\bfp_0,h,\delta)\ssa (h_1,P)$
		\lline[5] INPUT: $\bff, \bfp_0, h, F_1, \delta$ as described
			above subject to \refeq{lognormneg}.
		\lline[5] OUTPUT:  a polygonal path $P$ with step size $h_1$
			from time $0$ to $h$.  
		\lline[25]	Morever, $P$ lies in the
				$\delta$-tube of $\bfx$ in \refeq{bfxp0}.
		\lline
		\lline[10] $h_1 \ass
				\min\set{h,
					\frac{2\olmu\delta}{\|\bff^{[2]}(F_1)\|
						(e^{\olmu h}-1)}}$.
		\lline[10] $N\ass \floor{h/h_1}$
		\lline[10] $\wtbfy\ass \bfp_0$
		\lline[10] Initialize a list $P\ass [\bfp_0]$
		\lline[10] For $i=1$ to $N$
		\lline[15] 		$\wtbfy\ass \wtbfy + h_1\cdot \bff(\wtbfy)$
		\lline[15]		$P.append(\wtbfy)$
		\lline[10] $P.append(\wtbfy + (h-Nh_1)\cdot \bff(\wtbfy))$ 
		\lline[10] Return $(h_1, P)$
	}

	\bthmT[negative]{Correctness of {\sc Euler}}\ \\
		The algorithm ${\sc Euler}_\bff(\bfp_0,h,\delta)$
		is correct, i.e.,
		the output $P=[\bfp_0\dd \bfp_{N+1}]$
		lies within the $\delta$-tube of the solution
		$\bfx$ in \refeq{bfxp0}.
	\ethmT
	\bpf
		 Let 
		 	$t_i=
		 \begin{cases}
		 	ih_1, & i<N+1\\
		 	h,    & i=N+1
		 \end{cases}.$
		 
		 For any  \(i = 1 \dd N+1\) and  $t\in [t_{i-1},t_{i}]$,
		 let $\bfq=\bfp_i+(t-t_i)\bff(\bfp_i)$.
		 Then the theorem follows if we prove
		 	$\|\bfq - \bfx(t)\| \le \delta $.
		By \refLem{Euler-step},
		\beqarrys
		\|\bfq - \bfx(t)\|  
		&\leq& 
			\frac{\|\bff^{[2]}(F_1)\|h_1}{2\olmu}(e^{\olmu_1 t} - 1)  
		 \\  
		&\leq& \frac{(e^{\olmu_1 t} - 1)\delta}
			{e^{\olmu_1 	h} - 1} 
				& \text{(by the definition of \(h_1\))}\\  
		&\leq& \delta  
			& \text{(since \(t \le h\))}
		\eeqarrys
			
	\epf
	}
	
\ssectL[problems]{List of Problems and
		Local Experiments on Steps A and B} 
	\refTab{problems} is a list of problems used throughout
	this paper for our experiments.  In this subsection,
	we give ``local''
	(single-step) experiments on the effectiveness of
	our Steps A and B.  Later in \refSec{end-enclosure}, we
	will do ``global'' experiments based on our overall algorithm.
	We measure each technique by ratios denoted by $\sigma$,
	such that $\sigma>1$ shows the effectiveness of the technique.
	Note that the gains for local experiments may appear small
	(e.g., $1.0001$).  But in global $m$-step experiment, this translates
	to $(1.0001)^m$ which can be significant.
	
	\newcommand*{\myalign}[2]{\multicolumn{1}{#1}{#2}}
	\begin{table}[] \centering
		{\tiny
			\btable[l| l | l | l | l | l ]{
				Eg* & \myalign{c|}{\dt{Name}}
				& \myalign{c|}{$\bff(\bfx)$}
				& \myalign{c|}{\dt{Parameters}}
				& \myalign{c|}{\dt{Box} $B_0$}
				& \myalign{c}{\dt{Reference}}
				\\[1mm] \hline \hline
				Eg1 & Volterra 
				& $\mmatP{\phantom{-}ax(1-y) \\ -by(1-x)}$ 
				&  $\mmatP{a\\ b}=\mmatP{2\\ 1}$
				& $Box_{(1,3)}( 0.1)$
				& \cite{moore:diffEqn:09},
				\cite[p.13]{bunger:taylorODE:20}
				\\[2mm] \hline 
				Eg2 & Van der Pol 
				& $\mmatP{y \\ -c(1-x^2)y -x}$
				& $c=1$
				& $Box_{(-3,3)}( 0.1)$
				& \cite[p.2]{bunger:taylorODE:20}
				\\[1mm] \hline
				Eg3 & Asymptote
				& $\mmatP{x^2 \\ -y^2 + 	7x}$
				& N/A
				& $ Box_{(-1.5,8.5)}(0.01 )$
				& N/A
				\\[1mm] \hline
				Eg4 & Lorenz
				& $\mmatP{\sigma(y-x)\\ x(\rho-z)-y\\ xy-\beta z}$
				&
				$\mmatP{\sigma\\\rho\\\beta}=\mmatP{10\\ 28\\8/3}$
				& $ Box_{(15,15,36)}(0.001)$
				& \cite[p.11]{bunger:taylorODE:20}
				\\[1mm] \hline
		} }
		\caption{List of IVP Problems}
		\label{tab:problems}
	\end{table}
	
	First, in \refTab{StepA}, we compare our \stepA\ with
	the non-adaptive $\stepA_0$.  This non-adaptive
	$\stepA_0$ is basically the algorithm\footnote{
		We replace their $h_{j,0}$ by $H$, and
		$2h_{j,0}^k\bff\supn[k]([\wty_{j-1}])$ by $\bfveps$.
	} in
	\cite[p.458, Figure 1]{nedialkov-jackson-pryce:HOI:01}.  
	
	
	Let $(h_0,F_0)$ and $t_0$ be the admissible pair and computing time
	for
	$\stepA_0(E_0,H,\veps)$.
	Let $(h,F), t$ be the corresponding values for
	$\stepA(E_0,H,\veps)$.
	The performance of these 2 algorithms can be
	measured by three ratios:
		$$
		\rho \as \frac{\wmax(F)}{\wmax(F_0)}, \qquad 
		\cored{\sigma} \as \frac{h}{h_0}, \qquad 
		\tau \as \frac{t}{t_0}.
		$$
	The most important ratio is $\sigma$, which we want
	to be as large as possible and $>1$.
	A large $\sigma$ will make $\rho$  and $\tau$ to be $>1$,
	which is not good when viewed in isolation.  But such increases
	in $\rho$ and $\tau$, in moderation, is a good overall tradeoff.

	\begin{table*}[h!]
		{\tiny
			\centering
			\begin{tabular}{c|c|c|c|>{\color{red}}c|c|c}
				\hline
				Eg & $E_0$ & $\veps$ & H & $\sigma$ & $\tau$ & $\rho$ \\
				\hline
				\multirow{8}{*}{Eg1} & \multirow{4}{*}{$[0.9, 1.1],[2.9, 3.1]$} & \multirow{2}{*}{$0.1$} 
				& 1.0 & $38.5$ & $1.15$ & $3.05$ \\
				\cline{4-7}
				& & & 10 & $2.22 \times 10^6$ & $1.06$ & $3.84$ \\
				\cline{3-7}
				& & \multirow{2}{*}{$0.0001$} 
				& 1.0 & $56.8$ & $1.02$ & $5.26$ \\
				\cline{4-7}
				& & & 10 & $2.17 \times 10^6$ & $1.27$ & $3.62$ \\
				\cline{2-7}
				& \multirow{4}{*}{$[2, 4],[3, 5]$} & \multirow{2}{*}{$0.1$} 
				& 1.0 & $3.89\times 10^4$ & $1.03$ & $2.41$ \\
				\cline{4-7}
				& & & 10 & $1.89 \times 10^8$ & $1.18$ & $3.17$ \\
				\cline{3-7}
				& & \multirow{2}{*}{$0.0001$} 
				& 1.0 & $4.97\times 10^3$ & $1.01$ & $1.52$ \\
				\cline{4-7}
				& & & 10 & $2.65 \times 10^8$ & $1.45$ & $1.72$ \\
				\hline
				\multirow{8}{*}{Eg2} & \multirow{4}{*}{$[-3.1, -2.9],[2.9, 3.1]$} & \multirow{2}{*}{$0.1$} 
				& 1.0 & $5.12 \times 10^3$ & $14.6$ & $8.49$ \\
				\cline{4-7}
				& & & 10 & $3.50 \times 10^{12}$ & $10.2$ & $13.1$ \\
				\cline{3-7}
				& & \multirow{2}{*}{$0.0001$} 
				& 1.0 & $5.58 \times 10^3$ & $1.00$ & $8.49$ \\
				\cline{4-7}
				& & & 10 & $4.24 \times 10^{12}$ & $7.35$ & $10.4$ \\
				\cline{2-7}
				& \multirow{4}{*}{$[-4, -2],[3, 5]$} & \multirow{2}{*}{$0.1$} 
				& 1.0 & $3.69 \times 10^5$ & $1.08$ & $5.21$ \\
				\cline{4-7}
				& & & 10 & $1.90 \times 10^{14}$ & $1.19$ & $7.41$ \\
				\cline{3-7}
				& & \multirow{2}{*}{$0.0001$} 
				& 1.0 & $4.56 \times 10^5$ & $1.0$ & $2.58$ \\
				\cline{4-7}
				& & & 10 & $2.60 \times 10^{14}$ & $1.49$ & $3.16$ \\
				\hline
				\multirow{8}{*}{Eg3} & \multirow{4}{*}{$[-1.51, -1.49],[8.49, 8.51]$} & \multirow{2}{*}{$0.1$} 
				& 1.0 & $2.41 \times 10^4$ & $1.83$ & $65.2$ \\
				\cline{4-7}
				& & & 10 & $2.24 \times 10^9$ & $1.49$ & $30.5$ \\
				\cline{3-7}
				& & \multirow{2}{*}{$0.0001$} 
				& 1.0 & $3.23 \times 10^4$ & $1.005$ & $129$ \\
				\cline{4-7}
				& & & 10 & $2.28 \times 10^9$ & $1.81$ & $161$ \\
				\cline{2-7}
				& \multirow{4}{*}{$[-3.5, -3.4],[8.4, 8.5]$} & \multirow{2}{*}{$0.1$} 
				& 1.0 & $4.30 \times 10^4$ & $1.36$ & $21.1$ \\
				\cline{4-7}
				& & & 10 & $3.00 \times 10^9$ & $1.11$ & $26.2$ \\
				\cline{3-7}
				& & \multirow{2}{*}{$0.0001$} 
				& 1.0 & $4.36 \times 10^4$ & $1.20$ & $31.2$ \\
				\cline{4-7}
				& & & 10 & $3.05 \times 10^9$ & $1.20$ & $38.7$ \\
				\hline
				\multirow{8}{*}{Eg4} & \multirow{4}{*}{$[14.999, 15.001],[14.999, 15.001],[35.999, 36.001]$} & \multirow{2}{*}{$0.1$} 
				& 1.0 & $2.98 \times 10^3$ & $1.00$ & $59.9$ \\
				\cline{4-7}
				& & & 10 & $1.81 \times 10^8$ & $1.04$ & $80.5$ \\
				\cline{3-7}
				& & \multirow{2}{*}{$0.0001$} 
				& 1.0 & $3.83 \times 10^3$ & $1.01$ & $2050.88$ \\
				\cline{4-7}
				& & & 10 & $2.84 \times 10^8$ & $1.03$ & $1356.44$ \\
				\cline{2-7}
				& \multirow{4}{*}{$[12, 15],[13, 15],[34, 36]$} & \multirow{2}{*}{$0.1$} 
				& 1.0 & $1.63 \times 10^4$ & $1.04$ & $4.05$ \\
				\cline{4-7}
				& & & 10 & $8.70 \times 10^8$ & $1.08$ & $5.52$ \\
				\cline{3-7}
				& & \multirow{2}{*}{$0.0001$} 
				& 1.0 & $1.95 \times 10^4$ & $1.05$ & $2.05$ \\
				\cline{4-7}
				& & & 10 & $1.09 \times 10^9$ & $1.07$ & $2.51$ \\
				\hline
			\end{tabular}
		}
		\caption{\dt{Comparison of \stepA\ with $\stepA_0$.}
			Each row of the table is an experiment with one of our
			examples (Eg1, Eq2, etc), with the indicated values of 
			$(E_0,H,\veps)$. The key column is labeled $\sigma=h/h_0$,
			giving the ratio of the adaptive step size over
			the non-adaptive size.}
		\label{tab:StepA}
	\end{table*}

	\refTab{StepA} shows that 
	\stepA\ can dramatically increase the step size $h$
	without incurring a significant increase in computation time.
	So the adaptive version is highly effective and
	meaningful.

	In the \stepB\ experiments in \refTab{stepb},
	we look at two ratios:  
	\benum
	\item $\sigma_1 = \frac{\wmax(E_1)}{\wmax(E_2)}$ is
	the maximum width from the $C^r$-Lohner algorithm ($E_1$)
	divided by that of our \stepB\ ($E_2$).
	
	\item  $\sigma_2= $\(\frac{\wmax(DB_1)}{\wmax(B_1)}\) is
	the maximum width (\(DB_1\))
	from $\stepB_0$ (i.e., the Direct method) divided by that of
	(\(B_1\)) from our \(\stepB\).
	\eenum
	The effectiveness of our \stepB\ is indicated by
	when these ratios are greater than $1$.
	For each example, we provide an admissible triple and compute the
	logarithmic norm $\mu \ge \mu_2(J_{\bff}(F_1))$.

	\begin{table*}[]
		\centering
		{\tiny
			\begin{tabular}{|c|c|c|c|c|c|>{\color{red}}c|}
				\hline
				Eg* & $E_0$ & $F_1$ & $h$ & $\mu$ & $\sigma_1$ & $\sigma_2$ \\
				\hline
				
				\multirow{6}{*}{Eg1} 
				& \multirow{3}{*}{$Box_{(0.6,1.2)}(0.01)$} 
				& $(0.58,1.17)\pm(0.03,0.04)$ & 0.10 & -0.23 & 1.10 & 1.13 \\ \cline{3-7}
				& & $(0.575,1.135)\pm(0.065,0.075)$ & 0.22 & -0.03 & 1.16 & 1.41 \\ \cline{3-7}
				& & $(0.57,1.105)\pm(0.16,0.135)$ & 0.34 & 0.28 & 1.11 & 2.88 \\ \cline{2-7} 
				& \multirow{3}{*}{$Box_{(0.6,1.2)}(10^{-4})$} 
				& $(0.585,1.175)\pm(0.015,0.025)$ & 0.10 & -0.27 & 1.10 & 1.10 \\ \cline{3-7}
				& & $(0.57,1.115)\pm(0.08,0.095)$ & 0.33 & 0.09 & 1.15 & 5.10 \\ \cline{3-7}
				& & $(0.57,1.105)\pm(0.14,0.125)$ & 0.37 & 0.26 & 1.11 & 16.09 \\ 
				\hline
				
				\multirow{6}{*}{Eg2} 
				& \multirow{3}{*}{$Box_{(-3,3)}(0.1)$} 
				& $(-3.00,2.96)\pm(0.105,0.14)$ & 0.003 & 7.18 & 1.01 & 1.00 \\ \cline{3-7}
				& & $(-2.92,2.47)\pm(0.22,1.13)$ & 0.05 & 10.67 & 1.00 & 1.02 \\ \cline{3-7}
				& & $(-2.895,2.265)\pm(0.295,2.005)$ & 0.08 & 14.33 & 1.00 & 1.07 \\ \cline{2-7}
				& \multirow{3}{*}{$Box_{(-3,3)}(10^{-4})$} 
				& $(-2.985,2.925)\pm(0.015,0.075)$ & 0.006 & 6.03 & 1.01 & 1.03 \\ \cline{3-7}
				& & $(-2.895,2.35)\pm(0.185,1.57)$ & 0.085 & 11.79 & 1.00 & 2.25 \\ \cline{3-7}
				& & $(-2.895,2.265)\pm(0.295,2.005)$ & 0.09 & 12.66 & 1.00 & 2.57 \\ 
				\hline
				
				\multirow{6}{*}{Eg3} 
				& \multirow{3}{*}{$Box_{(-1.5,8.5)}(0.001)$} 
				& $(-1.495,8.475)\pm(0.015,0.035)$ & 0.0005 & -2.12 & 1.00 & 1.01 \\ \cline{3-7}
				& & $(-1.49,8.315)\pm(0.02,0.205)$ & 0.004 & -1.99 & 1.00 & 1.10 \\ \cline{3-7}
				& & $(-1.445,7.055)\pm(0.065,3.115)$ & 0.04 & 0.37 & 1.00 & 2.07 \\ \cline{2-7}
				& \multirow{3}{*}{$Box_{(-1.5,8.5)}(10^{-4})$} 
				& $(-1.495,8.475)\pm(0.005,0.025)$ & 0.0005 & -2.15 & 1.00 & 1.02 \\ \cline{3-7}
				& & $(-1.495,8.31)\pm(0.005,0.20)$ & 0.004 & -2.08 & 1.00 & 1.50 \\ \cline{3-7}
				& & $(-1.445,7.055)\pm(0.055,3.105)$ & 0.04 & 0.34 & 1.00 & 8.61 \\ 
				\hline
				
				\multirow{6}{*}{Eg4} 
				& \multirow{3}{*}{$Box_{(15,15,36)}(0.001)$} 
				& $(14.855,13.475,37.085)\pm(0.145,1.595,1.815)$ & 0.024 & 3.19 & 1.35 & 1.52 \\ \cline{3-7}
				& & $(14.74,12.885,37.23)\pm(0.28,2.325,2.27)$ & 0.027 & 3.67 & 1.36 & 1.80 \\ \cline{3-7}
				& & $(14.665,12.58,37.275)\pm(0.375,2.74,3.215)$ & 0.031 & 3.98 & 1.37 & 2.58 \\ \cline{2-7}
				& \multirow{3}{*}{$Box_{(15,15,36)}(10^{-4})$} 
				& $(14.855,13.475,37.085)\pm(0.145,1.595,1.815)$ & 0.020 & 3.19 & 1.33 & 1.79 \\ \cline{3-7}
				& & $(14.80,13.16,37.23)\pm(0.21,1.97,2.27)$ & 0.024 & 3.42 & 1.35 & 3.18 \\ \cline{3-7}
				& & $(14.665,12.58,37.275)\pm(0.375,2.74,3.215)$ & 0.031 & 3.98 & 1.37 & 13.52 \\ 
				\hline
			\end{tabular}
		}
		\caption{
			Comparison of \stepB\ with the Direct method and the $C^r$-Lohner
			algorithm.  
			The key column is $\sigma_2 = \frac{\wmax(DB_1)}{\wmax(B_1)}$,
			which reflects the ratio of the maximum width produced by the
			Direct method ($DB_1$) to that by \stepB\ ($B_1$), serving as a
			direct measure of their relative tightness.  
			We also report $\sigma_1$ which compares the maximum width from
			the $C^r$-Lohner algorithm  with that of the combined method
			based on \refCor{cor-1}. 
		}
		\label{tab:stepb}
	\end{table*}
	
	The data in the \refTab{stepb} show that intersecting either the
	$C^r$-Lohner method or the Direct method with the estimate from
	\refCor{cor-1} leads to tighter enclosures, with the improvement
	being especially pronounced for the Direct method.
	This effect becomes more noticeable as the step size increases.

		The \refTab{deltafull} compares various examples under a given
		$(E_0, H, F_1)$, showing the values of
		$h_1 = h\euler(H, \olM, \overline{\mu}, \delta)$
		computed for different choices of $\delta$ (see \refeq{h1}).
		It also reports the ratio of the maximum widths of the full
		enclosures obtained using \refLem{delta-distance} and
		\refeq{taylor-full}, respectively. 
	
	\begin{table*}[]
		\centering
		{\tiny
			\btable[c|c|c|c|c|c|c|>{\color{red}}c]{
				\hline
				\multirow{2}{*}{Eg*}
					& \multicolumn{3}{c|}{$(E_0,H,F_1)$}
					& \multirow{2}{*}{$\delta$}
					& \multirow{2}{*}{$h_1$}
					& \multirow{2}{*}{$\mu$}
					& \multirow{2}{*}{$\sigma$} \\\cline{2-4}
				& $E_0$ & $H$ & $F_1$ &  &  &  &  \\\hline
				
				\multirow{3}{*}{Eg1} 
				& \multirow{3}{*}{$(1.0,3.0) \pm (0.1,0.1)$} 
				& \multirow{3}{*}{0.1} 
				& \multirow{3}{*}{$(0.745,2.955) \pm (0.455,0.295)$}
				& 0.1 & 0.08 & 1.31 & 1.73 \\
				&  &  &  & 0.01 & 0.008 & 1.31 & 1.09 \\
				&  &  &  & 0.001 & 0.0008 & 1.31 & 1.01 \\\hline
				
				\multirow{3}{*}{Eg2} 
				& \multirow{3}{*}{$(-3.0,3.0) \pm (0.1,0.1)$} 
				& \multirow{3}{*}{0.05} 
				& \multirow{3}{*}{$(-2.92,2.40) \pm (0.28,0.80)$}
				& 0.1 & 0.019 & 9.57 & 1.62 \\
				&  &  &  & 0.01 & 0.0019 & 9.57 & 1.10 \\
				&  &  &  & 0.001 & 0.00019 & 9.57 & 1.01 \\\hline
				
				\multirow{3}{*}{Eg3} 
				& \multirow{3}{*}{$(-1.50,8.50) \pm (0.01,0.01)$} 
				& \multirow{3}{*}{0.04} 
				& \multirow{3}{*}{$(-1.445,6.635) \pm (0.165,1.975)$}
				& 0.1 & 0.0059 & -0.0026 & 2.48 \\
				&  &  &  & 0.01 & 0.00059 & -0.0026 & 1.75 \\
				&  &  &  & 0.001 & 0.000059 & -0.0026 & 1.14 \\\hline
				
				\multirow{3}{*}{Eg4} 
				& \multirow{3}{*}{$(15.000,15.000,36.000)
						\pm (0.001,0.001,0.001)$} 
				& \multirow{3}{*}{0.027} 
				& \multirow{3}{*}{$(14.736,12.800,37.279)
						\pm (0.365,2.301,2.442)$}
				& 0.1 & 0.0026 & 3.455 & 1.84 \\
				&  &  &  & 0.01 & 0.00026 & 3.455 & 1.74 \\
				&  &  &  & 0.001 & 0.000026 & 3.455 & 1.41 \\\hline
		}}
		\caption{Comparison of Full-Enclosures from
				\refLem{delta-distance} and \refeq{taylor-full}. 
			$\sigma \as \frac{\wmax(F_0)}{\wmax(F)}$
			where $F$ is the enclosure computed via
			\refLem{delta-distance}, and $F_0$ is the one obtained using
			\refeq{taylor-full}.}
		\label{tab:deltafull}
	\end{table*}
	
	The data in \refTab{deltafull} demonstrate that our method 
	described in \refLem{delta-distance} yields a better full 
	enclosure than the one obtained from \refeq{taylor-full}. It is 
	worth emphasizing that updating the full enclosure is important, 
	as it allows us to reduce the value of \lognorm, which in turn 
	enables further tightening of the end enclosure during subsequent 
	refinement steps.

%% file: inc/5comparison.tex
\sectL[xform]{Tighter Enclosures via Radical Transformation}
	In the previous section, we used the \lognorm\ in combination with
	the Taylor method to obtain tighter enclosures. However, the earlier
	approach has two main issues:
	\benum
	\item It may only reduce the maximum width of the enclosure, without
	considering the minimum width.
	
	For example, consider the ODE system $(x', y') = (7x, y)$, which
	consists of two independent one-dimensional subsystems. When
	analyzing this as a two-dimensional system, the logarithmic norm
	depends only on the component $x' = 7x$, since the logarithmic norm
	takes the maximum value.
	
	\item For methods like the Direct method -- which first track the
	midpoint and then estimate the range -- there is a potential problem:
	the tracked midpoint can deviate significantly from the true center
	of the solution set. This deviation may lead to considerable
	overestimation in the resulting enclosure.
	\eenum

	A radical map can be used to address these issues
	as suggested in our introduction.
	\ignore{%
	 For $n=1$, 
	 if the logarithmic norm is zero, then the
	midpoint of the tracked solution is equal to the center of
	exact solution's enclosure.
	particularly in the component with the
			"maximum width" ??
	Therefore,  we consider applying a suitable radical map so that the
	transformed IVP system has a logarithmic norm that is close to zero.
	}%
	
		Consider an admissible triple $(E_0, h, F_1)$.
	By the validity of $\ivp(E_0,h)$, the following	
	condition can be achieved if $E_0$ sufficiently shrunk:

	\beql{0ninolB1}
	\0\nin \olF_1 \as Box(\bff(F_1)) = \prod_{i=1}^n \olI_i.
	\eeql  
	
	This implies that there exists some $i =1\dd n$ such that
	$0 \notin \olI_i$. 
	We need such a condition because the radical map
	\refeq{essentially} is only defined if each $x_i>0$, which
	we can achieve by an affine transformation $\olpi$.
	Recall in \refSSec{keyIdea}
	that in the hard case, we compute the
	map $\pi = \whpi \circ \olpi$ in \refeq{circ}.
	Define the box $B_2$ and $\chb_{\max}$

	\beql{translation}
		B_2 \as Box(\olpi(F_1)) = \prod_{i=1}^n [1,\chb_i].
		\quad
		\chb_{\max} \as \max_{i=1\dd n} \chb_i.
	\eeql
	
	Using $\olpi$, we can introduce a new ODE system
		\beql{olbfy'}
			\olbfy' = \olbfg(\olbfy)
			\eeql
	with new differential variables $\olbfy=(y_1\dd y_n)$
	which are connected to $(\bfx,\bff)$ as follows:
	\beqarrys
		\olbfy &\as& \olpi(\bfx)
			&\text{(relation between $\olbfx$ and $\olbfy$)}\\ 
		\olbfg(\olbfy) &\as& J_{\olpi} \Bigcdot
					\bff(\olpi\inv(\olbfy))
			&\text{($\olbfg(\olbfy)$ is new algebraic function
				constructed from $\olpi$ and $\bff$)}\\ 
	\eeqarrys
	
	and
		\beql{olbfg}
		\olbfg(\olpi(F_1))\ge \1= [1\dd 1].
		\eeql
	Note that $(\pi(E_0), h, \pi(F_1))$ is
	an admissible triple in the $(\bfy,\bfg)$-space.
	
\bthmT[keylemma]{Radical Transform} \ \\
	\benum[(a)]
	\item For any $\bfd=(d_1\dd d_n)$, we have
	\beqarrys
	\mu_2 \big(J_{\bfg} (\pi(F_1))\big)
	&\le&
	\max\set{\tfrac{-(d_i+1)}{\chb_i}:
		i=1\dd n}\\
	&&
	+ \max_{i=1}^n \set{d_i} 
	\cdot
	\|J_{\olbfg}(\olpi(F_1))\|_2
	\cdot
	\max_{i=1}^n \set{\tfrac{(\chb_i)^{d_i+1}}{d_i}}.
	\eeqarrys
	\item If $d_1=\cdots=d_n=d$ then
	$$\mu_2 \left(J_{\bfg} (\pi(F_1))\right)
	\le
	-(d+1)\tfrac{1}{\chb_{\max}}
	+(\chb_{\max})^{d+1}
	\|J_{\olbfg}(\olpi(F_1))\|_2.
	$$
	\eenum
\ethml

\savespace{
	\bpf
	From \refLem{Jbfg}(b)
	we have for any $\bfp=(p_1\dd p_n) \in \olpi(F_1)$,
	\beql{splitJacobian1}
	J_{\bfg}(\whpi(\bfp))
	= A(\bfp) + P\inv(\bfp)
	\frac{\partial \olbfg}{\partial \bfx}(\bfp)P(\bfp)
	\eeql
	where
	$P(\bfp)=\diag\big(
	\tfrac{p_i^{d_i+1}}{d_i}: i=1\dd n \big)$
	and
	$A(\bfp) = \diag(a_1\dd a_n)$ with
	\beql{aid1}
	a_i \as -d_i(1 + \tfrac{1}{d_i}) p_i\inv
	\cdot (\olbfg(\bfp))_i. \eeql
	Thus, $A, P$ are diagonal matrices and $p_i\inv$ is well-defined
	since $\bfp\in B_2\ge \1$, \refeq{translation}.
	
	By \refLem{lognorm}(b) and \refeq{aid1}, we conclude that
	the form
	\beql{mudiag}
	\mu_2(A(\bfp))=	\mu_2(\diag(a_1\dd a_n))
	= \max\set{a_i: i=1\dd n}.
	\eeql
	
	From \refeq{olbfg}, we conclude that
	{\small
		\beqarrys
		\mu_2\big(J_{\bfg}(\whpi(\bfp))\big) 
		&=& \mu_2\left(A(\bfp) + P\inv(\bfp)\frac{\partial \olbfg}{\partial
			\bfx}(\bfp)P(\bfp)\right)\\
		&& \text{(by \refeq{splitJacobian1})} \\
		&\le& \mu_2(A(\bfp))
		+ \mu_2\left(P\inv(\bfp)\frac{\partial \olbfg}{\partial
			\bfx}(\bfp)P(\bfp)\right)\\
		&& \text{(by \refLem{lognorm}(a))} \\
		&\le& \mu_2(A(\bfp)) 
		+ \left\|P\inv(\bfp)\frac{\partial \olbfg}{\partial
			\bfx}(\bfp)P(\bfp)\right\|_2 \\
		&& \text{(by \refLem{lognorm}(b))} \\
		&\le& \max\set{\tfrac{-(d_i+1)}{\chb_i}:
			i=1\dd n}\\
		&&
		+\left\|P\inv(\bfp)\right\|\left\|\frac{\partial \olbfg}{\partial
			\bfx}(\bfp)\right\|\left\|P(\bfp)\right\|\\
		&& \text{(by \refLem{matrixnorm}(b))}\\
		&\le& \max\set{\tfrac{-(d_i+1)}{\chb_i}:
			i=1\dd n}\\
		&&
		+ \max_{i=1}^n \set{d_i} 
		\cdot
		\|J_{\olbfg}(\olpi(F_1))\|_2
		\cdot
		\max_{i=1}^n \set{\tfrac{(\chb_i)^{d_i+1}}{d_i}}.
		\eeqarrys
	}
	\epf
}

	Until now, the value of $\bfd$ in the radical map $\whpi$ was 
	arbitrary. We now specify $\bfd = \bfd(F_1)$.
	The definition of $\bfd$ is motivated by \refThm{keylemma}.
	The optimal choice of $\bfd$ is not obvious. So we make
	a simple choice by restricting
	$d_1=\cdots=d_n =d$.  In this case, we could choose the upper bound of $d$:

	\beql{d}
		\old(F_1) \as\max
			\Big\{1, \quad 2 \| J_{\olbfg}(\olpi(F_1)) \|_2 -1 \Big\}.
	\eeql

\bleml[Set-d]\
If $d \ge \old(F_1)$, we have:
\benum[(a)]
\item
	$\mu_2 \left(J_{\bfg} (\pi(F_1))\right)
	\le	 (-2+(\chb_{\max})^{d+2})
	\cdot \frac{\|J_{\olbfg}(\olpi(F_1))\|_2}{\chb_{\max}}.$
\item
	If
	$\log_2(\chb_{\max}) < \tfrac{1}{d+2}$
	then
	$\mu_2 \left(J_{\bfg} (\pi(F_1))\right)< 0$.
\eenum
\eleml

To use this lemma, we first check if choosing $d$ to be
$\old(F_1)$ satisfies
	$\mu_2 \left(J_{\bfg} (\pi(F_1))\right)< 0$.
 If so, we perform a binary search over $d \in [1, \old(F_1)]$ to 
 find an integer $d$ such that $\mu_2 \left(J_{\bfg} 
 (\pi(F_1))\right)$ is negative and as close to zero as possible. 
 Otherwise, $\pi=\id$.
	As seen in \refSSec{comparison}, this is a good strategy.

\savespace{
	\bpf
	\benum
	\item
	By \refThm{keylemma} we have
	{\small
		\beqarrays
		\mu_2\left(J_{\bfg} (\pi(F_1))\right)
		&\le& -(d+1)\frac{1}{\chb_{\max}}
		+(\chb_{\max})^{d+1}\|J_{\olbfg}(\olpi(F_1))\|_2\\
		&=& \Big(\tfrac{-(d+1)}
		{\|J_{\olbfg}(\olpi(F_1))\|_2}+(\chb_{\max})^{d+2} \Big)
		\cdot \tfrac{\|J_{\olbfg}(\olpi(F_1))\|_2}{\chb_{\max}}\\ 
		&&	 \text{(by factoring)} \\
		&\le&  \Big(-2+(\chb_{\max})^{d+2} \Big)
		\cdot \tfrac{\|J_{\olbfg}(\olpi(F_1))\|_2}{\chb_{\max}} \\
		&&
		\text{(By eqn.\refeq{d}, we have $(d+1)
			\ge 2(
			\|J_{\olbfg}(\olpi(F_1))\|_2)$)}. 
		\eeqarrays
	}
	\item
	Since $(\chb_{\max})^{d+2}<2$
	is equivalent to 
	$\log_2(\chb_{\max}) < \tfrac{1}{d+2}$,
	we conclude that
	$\mu_2 \left(J_{\bfg} (\pi(F_1))\right)< 0$.
	\eenum
	\epf
	
}

	Given an admissible triple $(E_0,h,F_1)$,
	we introduce a subroutine called \Transform$(\bff,F_1)$ to
	convert the differential equation
	$\bfx'=\bff(\bfx)$ to $\bfy'=\bfg(\bfy)$ according to above map
	$\pi$.   However, this transformation depends on the
	condition $\refeq{0ninolB1}$.  So we first define the
	following predicate \AvoidsZero($\bff,F_1$):

\Ldent\progb{
	\lline[0] 
	\AvoidsZero$(\bff,F_1) \ssa$ \true\ or \false.  
	\lline[5] INPUT: $F_1 \ib \intbox\RR^n$.
	\lline[5] OUTPUT: \true\ if and only if
	$\bf0 \notin \text{Box}(\bff(F_1))$.
}

Now we may define the transformation subroutine:

\Ldent\progb{\label{alg:transform}
	\lline[0] 
	\Transform$(\bff,F_1,\olmu_1) \ssa (\pi, \bfg, \olmu)$ 
	\lline[5] INPUT: $F_1\ib \intbox\RR^n$
				and $\olmu_1\ge \mu_2(J_\bff(F_1))$.
	\lline[5] OUTPUT: $(\pi, \mu,\bfg)$ where
	\lline[20] $\pi$ and $\bfg$ satisfy
					\refeq{circ} and \refeq{bfy'}.
	\lline[5] \myhlineX[3]{0.6}
	\lline[5] If (\AvoidsZero($\bff,F_1$)=\false\ \& $ \olmu_1\le 0$)
	\lline[10] Return ($\id,\bff, \olmu_1$)
	\lline[5] Compute $\olpi$ to satisfy \refeq{olbfg}
	\lline[5] Compute $\pi$ and $\bfg$ 
			according \refeq{circ} and \refeq{bfy'}.
	\lline[5] $\olmu \ass \mu_2(J_{\bfg}(\pi(B)))$.
	\lline[5] Return ($\pi,\bfg, \olmu$)
}
\ssect{Transformation of Error Bounds}
	We say that $B\ib\RR^n$ is a \dt{$\delta$-bounded end-enclosure}
	for an admissible triple $(E_0,h,F)$ if
		\beqarray
		\hspace*{0.5in} \text{(i)}
				&&	(E_0,h,F,B) \text{ is admissible, and} \notag \\
		\hspace*{0.5in} \text{(ii)}
				&& rad(B_0)\le rad(E_0)e^{\mu_2(J_\bff(F))\cdot h}
							+ \delta 		\label{eq:deltabounded}
		\eeqarray
		where $rad(S)$ is the radius of the circumball of $S$.
		Moreover, the $i$th stage $(E_{i-1},h_i,F_i,E_i)$
		of a scaffold $\stage$ is \dt{$\delta$-bounded} if
		$E_i$ is a $\delta$-bounded end-enclosure for
		$(E_{i-1},h_i,F_i)$.

	Given $\delta_x>0$, we want to compute a transformation
	$\delta_x\mapsto \delta_y$
	such that
	if $B$ is a $\delta_y$-bounded end-enclosure
			for $(\pi(E_0),h, \pi(F_1))$
	in the $(\bfy,\bfg)$-space, then $\pi\inv(B)$ is 
			a $\delta_x$-bounded end-enclosure
			of $(E_0,h,F_1)$ in the $(\bfx,\bff)$-space.
	The following lemma achieves this:
	
	\savespace{
		\bleml[error-bound-under-phi]
		Let $\bfp,\bfq\in B\ib\RR^n$ and $\phi\in C^1(F_1\to \RR^n)$, then 
		$\|\phi(\bfp)-\phi(\bfq)\|_2
		\le \|J_{\phi}(B)\|_2 \cdot \|\bfp-\bfq\|_2$
		\eleml
		\bpf
		\beqarrys
		\|\phi(\bfp)-\phi(\bfq)\|_2
		&\le&
		\|\phi(\bfq)+J_{\phi}(\xi)\Bigcdot(\bfp-\bfq) -\phi(\bfq)\|_2\\
		&& \text{(by Taylor expansion of $\phi(\bfp)$ at $\bfq$)}\\
		&=&\|J_{\phi}(\xi)\Bigcdot(\bfp-\bfq)\|_2\\
		&\le& \|J_{\phi}(\xi)\|_2 \cdot \|(\bfp-\bfq)\|_2\\
		&\le& \|J_{\phi}(B)\|_2   \cdot \|(\bfp-\bfq)\|_2,
		\eeqarrys
		where $\xi\in B$.
		\epf
	}
	
\bleml[error-bound-ode]
	\ \\
	Let $\bfy=\pi(\bfx)$ and 
	$$
	\mmatx[rcl]{
		\bfx 	&=&
		\ivp_{\bff}(\bfx_0,h,F_1),\\
		\bfy 	&=&
		\ivp_{\bfg}(\pi(\bfx_0),h,\pi(F_1)).
	}$$
	For any $\delta_x>0$ and any point $\bfp\in \RR^n$ satisfying 
	\beql{delta1}
	\|\pi(\bfp)-\bfy(h)\|_2
	\le \delta_y
	\as \frac{\delta_x}{\|J_{\pi\inv}(\pi(F_1))\|_2},
	\eeql
	we have 
	$$\|\bfp-\bfx(h)\|_2\le \delta_x.$$ 
\eleml
	\savespace{
		\bpf
		\beqarrys
		\|\bfp-\bfx(h)\|_2
		&=& \|\pi\inv(\pi(\bfp))-\pi\inv(\pi(\bfx(h)))\|_2\\
		&=& \|\pi\inv(\pi(\bfp))-\pi\inv(\bfy(h))\|_2\\
		&\le&  			  
		\|J_{\pi\inv}(\pi(F_1))\|_2
		\cdot \|\pi(\bfp)-\bfy(h)\|_2\\
		&& \text{(by \refLem{error-bound-under-phi})}\\
		&\le& \delta\qquad
		\text{(by condition \refeq{delta1}.)}
		\eeqarrys
		\epf
	}

	This bound transformation
	$\delta_\bfx\mapsto \delta_\bfy$
	in this lemma is encoded in the subroutine
	\refeq{transformBound}:
	
	\renewcommand{\alt}[2]{#2} 
	\renewcommand{\alt}[2]{#1} 
	\alt{ 
		\beql{transformBound}
			\hspace*{30mm}
			\includegraphics[width=0.45\columnwidth]
				{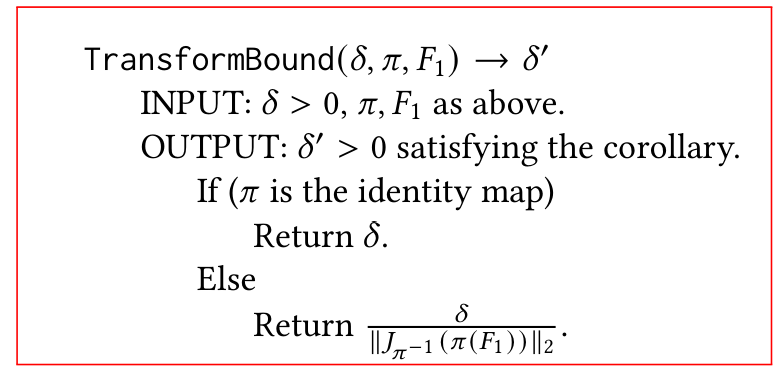}
		\eeql
	}{ 
	\Ldent\progb{
		\lline[0]  $\TransformBound(\delta,\pi,F_1) \ssa \delta'$
		\lline[5]	INPUT: $\delta>0$, $\pi, F_1$ as above.
		\lline[5]	OUTPUT: $\delta'>0$ satisfying the
							\refLem{error-bound-ode}.
		\lline[5] \myhlineX[3]{0.6}
		\lline[5]	If ($\pi$ is the identity map)
		\lline[10]		Return $\delta$.
		\lline[5]	Else
		\lline[10]		Return 
			$\frac{\delta}{\|J_{\pi\inv}(\pi(F_1))\|_2}$.
	}
	}

\ssectL[comparison]{Transformation of Enclosures}
		Let $(E_0, h, F_1)$ be an admissible triple
	that has been transformed into $(\pi(E_0),h,\pi(F_1))$.
	Thus we have a primal $(\bfx,\bff)$-space
	and a transformed $(\bfy,\bfg)$-space
	where $\bfy=\pi(\bfx)$.
	Our goal is to show how enclosures in the $\bfy$-space
	translate into enclosures in the $\bfx$-space.

	Let $\olmu^1$ be the \lognorm\ bound for $(\bff,F_1)$
	and
	$\olmu^2$ be the corresponding bound for $(\bfg,\pi(F_1))$.
	Given $\delta_x>0$, if we trace $m=m(E_0)$ to get a point $\bfQ$ such
	that $\|\bfQ-\bfx(h;m)\|\le \delta_x$ then
		\beql{end-enclosure0}
			E_1^{\std} \as Box_\bfQ\left(r_0 e^{\olmu^1 h} 
				+ \delta_x\right). \eeql 
	as an end-enclosure for $\ivp(E_0,h,F_1)$, $r_0$ radius
	of the circumball of $E_0$.  Using our $\pi$-transform we can
	first compute a point $\bfq$ such that 
	$\|\bfq-\bfy(h)\|\le \delta_y$ 
	and take its inverse, or we can take the inverse of
	the end-enclosure in $(\bfy,\bfg)$-space.  These two methods give
	us two end-enclosures:
		\begin{align}
			E_1^{\xform1}
			& \as Box_{\pi\inv(\bfq)}\left(r_0e^{\olmu^1 h} 
			+\delta_x\right), \nonumber \\
			E_1^{\xform2}
			& \as \pi\inv\left(Box_{\bfq}((r'_0+d_m) e^{\olmu^2 h} 
					+  \delta_y)\right), \nonumber\\
			E_1^{\xform}
			& \as E_1^{\xform 1}\cap E_1^{\xform2}
						\label{eq:end-enclosure12}
		\end{align}
	where $r'_0$ is the radius of the circumball of 
	$\pi(E_0)$ and $d_m\ge \|\pi(m(E_0))-m(\pi(E_0))\|$.
	To motivate these transforms, the following will analyze
	the situation in the special case $n=1$.

	\begT{Benefits of Transform ($n=1$)}{benefits}
	Consider the ODE $x' = x^e$ $(e\ne 0)$ for $e$ real with
	corresponding valid $\ivp(B_0,h)$
	where $B_0=0.2\pm 0.04$ and $h=1$.
	Apply the radical transform $y=x^{-d}$ for some real $d\ne 0$.
	Then we see that
		$y' = \frac{d}{dx}\Big(x^{-d}\Big)\cdot x'
			=-d y^{\tfrac{-e+1+d}{d}}.$
	Let $W(e,d)\as \frac{\wmax(E_1^{\std})}{\wmax(E_1^{\xform})}$
	denote the ratio of the widths of the end-enclosure
	using \refeq{end-enclosure0}
	and \refeq{end-enclosure12}.
	\refTab{wed} shows
	that the maximum value of $W(e,d)$ for a fixed $e\ne 1$ is achieved
	when $d=e-1$, i.e., $y'=-d$.

	\begin{table}[h]
	{\scriptsize
		\centering
		\begin{tabular}{r||rrrrr|rrrrr}
			$d\backslash e$ &  -2.5 & -2.0 & -1.5 & -1.0
						& -0.5 & 0.5 & \coblue{1.0} & 1.5 & 2.0 & 2.5 \\
			\midrule
			-3.5 &  \textcolor{red}{52.6123} & 1.0000 & 1.0000 & 1.0000
				& 1.0000 & 1.0000 & 1
				& 1.0000 & 1.0000 & 1.0000 \\
			-3.0 &  23.8482 & \textcolor{red}{22.1158} & 1.0000 & 1.0000
				& 1.0000 & 1.0000 & 1
				& 1.0000 & 1.0000 & 1.0000 \\
			-2.5 &  10.8027 & 10.2084 & \textcolor{red}{9.3113} & 1.0000
				& 1.0000 & 1.0000 & 1
				& 1.0000 & 1.0000 & 1.0000 \\
			-2.0 &  4.8901 & 4.7089 & 4.4287 & \textcolor{red}{3.9948}
				& 1.0000 & 1.0000 & 1
				& 1.0000 & 1.0000 & 1.0000 \\
			-1.5 &  2.2121 & 2.1707 & 2.1051 & 1.9992
				&\textcolor{red}{1.8276} & 1.0000 & 1
				& 1.0000 & 1.0000 & 1.0000 \\
			-1.0 &  1.0000 & 1.0000 & 1.0000 & 1.0000
				& 1.0000 & 1.0000 &  1
				& 1.0000 & 1.0000 & 1.0000 \\
			-0.5 &  1.0000 & 1.0000 & 1.0000 & 1.0000
				& 1.0000 & \textcolor{red}{1.5647} & 1
				& 1.0308 & 1.0168 & 1.0000 \\
			\hline
			0.5  &  1.0000 & 1.0000 & 1.0000 & 1.0000
				& 1.0000 & 1.0000 & 1
				& \textcolor{red}{1.0798} & 1.0464 & 1.0037 \\
			1.0  &  1.0000 & 1.0000 & 1.0000 & 1.0000
				& 1.0000 & 1.0000 & 1
				& 1.0611 & \textcolor{red}{1.0594} & 1.0075 \\
			1.5  &  1.0000 & 1.0000 & 1.0000 & 1.0000
				& 1.0000 & 1.0000 & 1
				& 1.0474 & 1.0492 & \textcolor{red}{1.0118} \\
		\end{tabular}
		\caption{
		This is a table of the ratios $W(e,d)$ using our 
		transform subroutines.
		The red entries are maximal for each column,
		and correspond to the choice $d=e-1$.
		Note that we exclude the column 
		for $e = 0$ since the ODE $x'=x^e=1$ is independent of $x$.
		We also excluded the row for
		$d=0$ as the radical transform $y=x^d=1$ makes
		$y$ independent of $x$. The column for $e=1$ is literally $1$
		(other values written ``$1.0000$'' are generally approximations).
		}
		\label{tab:wed}
		}
	\end{table}
	\eegT
\sssectL[comparison]{Local Experiments on Efficacy of Transformed Bounds} 
	We will compare $E_1^{\std}$ and $E_1^{\xform}$ using two
	independent ratios:
	\beql{rho}
	\rho(E_1^{\std}, E_1^{\xform}) \as 
	\Big( \tfrac{\wmax(E_1^{\std})}{\wmax(E_1^{\xform})}, 
	\tfrac{\wmin(E_1^{\std})}{ \wmin(E_1^{\xform}))}\Big).
	\eeql
	Our current experiments shows that the first ratio in 
	$\rho(E_1^{\std}, E_1^{\xform})$ is always less than
	the second ratio, and for simplicity, we only show the
	second ratio, which is denoted by
	$\sigma(E_1^{\std}, E_1^{\xform})$ in the
	last column of \refTab{transform}.

    \refTab{transform}
	compares a single step of our transform method with 
	the Standard method \refeq{end-enclosure0}.

	Each row represents a single experiment.
    The columns under $(E_0, F_1, h)$ represent an admissible triple.
	The column under $\olmu^1$ (resp.~ $\olmu^2$) represents
	the \lognorm\ bound of $F_1$ in the $(\bfx, \bff)$-space
	(resp.~ $\pi(F_1)$ in the $(\bfy,\bfg)$-space).
	The $d$ column refers to uniform exponent $\bfd=(d\dd d)$ 
	of our radical transform.  The last column
	$\sigma(E_1^{\std},E_1^{\xform})$
	is the most significant, showing the relative improvement
	of our method over $E_1^{\std}$ \refeq{end-enclosure0}.

\begin{table*}[h!]
	\centering
	{\tiny
		\btable[c|c|c|c|c|c|c|c]{
			Eg* & $E_0$ & $F_1$ & $h$ & $\olmu^1$
			& $d$ & $\olmu^2$ & $\sigma(E_1^{\std},E_1^{\xform})$ \\\hline
			
			\multirow{2}{*}{Eg1-a} & \multirow{2}{*}{ $Box_{(1,3)}(10^{-4})$}
			& $(0.95,2.95)\pm(0.05,0.05)$ 	& 0.00001 & 0.07 & 17 & -68.30 & $1.0000$ \\
			& & & & & 1 & -5.82 &  $1.0006$ \\\hline
			
			\multirow{2}{*}{Eg1-b} & \multirow{2}{*}{ $Box_{(1,3)}(10^{-4})$}
			& $(0.95,2.95)\pm(0.05,0.05)$ 	& 0.0028 & 0.07 & 17 & -68.30 & $1.0000$ \\
			& & & & & 1 & -5.82 &  $1.0000$ \\\hline

			\multirow{2}{*}{Eg2-a} & \multirow{2}{*}{$Box_{(3,-3)}(10^{-4})$}
			& $(2.95,-2.95)\pm(0.05,0.05)$ & 0.00086 & 5.90 & 23 & -140.80 & $1.0002$ \\
			& & & & & 1 & -11.00 & $1.0007$ \\\hline
			
			\multirow{2}{*}{Eg2-b} & \multirow{2}{*}{$Box_{(3,-3)}(10^{-4})$}
			& $(2.95,-2.85)\pm(0.05,0.15)$ & 0.01 & 5.93 & 23 & -370.14 & $1.0321$ \\
			& & & & & 1 & -9.20 & $1.0458$ \\\hline
			
			\multirow{2}{*}{Eg3-a} & \multirow{2}{*}{$Box_{(3,-3)}(10^{-4})$}
			& $(2.95,-2.95)\pm(0.05,0.05)$ & 0.001 & 9.75 & 20 & -177.05 & $1.0000$ \\
			& & & & & 1 & -9.87 & $1.0008$ \\\hline
			
			\multirow{2}{*}{Eg3-b} & \multirow{2}{*}{$Box_{(3,-3)}(10^{-4})$}
			& $(3.05,-2.80)\pm(0.15,0.20)$ & 0.02 & 10.64 & 20 & -163.12 & $1.0035$ \\
			& & & & & 1 & -9.39 & $1.0665$ \\\hline
			
			\multirow{2}{*}{Eg4-a} & \multirow{2}{*}{$Box_{(1.0,3.0,1.0)}(10^{-4})$}
			& $(0.95,2.95,0.95)\pm(0.05,0.05,0.05)$ & $0.001$ & 13.60 & 58 & -22.10 & $1.0102$ \\
			& & & & & 1 & -2.97 & $1.0186$ \\\hline
			
			\multirow{2}{*}{Eg4-b} & \multirow{2}{*}{$Box_{(1.0,3.0,1.0)}(10^{-4})$}
			& $(1.20,3.30,0.95)\pm(0.30,0.40,0.05)$ & $0.02$ & 13.62 & 10 & -6.01 & $1.3286$ \\
			& & & & & 1 & -3.01 & $1.3933$ \\\hline
		}
	}
	\caption{Comparison of our transform method with $E_1^{\std}$
		\refeq{end-enclosure0}. The value $\delta$ is fixed
		at $10^{-7}$ throughout.}
	\label{tab:transform}
\end{table*}

	 \refTab{transformh} further investigates the impact of the step size $h$ on the improvement ratio. In this experiment, the initial box $E_0$ is fixed, while $h$ is gradually increased (from $0.00001$ to $0.6$), and the corresponding changes in $\sigma$ are observed.

\begin{table*}[]
	\centering
	\begin{tabular}{|c|c|c|c|c|c|c|c|c|c|}
		\hline
		& $E_0$  & $0.00001$  &$0.0001$  & $0.001$  & $0.01$
			& $0.1$ & $0.2$ & $0.4$ & $0.6$  \\
		\hline
		Eg1	& $Box_{(1,3)}(10^{-4})$ & \cored{1.0006} & 1.0000  
			& 1.0000 & 1.0000 &  1.0000 &  1.0000 & 1.0000 &1.0000   \\
		\hline
		Eg2	&$Box_{(3,-3)}(10^{-4})$  & 1.0001  & 1.0008 & 1.005
			& 1.055 & 2.628  & 6.227 & 32.772&  192.823  \\
		\hline
		Eg3	&$Box_{(3,-3)}(10^{-4})$& 1.0005 & 1.0013 &  1.003
			&1.032  &  1.706 & 2.517& 7.923& 17.892  \\
		\hline
		Eg4	& $Box_{(1,3,1)}(10^{-4})$& 1.0006 & 1.0015 & 1.016
			& 1.156 &   1.737 & 3.022 &  9.027&  27.283 \\
		\hline
	\end{tabular}
	\caption{Comparison of our transform method with $E_1^{\std}$
		\refeq{end-enclosure0}. Under Increasing Step Sizes.}
	\label{tab:transformh}
\end{table*}

	From the experimental results, we can conclude the following:  
	\\ 1. Applying the transformation consistently yields a tighter
	end-enclosure.  Moreover, this improvement appears to
	grow exponentially.
	\\ 2. When the IVP system exhibits significantly faster growth in one
	coordinate direction over a certain step size range, the benefit of
	applying the transformation becomes increasingly pronounced as the
	step size grows. This is clearly observed in examples such as eg2,
	eg3, and eg4.
	 The case of eg1 with a loop trajectory (see
	 \refFig{Volterra-21-13}), when the step size is small (e.g., 
		$h=0.00001$), the system is in the positive zone region, and
		the transformation has a slight
		noticeable effect. However, for larger step
		sizes, the trajectory enters the negative zone,
		where the transformation loses its effectiveness. 
\ignore{%
\ssect{Some Examples}
    We will use the following running examples in our experiments:

		\begin{Example}[ 0.]
            This is the trivial $x'=ax$ with solution $x(t)=e^{at}$.
            Clearly, the logarithmic norm of this system is
            negative iff $a<0$.
        \end{Example}

	\begin{Example}[ 1. Volterra System]
            This is the Predator-Prey equation for
            $\bfx\in \ivp(E_0,1)$ where
            \beql{volterra}
			\bff=\mmat{x \\ y} '
			= \mmat{\phantom{-}ax(1-y) \\ -by(1-x)} ,
                    \qquad E_0 = Box_{(1,3)}( 0.1),
			\eeql
		with $a>1$ and $0<b<1$.  We choose $(a,b)=(2,1)$ and
		$E_0=Box_{(1,3)}(0.1)$ to closely
    track the AWA examples in
	\cite{moore:diffEqn:09} and
    \cite[p.13]{bunger:taylorODE:20}. 
	Thus $J_{\bff} =\mmat{2(1-y) & -2x\\ y & x-1}$ and the eigenvalues of
	$\half(J_{\bff}+ J_{\bff}^{\tr})$ are
            $$\half(1+x-2y \pm \sqrt{\Delta(x,y)}), \qquad \text{where } 
            \Delta(x,y)\as 5(x^2+y^2)-6(x+2y)+9$$
	  Note that
	  $\Delta(x,y)=5(x-\tfrac{3}{5})^2 + 5(y-\tfrac{6}{5})^2$ is
	  positive. Thus $\sqrt{\Delta(\bfq)}=\sqrt{5}\|\bfq-\bfp_0\|$
	  where $\bfp_0=(\tfrac{3}{5},\tfrac{6}{5})$.
	  Let $\ell(x,y)\as 1+x-2y$.  Thus $\ell(\bfq)> 0$ iff
	  $\bfq$ lies below the line $1+x-2y=0$.
	  \refFig{Volterra-21-13} shows
	  the trace of $\bfx(t)$ with $\bfx(0)=(3,1)$ and $t\in [0, 5.48]$
	  representing one complete
      cycle as computed by MATLAB; Bunger \cite{bunger:taylorODE:20}
      said that AWA cannot complete this computation 
	
      \FigEPS{Volterra-21-13}{0.2}{Solution to the Volterra system
	  	with $(a,b)=(1,2)$, and $\bfx(0)=(1,3)$ is shown in blue.
		The points $\bfq$ with negative
		$\mu_2(J_\bff(\bfq))$ lie above the green parabola as defined by
		the red line and red dot.}
	 
	We have $\mu_2(J_\bff(\bfq)) = \half(\ell(\bfq)+\sqrt{\Delta(\bfq)})$,
	and this can only be negative when $\ell(\bfq)<0$, i.e., when $\bfq$
	lies above the line $\ell(x,y)=0$.  The algebraic locus of the curve
	$\mu_2(J_\bff(\bfq))=0$ is given by the equation
	$|\ell(\bfq)|^2=\Delta(\bfq)$.
    \ignore{
      We have $\sqrt{5}|\ell(\bfq)|$ is the distance
	  of $\bfq$ from the line $\ell=0$.}   This equation is a parabola
	  defined by the line $\ell(x,y)=0$ and the point $\bfp_0$.
 
    \end{Example}

    \begin{Example}[ 2. Van der Pol System]
		\beql{vanderpol}
			\bff=\mmat{x \\ y} '
			= \mmat{y \\ -x^2y + y-x} ,
                    \qquad E_0 = Box_{(-3,3)}( 0.1).
			\eeql
		It could be checked that the logarithmic norm of this system is
		always positive.

        \NOignore{
              $J_f = \mmat{0 & 1 \\ -2xy -1 & 1-x^2}$
              and the two eigenvalues
              of $(J_f+ J_f\tr)/2$ are
              
              $\mmat{ 1-x^2 + \sqrt{(1-x^2)^2 + 4x^2y^2}\\
                     1-x^2 - \sqrt{(1-x^2)^2 + 4x^2y^2}
                         }$,
            and the larger eigenvalue is always positive.
        }
	\end{Example}
	
	\begin{Example}[ 3.]
		Consider the system  
		\[
		\bff=\mmat{x \\ y} '
		= \mmat{x^2 \\ -y^2 + 	7x} , 
            \qquad E_0 = [-2.0, -1.0] \times [8.0, 9.0].
		\]
		The logarithmic norm of this system is initially negative but
		becomes positive. The solution to the $x$-component is $x(t)=
		\tfrac{1}{x_0\inv -t}$ which is negative for all $t>0$ (assuming
		$x_0<0$). Thus the $y$-axis is an asymptote of the solution curve
		(red curve in \refFig{neg-pos-system}). If the end enclosure
		$F_1=Box_{\bfp_1}(w_x, w_y)$ then $w_y/w_x\to \infty$. We have
		$J_\bff=\mmat{2x& 0\\ 7&-2y}$, and the eigenvalues of
		$\half(J_\bff+J_\bff{\tr})$ are
        $(x-y) + \sqrt{(x+y)^2 + (7/2)^2}$.
		        The phase portrait of this system is seen in
		\refFig{neg-pos-system}. The solution
		$\bfx$ with $\bfx(0)=(-1.5, 8.5)$ is shown in red.
	
		\FigEPS{neg-pos-system}{0.3}{Phase Portrait of Example 4}
	
    We check that for any point $\bfp\in\RR^2$ outside
    of the second quadrant $\set{x<0, y>0}$, we have
    the property
    $\mu_2(J_\bff(\bfp))>0$.   Moreover, for
     any solution $\bfx(t)\in \ivp(E_0,1)$ there is a moment 
        $0< t_1<1$ such that
        $\mu_2(J_\bff(x(t)))$ is negative if $t<t_1$
        and positive if $t>t_1$.  
	\end{Example}

     \begin{Example}[ 4. Lorenz System]

     {\small\beql{lorenz}
       \bff = \mmat{x\\y\\z}' = 
            \mmat{\sigma(y-x)\\ 
                    x(\rho-z)-y\\
                    xy-\beta z},
        \qquad
        E_0 = Box_{(15,15,36)}(0.1),
        \quad
        (\sigma,\rho,\beta)=(10,28,8/3).  
          \eeql}

     We have
     {\small $J_\bff=\mmat{-10 & 10 & 0\\
                    28-z & -1 & -x\\
                    y  & x  & -8/3}
     $} and
    {\small $\half(J_\bff+J_\bff{\tr})=\mmat{-10 & 19-\half z & \half y\\
                    19-\half z & -1 & 0\\
                    \half y  & 0  & -8/3}
     $}.  
	 We could no longer easily compute global bounds on log norms,
	 but only obtain bounds on $\olmu =\mu_2(J_\bff(F_1))$ for given
     admissible pair $(h,F_1)$ for $E_0$
 	E.g. for {\small $(h,F_1)=(0.01, 
        [14.80, 15.12]\times [13.34, 15.10] \times [35.75, 37.45])$}
    we obtain the bound
    $\olmu\le 4.33$.
     \end{Example}

	\ignore{\begin{Example}[ 4.]
	 	Consider the IVP where   
	 	\[
	 	\bff=\mmat{x \\ y} '
	 	= \mmat{x^2-2x+4 \\ -y^2 + 2y-7x-4} ,
                 \qquad
                E_0 = [-2.1, -1.9] \times [6.9, 7.1].
	 	\]
        \end{Example}
    }
}%

%% file: inc/6refine.tex

\sectL[extend-refine]{ \Extend\ and \Refine\ Subroutines}
	In this section, we present algorithms for
	\Extend\ and \Refine\
	as specified in \refeqs{extend}{refine}.
	But we first need to describe the refinement substructure
	inside the scaffold; this was deferred from
	\refSSec{scaffold}.
	
\ssect{Refinement Substructure}

	Let $\stage=(\bfE,\bft,\bfF,\bfG)$ be an $m$-stage scaffold
		as in \refeq{Ei-1}).  
	So $\bft=(t_0<t_1<\cdots< t_m)$ with the $i$-th quad is 
		$(E_{i-1},\Delta t_i, F_i, \cored{G_i})$.
	Here are the details\footnote{
		The first 2 components are computed only once; 
		the remaining components (in red font) are 
		updated with each phase of the refinement.
	}
	of the $i$th refinement substructure $G_i$:
		\beql{Gi}
			G_i = G_i(\stage) =((\pi_i,\bfg_i),
				\cored{ (\ell_i, \ol\bfE_i, \ol\bfF_i),
				\ol\bfmu^1_i,\ol\bfmu^2_i,
				\delta_i, h\euler_i})\eeql 
	where:
	\begin{itemize}[leftmargin=1.5cm]
	\item[{[G0]}] The pair $(\pi_i, \bfg_i)$
		represents the radical transformation of the $i$th stage.
		More precisely, $\pi_i$ is the map $\pi$
		in \refeq{circ} that transforms
		the original ODE $\bfx'=\bff(\bfx)$ to $\bfy'=\bfg_i(\bfy)$
		where $\bfy=\pi_i(\bfx)$ (see \refeq{bfy'}).
	\item[{[G1]}]
		The triple $(\ell_i, \ol\bfE_i, \ol\bfF_i)$ is called 
		the $i$th \dt{mini-scaffold} where
		 $\ell_i\ge 0$ called the \dt{level}.  The level starts
		 from $0$ and is incremented with each ``phase'' of \Refine,
		 and across multiple \Refine's. 
		 In other words, it is never reset.
	\item[{[G2]}]
		The time span $[t_{i-1},t_i]$ is subdivided
		into $2^{\ell_i}$ ministeps, each of ministep size
				$$\whh_{i} \as \Delta t_i/2^{\ell_i}.$$
	\item[{[G3]}]
		$\ol\bfE_i, \ol\bfF_i$ are arrays of length $2^{\ell_i}$
		representing these $2^{\ell_i}$ admissible quads
				$$(\ol\bfE_i[j-1],\whh_i, \ol\bfF_i[j], \ol\bfE_i[j]),
				\qquad (j=1\dd 2^{\ell_i})$$
		of stepsize $\whh_i$, where $\ol\bfE_i[0]=E_i(\stage)$
		(cf.~\refeq{Ei-1}).
		These arrays are updated in one of two ways:
		they are usually updated using
		Bisection, and periodically, by the Euler-tube Method.
	\item[{[G4]}]
		$\ol\bfmu^1_i,\ol\bfmu^2_i$
		are arrays of length $2^{\ell_i}$.
		For $j=1\dd 2^{\ell_i}$, each
			$\olbfmu^1[j]$ is an upper bound
			on $\mu_2(J_{\bff}(\ol\bfF_i[j])$.  
			Then $\olbfmu^2[j]$ is simply the 
			transformed bound $\mu_2(J_{\bfg}(\pi(\ol\bfF_i[j]))$

	\item[{[G5]}]
	Finally, the pair $(\delta_i, h\euler_i)$ is used to trigger the
	periodic invocation of the Euler Tube method. These quantities are
	computed from \refeq{h1} see \refLem{eulerStep} and are chosen so
	that, if the mini-step size $\whh_{i}$ does not exceed $h\euler_i$,
	the Euler Tube method constructs a $\delta_i$-tube that contains the
	exact solution over the mini-step. In this case we invoke the Euler
	Tube method to update the enclosures using the $\delta_i$-bounds.
	Immediately after the call we halve $\delta_i$ and recompute
	$h\euler_i$ to determine the next trigger threshold.
	We perform these constructions in the transformed space of $\bfy'=\bfg(\bfy)$, since we use  \refeq{end-enclosure12} to get the tighter enclosures.

	\end{itemize}
	
	Remarks on Design: the strict separation of the scaffold stages
	(as represented by the fixed sequence
	$(t_0<t_1<\cdots<t_m)$) from its ministeps
	is motivated by the fact that the $i$th stage 
	depends on the parameters $(\pi_i, \bfg_i)$
	in [G0].  We need to compute $\pi_i$ and $\bfg_i$ using
	symbolic methods\footnote{
			A numerical approach via automatic differentiation is,
			in principle possible, but it gives extremely poor bounds.
	}
	using symbolic manipulations of the expressions in $\bff$.
	Since this computation is expensive, we avoid 
	splitting up stages into smaller stages.
	The	trigger mechanism in [G5] allows us to
		combine the different powers of the Bisection
		and Euler Tube method (as each method is individually
		capable of reaching our termination goals).
		Bisection is unconditionally applicable to give 
		enclosure bounds, but gives poorer bounds than
		Euler Tube.  On the other hand, Euler
		Tube requires ministep size to be less than
		$h\euler_i$, which may be pessimistically small.
	
\ignore{ OLD:
	we compute transformation parameters $\pi_i$ and $\bfg_i$.
	It turns out that to compute $\pi_i$ and $\bfg_i$, it is
	necessary\footnote{
		A numerical approach via automatic differentiation is
		in principle possible, but this gives extremely poor bounds.
		Instead, we need symbolic manipulations
		of the expressions in $\bff$.
	}
	using symbolic methods. Since this computation is expensive,
	we do not refine stages by splitting a stage into two or
	more stages.  Instead, we use
	a ``light-weight'' approach encoded in the
	refinement substructure $G_i$
	that does not recompute $\pi_i$ and $\bfg_i$.
	Specifically, the
	time interval $\Delta t_i$ is uniformly subdivided into
	$2^{\ell_i}$ \dt{mini-steps} where $\ell_i$ is the \dt{level}.
	For each mini-step, we compute the full enclosure
	$\bfF_i$, end enclosure $\bfE_i$, and their associated logarithmic
	norms (\lognorm) $\ol\bfmu_1$ (in the original $\bfx$-space)
	and $\ol\bfmu_2$ (in transformed $\bfy$-space).
	Here are the details of $G_i$:
		\beql{Gi}
			G_i \as G_i(\stage) =(\pi_i,\bfg_i;
					\cored{\ol\bfmu^1_i,\ol\bfmu^2_i,\delta_i,
				h\euler_i, (\ell_i, \bfE_i, \bfF_i)})\eeql 
	where $\bf{\ol\bfmu}^1_i,\bf{\ol\bfmu}^2_i, \bfE_i, \bfF_i$
	are arrays of length $2^{\ell_i}$ and the parameters in red are
	extra data needed by the \Refine\ subroutine below.
	We call $(\ell_i, \bfE_i, \bfF_i)$ the \dt{mini-scaffold}.
	}

	\ignore{
		\dt{On the $\ell\bf{Flag}$:}
		It turns out that the value of $h\euler_i$ is non-increasing
		over time; we set the boolean flag $\ellFlag_i$ to be
		$\false$ if and only if 
		$h\euler_i$ is unchanged in the current phase.
		According to our code for $\stage.\Refine$,
		we see that in the next phase, we do not do any refinement except
		to update the value $h\euler_i$.
		In particular $\ell_i$ is not incremented.
	}
\ssect{\Extend\ Subroutine}
	We now implement the 
	$\stage.\Extend()$ subroutine whose 
	header \refeq{extend} was described in the overview:
	
	{\scriptsize
		\Ldent\progb{
			\lline[0] $\stage.\Extend(\veps_0,H)$ 
			\lline[5] INPUT:  $m$-stage scaffold $\stage$,
					$\veps>0, H>0.$
			\lline[5] OUTPUT:  $\stage'$ is a $m+1$-stage 
					extension $\stage$ such that
			\lline[10] $\Delta t_{m+1}(\stage')\le H$, and 
					$(\Delta t_{m+1}(\stage'), E_{m+1}(\stage')$
					is an $\veps$-admissible pair for $E_m(\stage)$.
			\lline[5] \myhlineX[3]{0.6}
			\lline[5]  $(\wh{h},F_1)\ass \stepA(E_m(\stage),\veps_0,H)$.
			\lline[5] $\olmu_1\ass \mu_2(J_{\bff}(F_1))$. 
			\lline[5]  $E_1\ass \stepB(E_m(\stage),\wh{h},F_1)$. 
			\lline[5]  $(\olmu_2,\pi,\bfg)
						\ass \Transform(\bff,F_1,\olmu_1)$.
			\lline[5]  $\delta'_1\ass
						\TransformBound(\veps_0, \pi, F_1)$
			\lline[5]  $h_1\ass h\euler(\wh{h},\|\bfg^{[2]}(\pi(F_1))\|,
						\olmu_2,\delta'_1)$.
						\Commentx{ See \refeq{h1}.}
			\lline[5] 	Let $\stage_{m+1}\ass (t_m+h_1,E_1,F_1)$ and
						$G_{m+1}\ass ((\pi,\bfg),
							\cored{(0,(E_m(\stage),E_1),(F_1)),
								\olmu_1,\olmu_2,\veps_0, h_1})$.
			\lline[5] 	Return $\stage; (\stage_{m+1},G_{m+1})$.
	}}

	It is a straightforward invocation of \stepA\ to 
	add a new stage to $\stage$, and \stepB\ to compute
	a end-enclosure.  NEEDS more explanation...

\ignore{
	correctness lemma ... BUT add some words!

	This is NEW!!! Explain what is at issue...
	What does it mean for $G_i$ to be correct?
\bleml[extend]
	ROUGH:
	The subroutine $\stage.\Extend(\veps_0)$ is correct:
		the $m$-stage scaffold is transformed to a
		$(m+1)$-stage scaffold with the parameter
		$h_1$ correctly initialized.
		WHAT DOES it mean?  Well, $h_1$ is the promise that
		if the Euler step size is less than $h_1$, then
		the result would be $\veps_0$-bounded.
\eleml
}

\ssect{\Refine\ Subroutine}\label{ssec:refine}
	Finally we implement the 
	$\stage.\Refine(\veps_0)$ subroutine with the
	header \refeq{refine} in the overview.
	It ensures that
	the end-enclosure of $\stage$ has max-width $\le \veps_0$.
	Each iteration of the main loop of \Refine\ is called a \dt{phase}.
	If $\stage$ has $m$ stages, then in each phase, 
	we process stages $i=1\dd m$ in this order.
	Recall the\footnote{
		Viewing the $i$th stage as a \bigStep,
		the mini-scaffold represent \smallStep s of 
		the $i$th stage. 
	} 
	mini-scaffold $(\ell_i, \ol\bfE_i,\ol\bfF_i)$ of stage $i$ above.
	This mini-scaffold has a uniform 
	time step of size $(\Delta t_i)\cdot 2^{-\ell_i}$.
	See \refFig{d-q} for illustration. 
	We will call the following
	$\stage.\Bisect(i)$ to perform a bisection of each mini-step:
	%
	\renewcommand{\alt}[2]{#1} 
	\renewcommand{\alt}[2]{#2} 
	\alt{ 
		\beql{bisect}
			\hspace*{30mm}
			\includegraphics[width=0.45\columnwidth]{figs/bisect}
		\eeql
	}{ 
	\Ldent\progb{
		\lline[0] $\stage.\Bisect(i)$  
		\lline[5] INPUT: $i$ refers to the $i$th stage of $\stage$.
		\lline[20]  Let
				$((\pi_i,\bfg_i),
				(\ell_i, \ol\bfE_i, \ol\bfF_i),
				\olbfmu^1_i,\olbfmu^2_i, \delta_i, \wh{h}_i)
					\ass
							G_i(\stage)$
		\lline[25] be the $i$th refinement substructure
		\lline[5] OUTPUT:  each mini-step of the $i$th stage
		\lline[20] is halved and $\olbfmu^1_i, \bfE_i, \bfF_i$
						are updated.
		\lline[5] \myhlineX[3]{0.6}
		\lline[5]  Initialize three new vectors
					$\bfmu=[]$,
					$\bfE'=[\bfE_i[0]]$ and $\bfF'=[]$.
		\lline[5]  $h \ass (\Delta t_i)2^{-\ell_i-1}$
		\lline[5]  For $j=1\dd 2^{\ell_i}$,
		\lline[10]	\commenT{First half of $j$ step}
		\lline[13] 	$tmpF_1 \ass \sum_{i=0}^{k-1} \left(
					[0,h]^i \bff^{[i]}(\bfE'.back  ())
								+ [0,h]^k \bff^{[k]}(\bfF_i[j]) \right).$
		\lline[13] $\mu'\ass \mu_2(J_{\bff}(tmpF_1))$;
						$\bf{\bfmu}.\text{push\_back}(\mu')$.
		\lline[13]  	$E\ass \stepB(\bfE'.back  (), h, tmpF_1)$.
		\lline[13]  	$\bfE'.\text{push\_back}(E)$;
						  	$\bfF'.\text{push\_back}(tmpF_1)$;
		\lline[10]	\commenT{Second half of $j$ step}
		\lline[13]		$tmpF_2 \ass \sum_{i=0}^{k-1} \left(
					[0,h]^i \bff^{[i]}(E)
							+ [0,h]^k \bff^{[k]}(\bfF_i[j]) \right).$ 
		\lline[13] 	 $\mu'\ass \mu_2(J_{\bff}(tmpF_2))$;
						$\bf{\bfmu}.\text{push\_back}(\mu')$.
		\lline[13]  	$E\ass \stepB(E, h, tmpF_2)$
		\lline[13]  $\bfE'.\text{push\_back}(E)$;
						$\bfF'.\text{push\_back}(tmpF_2)$;
		\lline[5]  $(\ol\bfmu^1_i,\ell_i, \ol\bfE_i, \ol\bfF_i)
						\ass (\bfmu,\ell_i+1, \bfE', \bfF')$
					\Commentx{mini-scaffold is updated}
		}
	}
	In the above code, we view
		$\bfmu$, $\bfE'$ and $\bfF'$ as
	vectors in the sense of \ttt{C++}.
	We append an item $E$ to the end of vector $\bfE'$ by calling
	$\bfE'.\text{push\_back}(E)$ and $\bfE'.\text{back()}$
	returns the last item.
	When $(\Delta t_i)2^{-\ell_i}$ is less than the bound
	in \refeq{h1}, instead of bisection,
	we can use the \eulertube\ subroutine (to be described next)
	to update the data
		$E_i(\stage), F_i(\stage), \ol\bfmu^1, \ol\bfmu^2$

	{\scriptsize
		\Ldent\progb{
			\lline[0] $\stage.\eulertube(i)$
			\lline[5] INPUT: $i$ refers to the $i$th stage of $\stage$.
			\lline[5] OUTPUT: 
					refine the $i$th stage using \refLem{delta-distance},
			\lline[20] such that the $i$th stage is
					$\delta(G_i(\stage))$-bounded
					(see \refeq{deltabounded})
			\lline[10]
			(Note: $E_i(\stage), F_i(\stage), G_i(\stage)$ are modified)
			\lline[5] \myhlineX[3]{0.6}
			\lline[5]     	Let $(\pi,\bfg,\ol\bfmu^1,\ol\bfmu^2,
			\delta, \wh{h  }, \ell, 
			\bfE, \bfF)$ be $G_i(\stage)$
			\lline[5] 			Let $Ball_\bfp(r_0)$ be the
			circumscribing ball of $\bfE[0]$
			\lline[10]          and $Ball_{\bfp'}(r'_0)$ be the
			circumscribing ball of $\pi(\bfE[0])$.
			\lline[5]			$\bfq\ass \pi(\bfp)$, $d\ass 
			\|\bfq-m(Box(\pi(\bfE[0])))\| $.
			\lline[5] Let $H$ be the step size each mini-step of $\stage[i]$.
			\lline[5]			For ($j=1\dd 2^{\ell}$)
			\lline[10]	$\bfq\ass \bfq+\bfg(\bfq)H$,
			\lline[10]				$\delta_1
			\ass \TransformBound(\delta, \pi, \bfF[j])$.
			\lline[10]          $\ol\bfmu^2[j]
			\ass \mu_2(J_{\bfg}(\pi(\bfF[j])))$. 
			\lline[10]{ 		$r_1\ass r_0e^{j\ol\bfmu^1[j] H}+\delta$;
				$r'_1\ass (r'_0+d)e^{j\ol\bfmu^2[j] H}+\delta_1$}
			\lline[10] 				$B\ass Box(	r_1)$;
			$B'\ass  Box(r'_1)$.
			\lline[10]  			$\bfF[j]\ass \bfF[j]
			\cap \pi\inv(Box(center(\pi(\bfE[j-1]))+B',\bfq+B')$
			\Commentx{Full-enclosure for mini-step}
			\lline[10]        $\ol\bfmu^1[j]\ass \mu_2(J_{\bff}(\bfF[j]))$.
			\lline[10]				$\bfE[j]\ass \bfE[j]\cap    
			\pi\inv(\bfq+B')	\cap (\pi\inv(\bfq)+B)$.
			\Commentx{End-enclosure for mini-step}
	}}
	
	Observe that \SubrSeven\ performs
	all its Euler computation in transformed space, and
	only pulls back the enclosures back to primal space.
	It turns out that \SubrSeven\ is extremely efficient
	compared to \Bisect, and moreover, it ensures
	that the $i$th stage is now $\delta_i$-bounded.
	
	We are ready to describe the \Refine\ subroutine:

	\begin{figure}
		\centering
		\includegraphics[width=0.7\linewidth]{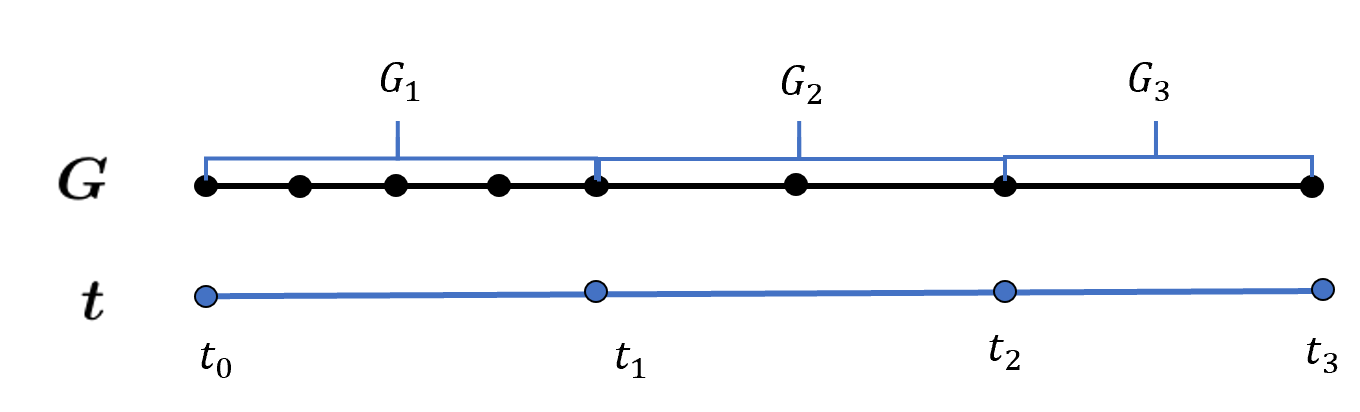}
		\caption{$3$-stage scaffold $\stage$ with $\ell_1=2$,
			$\ell_2=1$ and $\ell_3=0$ in $G$.}
		\label{fig:d-q}
	\end{figure}

%
	

{\scriptsize
	\Ldent\progb{
		\lline[0]  $\stage.\Refine(\veps_0)$
		\lline[5]  INPUT:  $m$-stage scaffold $\stage$ and $\veps_0>0$.
		\lline[5]  OUTPUT: $\stage$ remains an $m$-stage scaffold
		\lline[25]           that satisfies $\wmax(E_{m})\le \veps_0$.
		\lline[5]  \myhlineX[3]{0.6}
		\lline[5]  $r_0\ass \wmax(E_m(\stage))$.
		\lline[5]  While ($r_0> \veps_0$)
		\lline[10]    For ($i=1\dd m$) \Commentx{Begin new phase}
		\lline[15]      Let $\ell \ass G_i(\stage).\ell$ and
		$\Delta t_i \ass \bft(\stage).t_i - \bft(\stage).t_{i-1}$.
		\lline[15]      $H\ass (\Delta t_i)2^{-\ell}$.
		\lline[15]      If ($H>\wh{h}$)
		\lline[20]        $\stage.\Bisect(i)$
		\lline[15]      Else
		\lline[20]        $\stage.\SubrSeven(i)$
		\lline[20]        Let $G_i(\stage)$ be
			$((\pi,\bfg),\cored{ (\ell,\ol\bfE, \ol\bfF),
					\ol\bfmu^1,\ol\bfmu^2,\delta, \wh{h}})$
			after the Euler Tube event.
		\lline[20]        \commenT{We must update $\delta$ and $\wh{h}$
					for the next Euler Tube event.}
		\lline[20]        $G_i(\stage).\delta\ass \delta/2$
					\Commentx{(Updating the $\delta$ target is easy)}
		\lline[20]        \commenT{Updating $\wh{h}$ to reflect the new
				$\delta$ requires six preparatory steps: (1)--(6)}
		\lline[20]        (1)\quad $E_i(\stage)\ass \bfE[2^{\ell}]$
		\Commentx{End-enclosure for stage}
		\lline[20]        (2)\quad $F_i(\stage)\ass
					\bigcup_{j=1}^{2^\ell} \bfF[j]$
				\Commentx{Full-enclosure for stage}
		\lline[20]       (3)\quad $\mu^2_2\ass \max\{\ol\bfmu^2[j]\;:\;
			j=1\dd 2^\ell\}$
		\lline[20]        (4)\quad $\delta'_1\ass
			\TransformBound(\delta,\pi, F_i(\stage))$
			(cf.~\refeq{transformBound}).
		\lline[20]        (5)\quad $\ol{M}\ass
			\big\|\bfg^{[2]}(\pi(F_i(\stage)))\big\|_2$
		\lline[20]        (6)\quad $htmp \ass \min\big\{\Delta t_i,\;
			h_{\mathrm{euler}}(\wh{h},\ol{M},\mu^2_2,\delta'_1)\big\}$
			\Commentx{(see Euler Tube lemma, \refLem{eulerStep})}
		\lline[20]        $G_i(\stage).\wh{h}\ass htmp$
				\Commentx{Finally, update $\wh{h}$}
		\lline[10]    \hspace*{-3.0mm}\Commentx{End of For-Loop}
		\lline[10]    $E_0(\stage)\ass \half E_0(\stage)$
		\lline[10]    $r_0\ass\wmax(E_m(\stage))$
		\lline[5]     \hspace*{-3.0mm}\Commentx{End of While-Loop}
		\lline[5]     Return $\stage$
	}
	}

	\bthml[refine]
	The subroutine $\stage.\Refine(\veps_0)$ is correct.
	In particular, it halts.
	\ethml
		
	This proof is a bit subtle, and can be illustrated
	by the following \dt{phase-stage} diagram:
	\beql{phase-stage}
		\hspace*{30mm}
		\includegraphics[width=0.45\columnwidth]{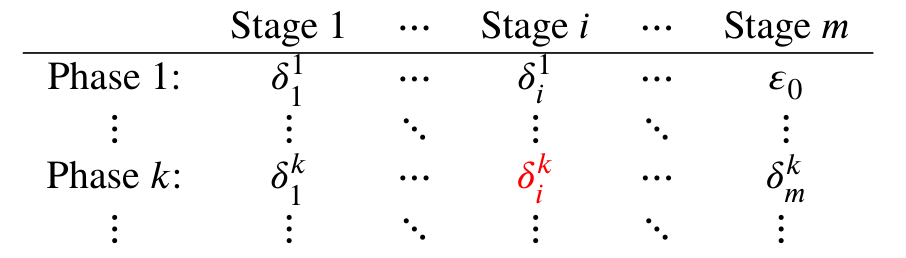}
		\eeql
	Each phase will refine the stages $1,2\dd m$ (in this order).
	The $i$th stage in phase $k$ has a ``target bound''
		$\delta^k_i>0$ stored as $G_i(\stage).\delta$.
	For $k=1$, $\delta^1_i$ is $\veps_0$ if $i=m$
		and otherwise inherited from $(m-1)$-stage substructure
		before $\Extend(\veps_0,H)$.
	For $k>1$, $\delta^k_i$ halved after a call to
		\eulertube\ subroutine (see $\Refine(\veps)$).
	The proof in the
	appendix will show that $\lim_{k\to\infty}\delta^k_i=0$ for each
		$i=1\dd m$, which will contradict the non-halting of \Refine.
	\ignore{ 
			\btable{
				& Stage $1$ & $\cdots$ & Stage $i$
								& $\cdots$ & Stage $m$\\\hline
			Phase $1$: & $\delta^1_1$ & $\cdots$ &	$\delta^1_i$
								& $\cdots$ & $\veps_0$	\\
				$\vdots$ & $\vdots$	& $\ddots$ &$\vdots$
								& $\ddots$&	$\vdots$ \\
			Phase $k$: & $\delta^k_1$	& $\cdots$ & \cored{$\delta^k_i$}
								& $\cdots$ & $\delta^k_m$\\
				$\vdots$ & 	$\vdots$	& $\ddots$ & $\vdots$
								& $\ddots$ & $\vdots$	
			}
	}%
	
\ignore{
	\textbf{	Part 1: Standard IVP Solver Structure (The Extend
	Subroutine)}\\
	The first part operates similarly to conventional IVP solvers,
	iterating modified stepA and stepB procedures. We refer to each
	iteration as a major step or stage, indexed by $i$, corresponding to
	$\stage[i]$. This process is encapsulated in the `\Extend` subroutine:
	repeatedly calling `\Extend` executes the sequence of stepA and stepB
	operations to solve the IVP. During this phase, we initialize
	$G_{m+1}$ for use in Part 2. Importantly, the components $\pi$ and
	$\bfg$ within $G_{m+1}$ remain fixed during Part 2. This design
	choice minimizes symbolic computations involving $\bfg$, which are
	computationally expensive in our current implementation.
	\\
	\textbf{Part 2: Enclosure Refinement (The \Bisect\ Subroutine  Lemma
	Application)}\\
	The second part refines the enclosures obtained for each stage (i)
	from Part 1. As discussed in Section StepB, we employ two methods for
	calculating end-enclosures: the \stepB algorithm using
	\refeq{taylor-full} (computing full enclosures), and the method from
	\refLem{delta-distance}. Comparative results in \reftab{deltafull}
	indicate the superiority of the \refLem{delta-distance} method.
	However, it requires the step size to be below a fixed threshold
	$h_i$ (defined in \refeq{h1}), initialized during \Extend.\\
	Therefore, our refinement process proceeds as follows:
	\\ 1.  Initial Bisection: We apply the \Bisect\ subroutine to tighten
	the end-enclosure. This subdivides the major step into mini-steps,
	indexed by j.
	\\ 2.  Mini-step Computation: For each mini-step j, we compute the
	end-enclosure and full-enclosure using \stepB and
	\refeq{taylor-full}. Concurrently, we compute and store the
	logarithmic norm value for each full-enclosure in $\bfmu^1$, as
	required by \stepB.
	\\ 3.  Threshold Application Transformation: Once bisection reduces
	the step size below $h_i$, we switch to the \refLem{delta-distance}
	method to compute the end-enclosure and full-enclosure. At this
	point, we also apply a transformation step to further tighten the
	enclosures and update $\ol\bfmu^2$.
	\\ 4.  Step Size Update: Finally, we update $h_i$. For each mini-step
	j, we compute an Euler step size based on \refeq{h1}. The new $h_i$
	is set to the minimum of these values. Since \refeq{h1} depends on
	$\delta$, we may iteratively bisect $\delta$ to achieve a smaller
	$h_i$.
}


%% file: inc/7mainAlgo.tex

\sectL[end-enclosure]{The Main Algorithm and Experiments}
	The following is our algorithm to
	solve the \endEncIVP\ problem of \refeq{endEncProb}:
	
	{\small
	\Ldent\progb{\label{alg:endEnc}
		\lline[0]  $\endEncAlgo_\bff(B_0, \veps_0) \ssa(\ulB_0, \olB_1)$:
		\lline[5] INPUT: $\veps_0>0, B_0\in \intbox\RR^n$
		\lline[20] such that $\ivp_\bff(B_0,1)$ is valid
		\lline[5] OUTPUT:	$\ulB_0,\olB_1 \in \intbox\RR^n$,
							$\ulB_0\ib B_0$, $\wmax(\olB_1)<\eps$
		\lline[20] 			and $\olB_1$ is an end-enclosure of
							$\ivp(\ulB_0,1)$
		\lline[5] \myhlineX[3]{0.6}
		\lline[5] $\stage\ass Init(\bff,B_0,\veps_0)$
				\Commentx{Initialize a $0$-stage scaffold}
		\lline[5]  $t\ass 0$
		\lline[5] While $t<1$ 
		\lline[10] $\stage.\Extend(\veps_0,1-t)$
		\lline[10] $\stage.\Refine(\veps_0)$
		\lline[10] $t\ass \bft(\stage).\text{back()}$
		\lline[5] Return $(E_0(\stage), \bfE(\stage).\text{back()})$
	}}

	\bthml[correct-main]
			Algorithm $\endEncAlgo(B_0, \veps_0)$ halts,
			provided the interval computation of $\stepA$
			is isotonic.  The output is also correct.
	\ethml

\ssectL[implement]{Implementation and Limitations}
	We implemented \ourAlgo\ in \Cpp.
	Our implementation follows the explicitly described
	subroutines in this paper.  There are no hidden hyperparameters
	(e.g., our step sizes are automatically adjusted).
	Our eventual code will be open-sourced in \cite{core:home}.
	Implementation of the various numerical formulas
	such as Taylor forms implicitly call interval methods
	as explained in \refSSec{implicit}.
	The radical transform requires symbolic computation
	(\refSec{extend-refine})
	which we take from the \ttt{symEngine} library
	(\myHrefx{https://symengine.org/}).

	\dt{Limitations.}
	The main caveat is that we use machine arithmetic (IEEE standard).
	There are two main reasons. First, 
	this is necessary to have fair comparisons to existing software
	and we rely on library routines based on machine arithmetic.
	In principle, we can implement
	our algorithm using arbitrary precision number types
	(which will automatically get a hit in performance
	regardless of needed precision).

\ssectL[experiment]{Global Experiments}

	In previous sections, we had tables of
	experimental results evaluating ``local'' (1-step) operations.
	In this section, we show three tables (A, B, C) that solve
	complete IVP problems over time $t\in [0,1]$ (with one exception
	in Table C).
	Table A compares our \stepA/\stepB\ with standard
	methods.  Table B is an internal evaluation
	of our transform and \Bisect/\eulertube\ heuristic.
	Table C compares our complete algorithm with
	existing validated IVP software after the initial subbox
	$\ulB_0$ has been fixed.
	The problems are from \refTab{problems}.
	We informally verify our outputs by tracing
	points using \capdCr's code to see that their outputs
	are within our end-enclosures.
	Timings are taken on a laptop with a
	13th Gen Intel Core i7-13700HX 2.10 GHz processor and 16.0 GB RAM.

	Our tables indicate two kinds of error conditions:
	\cored{Timeout}
	and \cored{No Output}.
	The former means that the code took more than 1 hour to run;
	the latter means that the code stopped without producing an output.

	In each table, we compare our main enclosure algorithm,
	denoted \Ours, with some algorithm $X$, where
	$X$ may be a variant of \Ours\ or other validated IVP software.
	We define
	\dt{speedup over $X$} as
	\[
		\sigma(X)\as
		\frac{\mbox{\scriptsize Time}(X)}
		{\mbox{\scriptsize Time}(\mbox{\scriptsize Ours})}.
	\]


	\dt{TABLE A:}
	We compare the relative computation times
	of \Ours\ against $\Ours/\stepA_0$ and $\Ours/\stepB_0$.
	Here $\Ours/\stepA_0$ denotes the algorithm where
	we replace $\stepA$ by $\stepA_0$ in $\Ours$.
	Similarly for $\Ours/\stepB_0$.
	Recall that $\stepA_0$ is the non-adaptive StepA in
	\refSSec{stepA},
	and $\stepB_0$ is the direct method of \refeq{direct}.

	The speedup $\sigma(\Ours/\stepB_0)$
	is a useful measure of the relative performance of $\stepB$
	and $\stepB_0$ when $\Ours$ and $\Ours/\stepB_0$
	have approximately the same number of phases.
	We therefore also include the statistic
	\[
		\rho(\Ours/\stepB_0)\as
		\frac{\mbox{\scriptsize phases(Ours/stepB}_0)}
		{\mbox{\scriptsize phases(Ours)}}.
	\]

	We conclude from Table A that our \stepA\ significantly improves
	efficiency ($\sigma(\Ours/\stepA_0)$).
	Moreover, \stepB\ also yields a noticeable performance
	gain and can reduce the number of required phases
	($\sigma(\Ours/\stepB_0)$ and $\rho(\Ours/\stepB_0)$).

\begin{table*}[] 
	\centering
	{\tiny
		\begin{tabular}{c|c|c|c|>{\color{red}}c|>{\color{red}}c|c}
			Example & $E_0$ & $\veps$ & $\Time(\Ours)$
			& $\sigma(\Ours/\stepA_0)$
			& $\sigma(\Ours/\stepB_0)$
			& $\rho(\Ours/\stepB_0)$
			\\\hline\hline

			\multirow{4}{*}{Eg1}
			& \multirow{4}{*}{$Box_{(1,3)}(0.1)$}
			& 0.1     & 0.018  & 5.44  & 1.00 & $4/4$ \\
			&         & 0.01    & 0.073  & 1.57  & 1.08 & $5/5$ \\
			&         & 0.001   & 0.643  & 2.465 & 1.75 & $8/7$ \\
			&         & 0.0001  & 12.803 & 1.49  & 1.03 & $9/9$
			\\\hline\hline

			\multirow{4}{*}{Eg2}
			& \multirow{4}{*}{$Box_{(-3,3)}(0.1)$}
			& 0.1     & 0.031  & 653.22  & 1.00  & $4/4$ \\
			&         & 0.01    & 0.086  & 334.875 & 1.025 & $7/7$ \\
			&         & 0.001   & 1.437  & $>1000$ & 1.074 & $10/10$ \\
			&         & 0.0001  & 11.491 & $>1000$ & 1.102 & $13/13$
			\\\hline\hline

			\multirow{4}{*}{Eg3}
			& \multirow{4}{*}{$Box_{(-1.5,8.5)}(0.01)$}
			& 10.0 & 0.043 & $>1000$ & 1.03 & $1/1$ \\
			&      & 5.0   & 0.027   & $>1000$ & 1.04 & $1/1$ \\
			&      & 1.0   & 0.022   & $>1000$ & 1.06 & $1/1$ \\
			&      & 0.1   & 2.159   & $>1000$ & 5.81 & $3/3$
			\\\hline\hline

			\multirow{4}{*}{Eg4}
			& \multirow{4}{*}{$Box_{(15,15,36)}(0.001)$}
			& 10.0 & 0.142 & $>1000$ & 1.24 & $1/1$ \\
			&      & 5.0   & 0.130   & $>1000$ & 2.24 & $5/1$ \\
			&      & 1.0   & 0.122   & $>1000$ & 14.38 & $7/1$ \\
			&      & 0.1   & 0.205   & $>1000$ & $>1000$
			& \cored{Timeout}
			\\\hline
		\end{tabular}
	}
	\\[2mm]
	\dt{Table A}:
	Comparison of \stepA\ and \stepB\ with $\stepA_0$
	and the direct method, all run with $order=20$.
	\label{tab:combined}
	\label{tab:A}
\end{table*}


	\dt{TABLE B:}
	We conduct experiments to show the benefits of various
	techniques used in our algorithm.
	The algorithms $X$ being compared in Table B differ
	from \Ours\ only in the use of variants of
	the \texttt{Refine} subroutine.

	Specifically,
	$X=\Standard$ uses \(E_1^{\std}\) in
	\refeq{end-enclosure0}
	to compute the end-enclosure,
	without performing our \Bisect\ subroutine.
	$X=\StandardT$ is similar, except that it uses
	\(E_1^{\text{xform}}\)
	in \refeq{end-enclosure12}.
	Similarly,
	$X=\OurNoT$ represents a variant of our algorithm with the transform
	step disabled (i.e., the transformation \(\pi\) is set to the
	identity map).
	Finally, $X=\OurNoSubr$ is our algorithm with
	\SubrSeven\ disabled.

	We run all experiments with order $k=3$ because with high order
	(e.g., $k=20$), the number of stages is too small
	to clearly expose the effects of these refinement techniques.

\begin{table*}[h!] 
	{\tiny
		\centering
		\begin{tabular}{c|c|c|c||c|c|c|c}
			\hline
			\textbf{Eg} & $\veps$ & $B_0$ & \textbf{Method}
			& $\ulB_0$ & $\olB_1$ & $\#(\miniStep)$
			& \textbf{Time(s)} \\
			\hline

			\multirow{5}{*}{$x'=x^2$}
			& \multirow{5}{*}{$0.01$}
			& \multirow{5}{*}{$[0.8,0.9]$}
			& \Standard\ & ${0.8495\pm0.0005}$ & ${5.6665\pm0.0045}$
			& $20861$ & 45.630 \\
			& & & \StandardT\ & ${0.8495\pm0.0005}$ & ${5.6665\pm0.0045}$
			& $18$ & \coblue{11.854} \\
			& & & \OurNoT\ & ${0.8495\pm0.0005}$ & ${5.6665\pm0.0045}$
			& $297$ & 15.073 \\
			& & & \OurNoSubr\ & ${0.8495\pm0.0005}$ & ${5.6665\pm0.0045}$
			& $31$ & 71.846 \\
			& & & \Ours\ & ${0.8495\pm0.0005}$ & ${5.6665\pm0.0045}$
			& $16$ & \cored{5.966} \\
			\hline

			\multirow{5}{*}{$x'=x^2$}
			& \multirow{5}{*}{$0.001$}
			& \multirow{5}{*}{$[0.98,0.99]$}
			& \Standard\ & ${0.985\pm0.0005}$ & ${65.66665\pm0.00035}$
			& $128890$ & 2104.03 \\
			& & & \StandardT\ & ${0.985\pm0.0005}$ & ${65.66665\pm0.00035}$
			& $103$ & \cored{31.898} \\
			& & & \OurNoT\ & ${0.985\pm0.0005}$ & ${65.66665\pm0.00035}$
			& $22763$ & 85.051 \\
			& & & \OurNoSubr\ & ${0.985\pm0.0005}$ & ${65.66665\pm0.00035}$
			& $72489$ & 327.89 \\
			& & & \Ours\ & ${0.985\pm0.0005}$ & ${65.66665\pm0.00035}$
			& $105$ & \coblue{49.861} \\
			\hline

			\multirow{5}{*}{Eg1}
			& \multirow{5}{*}{$0.01$}
			& \multirow{5}{*}{$Box_{(1,3)}(0.1)$}
			& \Standard\ & $(0.995,2.995)\pm(0.005,0.005)$
			& $(0.077,1.4635)\pm(0.001,0.0035)$
			& 1454 & 13.877 \\
			& & & \StandardT\ & $(0.995,2.995)\pm(0.005,0.005)$
			& $(0.077,1.4635)\pm(0.001,0.0035)$
			& 1888 & 15.370 \\
			& & & \OurNoT\ & $(0.995,2.995)\pm(0.005,0.005)$
			& $(0.077,1.4635)\pm(0.001,0.0035)$
			& 47 & \coblue{9.386} \\
			& & & \OurNoSubr\ & $(0.995,2.995)\pm(0.005,0.005)$
			& $(0.077,1.4635)\pm(0.001,0.0035)$
			& 47 & 9.415 \\
			& & & \Ours\ & $(0.995,2.995)\pm(0.005,0.005)$
			& $(0.077,1.4635)\pm(0.001,0.0035)$
			& 47 & \cored{9.232} \\
			\hline

			\multirow{5}{*}{Eg2}
			& \multirow{5}{*}{$0.1$}
			& \multirow{5}{*}{$Box_{(-3,3)}(0.1)$}
			& \Standard\ & $(-2.995,3.0)\pm(0.025,0.025)$
			& $(-2.13,0.56)\pm(0.05,0.02)$
			& 997 & 15.343 \\
			& & & \StandardT\ & $(-2.995,3.0)\pm(0.025,0.025)$
			& $(-2.13,0.56)\pm(0.05,0.02)$
			& 1367 & 16.540 \\
			& & & \OurNoT\ & $(-2.995,3.0)\pm(0.025,0.025)$
			& $(-2.13,0.56)\pm(0.05,0.02)$
			& 14 & \coblue{14.785} \\
			& & & \OurNoSubr\ & $(-2.995,3.0)\pm(0.025,0.025)$
			& $(-2.13,0.56)\pm(0.05,0.02)$
			& 14 & 15.119 \\
			& & & \Ours\ & $(-2.995,3.0)\pm(0.025,0.025)$
			& $(-2.13,0.56)\pm(0.05,0.02)$
			& 14 & \cored{14.590} \\
			\hline

			\multirow{5}{*}{Eg3}
			& \multirow{5}{*}{$0.1$}
			& \multirow{5}{*}{$Box_{(-1.5,8.5)}(0.01)$}
			& \Standard\ & $(-1.495,8.495)\pm(0.005,0.005)$
			& $(-0.595,-6.685)\pm(0.005,0.045)$
			& 2908 & 25.710 \\
			& & & \StandardT\ & $(-1.495,8.495)\pm(0.005,0.005)$
			& $(-0.595,-6.685)\pm(0.005,0.045)$
			& 3223 & \coblue{23.110} \\
			& & & \OurNoT\ & $(-1.495,8.495)\pm(0.005,0.005)$
			& $(-0.595,-6.685)\pm(0.005,0.045)$
			& 35 & 23.773 \\
			& & & \OurNoSubr\ & $(-1.495,8.495)\pm(0.005,0.005)$
			& $(-0.595,-6.685)\pm(0.005,0.045)$
			& 1034 & 455.548 \\
			& & & \Ours\ & $(-1.495,8.495)\pm(0.005,0.005)$
			& $(-0.595,-6.685)\pm(0.005,0.045)$
			& 25 & \cored{11.795} \\
			\hline

			\multirow{5}{*}{Eg4}
			& \multirow{5}{*}{$5$}
			& \multirow{5}{*}{$Box_{(15,15,36)}(0.001)$}
			& \Standard\ & $(15,15,36)\pm(0.0005,0.0005,0.0005)$
			& $(-6.94,1.81,35.52)\pm(1.64,2.33,2.5)$
			& 26 & 294.019 \\
			& & & \StandardT\ & $(15,15,36)\pm(0.0005,0.0005,0.0005)$
			& $(-6.94,1.81,35.52)\pm(1.64,2.33,2.5)$
			& 26 & 166.969 \\
			& & & \OurNoT\ & $(15,15,36)\pm(0.0005,0.0005,0.0005)$
			& $(-6.945,2.99,35.14)\pm(1.005,2.15,1.88)$
			& 61 & \coblue{146.360} \\
			& & & \OurNoSubr\ & $(15,15,36)\pm(0.0005,0.0005,0.0005)$
			& $(-6.94,1.81,35.52)\pm(1.64,2.33,2.5)$
			& 26 & 159.028 \\
			& & & \Ours\ & $(15,15,36)\pm(0.0005,0.0005,0.0005)$
			& $(-6.945,2.99,35.14)\pm(1.005,2.15,1.88)$
			& 61 & \cored{123.64} \\
			\hline
		\end{tabular}
	}
	\\[2mm]
	\dt{Table B}:
	The global effects of radical transform and
	bisection for our algorithm, all run with $order=3$.
	\label{tab:B}
\end{table*}

	We conclude from Table B that the transform method
	improves efficiency overall
	(compare \Standard\ vs.~\StandardT,
	and \Ours\ vs.~\OurNoT).
	In certain cases, such as $x'=x^2$ and Eg3,
	the transform method can significantly improve performance.

	For the two-dimensional examples Eg1, Eg2, and Eg3,
	the comparison between \StandardT\ and \Ours\
	shows that our \Bisect\ subroutine can enhance
	the overall performance of the algorithm.
	Also, comparing \OurNoSubr\ with \Ours\
	shows that \SubrSeven\ can significantly improve performance
	when there are many mini-steps
	(e.g., $x'=x^2$ and Eg3).


	\dt{TABLE C:}
	The time span is $t\in[0,1]$ in all experiments
	except for Eg1-b, where $t\in[0,5.5]$, corresponding to
	one full loop.
	This is an example that AWA cannot solve
	\cite[p.~13]{bunger:taylorODE:20}.

	For each example with order $k=20$, our algorithm first computes
	a scaffold $\stage(\veps_0)$ for an initial value of $\veps_0$.
	Subsequently, this scaffold is refined using smaller values
	$\veps_i$ ($i=1,2,\ldots$) to obtain
	$\stage(\veps_1),\stage(\veps_2),\ldots$.
	The total number of mini-steps in all stages of
	$\stage(\veps_i)$ is shown in column
	$\#(\miniStep)$, and the timing for each refinement
	is the incremental time.
	Thus, once a scaffold has been constructed,
	our \Refine\ procedure provides finer precision control
	with relatively low additional cost.

	There is an important distinction in the comparison with
	\capdCr\ and \vnodelp.
	Our \endEncIVP\ problem starts from the original box $B_0$
	and requires the algorithm itself to determine a subbox
	$\ulB_0\subseteq B_0$ such that the resulting end-enclosure
	satisfies
	\[
		w_{\max}(\olB_1)<\veps.
	\]
	In contrast, for each \capdCr\ and \vnodelp\ experiment
	in Table C, we directly supply the corresponding $\ulB_0$
	already computed by \Ours\ or \Refine.
	Thus, these solvers do not perform the search for a suitable
	$\ulB_0$; their reported times measure only the validated
	propagation of an already selected initial box.
	This design separates the cost of finding $\ulB_0$
	from the cost of propagating it to the final time.

\begin{table*}[h!] 
	{\tiny
		\centering
		\begin{tabular}{c|c|c|c|c|c|c|>{\color{red}}c}
			\hline
			\textbf{Case} & \textbf{Method} & $\veps$
			& $\ulB_0$ & $\olB_1$
			& $\#(\miniStep)$ & \textbf{Time(s)} & $\sigma(X)$ \\
			\hline

			\multirow{9}{*}{Eg1-a}
			& \Ours\ & $1.0$
			& $Box_{(1.0,3.0)}(0.1)$
			& $(0.08,1.46)\pm(0.06,0.16)$
			& 7 & 0.010 & 1 \\
			& \Refine & $0.05$
			& $Box_{(1.0,3.0)}(0.05)$
			& $(0.08,1.46)\pm(0.02,0.05)$
			& 13 & 0.009 & N/A \\
			& \Refine & $0.03$
			& $Box_{(1.0,3.0)}(0.025)$
			& $(0.08,1.46)\pm(0.006,0.02)$
			& 25 & 0.017 & N/A \\
			& \capdCr\ & N/A
			& $Box_{(1.0,3.0)}(0.1)$
			& $(0.08,1.46)\pm(0.025,0.093)$
			& 7 & 0.00043 & 0.043 \\
			& \capdCr\ & N/A
			& $Box_{(1.0,3.0)}(0.05)$
			& $(0.08,1.46)\pm(0.0081,0.038)$
			& 7 & 0.00041 & 0.041 \\
			& \capdCr\ & N/A
			& $Box_{(1.0,3.0)}(0.025)$
			& $(0.08,1.46)\pm(0.0034,0.018)$
			& 7 & 0.00041 & 0.041 \\
			& \vnodelp\ & N/A
			& $Box_{(1.0,3.0)}(0.1)$
			& $(0.08,1.46)\pm(0.024,0.088)$
			& 8 & 0.100 & 10.0 \\
			& \vnodelp\ & N/A
			& $Box_{(1.0,3.0)}(0.05)$
			& $(0.08,1.46)\pm(0.0080,0.037)$
			& 7 & 0.087 & 8.7 \\
			& \vnodelp\ & N/A
			& $Box_{(1.0,3.0)}(0.025)$
			& $(0.08,1.46)\pm(0.0034,0.017)$
			& 7 & 0.087 & 8.7 \\
			\hline

			\multirow{9}{*}{Eg1-b}
			& \Ours\ & $3.3$
			& $Box_{(1.0,3.0)}(0.0125)$
			& $(0.95,3.00)\pm(0.20,0.30)$
			& 294 & 0.404 & 1 \\
			& \Refine\ & $0.15$
			& $Box_{(1.0,3.0)}(0.00625)$
			& $(0.95,3.00)\pm(0.10,0.14)$
			& 587 & 0.326 & N/A \\
			& \Refine\ & $0.07$
			& $Box_{(1.0,3.0)}(0.00313)$
			& $(0.95,3.00)\pm(0.05,0.7)$
			& 1173 & 0.638 & N/A \\
			& \capdCr\ & N/A
			& $Box_{(1.0,3.0)}(0.0125)$
			& $(0.95,3.00)\pm(0.16,0.14)$
			& 28 & 0.0015 & 0.0037 \\
			& \capdCr\ & N/A
			& $Box_{(1.0,3.0)}(0.00625)$
			& $(0.95,3.00)\pm(0.040,0.021)$
			& 27 & 0.0014 & 0.0035 \\
			& \capdCr\ & N/A
			& $Box_{(1.0,3.0)}(0.00313)$
			& $(0.95,3.00)\pm(0.016,0.0058)$
			& 27 & 0.0014 & 0.0035 \\
			& \vnodelp\ & N/A
			& $Box_{(1.0,3.0)}(0.0125)$
			& $(0.95,3.00)\pm(0.097,0.061)$
			& 31 & 0.382 & 0.95 \\
			& \vnodelp\ & N/A
			& $Box_{(1.0,3.0)}(0.00625)$
			& $(0.95,3.00)\pm(0.033,0.015)$
			& 30 & 0.367 & 0.91 \\
			& \vnodelp\ & N/A
			& $Box_{(1.0,3.0)}(0.00313)$
			& $(0.95,3.00)\pm(0.015,0.0052)$
			& 30 & 0.367 & 0.91 \\
			\hline

			\multirow{9}{*}{Eg2}
			& \Ours\ & $1.0$
			& $Box_{(-3.0,3.0)}(0.1)$
			& $(-2.14,0.57)\pm(0.28,0.28)$
			& 10 & 0.016 & 1 \\
			& \Refine & $0.1$
			& $Box_{(-3.0,3.0)}(0.05)$
			& $(-2.14,0.57)\pm(0.08,0.04)$
			& 19 & 0.012 & N/A \\
			& \Refine & $0.05$
			& $Box_{(-3.0,3.0)}(0.025)$
			& $(-2.14,0.57)\pm(0.03,0.01)$
			& 37 & 0.023 & N/A \\
			& \capdCr\ & N/A
			& $Box_{(-3.0,3.0)}(0.1)$
			& $(-2.14,0.57)\pm(0.29,0.29)$
			& 11 & 0.00049 & 0.031 \\
			& \capdCr\ & N/A
			& $Box_{(-3.0,3.0)}(0.05)$
			& $(-2.14,0.57)\pm(0.12,0.062)$
			& 10 & 0.00042 & 0.026 \\
			& \capdCr\ & N/A
			& $Box_{(-3.0,3.0)}(0.025)$
			& $(-2.14,0.57)\pm(0.052,0.023)$
			& 10 & 0.00041 & 0.026 \\
			& \vnodelp\ & N/A
			& $Box_{(-3.0,3.0)}(0.1)$
			& $(-2.15,0.58)\pm(0.80,2.4)$
			& 14 & 0.182 & 11.4 \\
			& \vnodelp\ & N/A
			& $Box_{(-3.0,3.0)}(0.05)$
			& $(-2.14,0.57)\pm(0.17,0.092)$
			& 11 & 0.138 & 8.6 \\
			& \vnodelp\ & N/A
			& $Box_{(-3.0,3.0)}(0.025)$
			& $(-2.14,0.57)\pm(0.060,0.026)$
			& 10 & 0.124 & 7.7 \\
			\hline

			\multirow{9}{*}{Eg3}
			& \Ours\ & $1.0$
			& $Box_{(-1.5,8.5)}(0.01)$
			& $(-0.6,-6.69)\pm(0.00,0.19)$
			& 10 & 0.012 & 1 \\
			& \Refine & $0.06$
			& $Box_{(-1.5,8.5)}(0.005)$
			& $(-0.6,-6.69)\pm(0.0008,0.06)$
			& 19 & 0.012 & N/A \\
			& \Refine & $0.03$
			& $Box_{(-1.5,8.5)}(0.0025)$
			& $(-0.6,-6.69)\pm(0.0004,0.02)$
			& 37 & 0.022 & N/A \\
			& \capdCr\ & N/A
			& $Box_{(-1.5,8.5)}(0.01)$
			& $(-0.60,-6.69)\pm(0.0017,0.19)$
			& 16 & 0.00055 & 0.046 \\
			& \capdCr\ & N/A
			& $Box_{(-1.5,8.5)}(0.005)$
			& $(-0.60,-6.69)\pm(0.00081,0.091)$
			& 16 & 0.00055 & 0.046 \\
			& \capdCr\ & N/A
			& $Box_{(-1.5,8.5)}(0.0025)$
			& $(-0.60,-6.69)\pm(0.00041,0.045)$
			& 16 & 0.00055 & 0.045 \\
			& \vnodelp\ & N/A
			& $Box_{(-1.5,8.5)}(0.01)$
			& $(-0.60,-6.69)\pm(0.0017,0.19)$
			& 19 & 0.227 & 18.9 \\
			& \vnodelp\ & N/A
			& $Box_{(-1.5,8.5)}(0.005)$
			& $(-0.60,-6.69)\pm(0.00081,0.092)$
			& 19 & 0.227 & 19.0 \\
			& \vnodelp\ & N/A
			& $Box_{(-1.5,8.5)}(0.0025)$
			& $(-0.60,-6.69)\pm(0.00041,0.045)$
			& 19 & 0.227 & 18.9 \\
			\hline

			\multirow{9}{*}{Eg4}
			& \Ours\ & $4.5$
			& $Box_{(15.0,15.0,36.0)}(0.001)$
			& $(-6.94,2.99,35.14)\pm(0.09,0.15,0.15)$
			& 23 & 0.053 & 1 \\
			& \Refine & $0.6$
			& $Box_{(15.0,15.0,36.0)}(0.0003)$
			& $(-6.94,2.99,35.14)\pm(0.05,0.06,0.06)$
			& 89 & 0.161 & N/A \\
			& \Refine & $0.03$
			& $Box_{(15.0,15.0,36.0)}(0.0001)$
			& $(-6.94,2.99,35.14)\pm(0.02,0.02,0.02)$
			& 177 & 0.203 & N/A \\
			& \capdCr\ & N/A
			& $Box_{(15.0,15.0,36.0)}(0.001)$
			& $(-6.95,3.00,35.14)\pm(0.033,0.012,0.042)$
			& 25 & 0.0019 & 0.036 \\
			& \capdCr\ & N/A
			& $Box_{(15.0,15.0,36.0)}(0.0003)$
			& $(-6.95,3.00,35.14)\pm(0.0095,0.0031,0.012)$
			& 25 & 0.0018 & 0.034 \\
			& \capdCr\ & N/A
			& $Box_{(15.0,15.0,36.0)}(0.0001)$
			& $(-6.95,3.00,35.14)\pm(0.0032,0.00098,0.0040)$
			& 25 & 0.0018 & 0.034 \\
			& \vnodelp\ & N/A
			& $Box_{(15.0,15.0,36.0)}(0.001)$
			& $(-6.95,3.00,35.14)\pm(0.033,0.012,0.042)$
			& 30 & 0.571 & 10.8 \\
			& \vnodelp\ & N/A
			& $Box_{(15.0,15.0,36.0)}(0.0003)$
			& $(-6.95,3.00,35.14)\pm(0.0095,0.0031,0.012)$
			& 30 & 0.570 & 10.8 \\
			& \vnodelp\ & N/A
			& $Box_{(15.0,15.0,36.0)}(0.0001)$
			& $(-6.95,3.00,35.14)\pm(0.0032,0.00098,0.0040)$
			& 30 & 0.572 & 10.8 \\
			\hline
		\end{tabular}
	}
	\\[2mm]
	{\dt{Table C}:
	Experiments on \ourAlgo\ and \Refine,
	compared with \capdCr\ and \vnodelp.
	For each comparison, \capdCr\ and \vnodelp\ are initialized
	directly with the same $\ulB_0$ produced by \Ours\ or \Refine;
	their reported times therefore measure validated propagation
	from an already selected initial box and do not include the
	cost of determining such a $\ulB_0$.
	All methods use $order=20$
	(tolerance $10^{-12}$ for \capdCr\ and \vnodelp).
	For \capdCr\ and \vnodelp, $\#(\miniStep)$ is the number
	of accepted integration steps and Time is the median of 51 runs.
	$\sigma(X)\as
	\frac{\mbox{\scriptsize Time}(X)}
	{\mbox{\scriptsize Time}(\mbox{\scriptsize Ours})}$.
	}
	\label{tab:main}
	\label{tab:C}
\end{table*}


	\dt{Interpretation of Table C.}
	Table C reveals an important distinction between the
	full computational task solved by \ourAlgo\ and the
	validated propagation of a fixed initial box.
	Once a suitable $\ulB_0$ has already been determined,
	existing validated solvers can propagate it very efficiently.
	In particular, \capdCr\ is extremely fast in these experiments;
	\vnodelp\ is also competitive in some cases.

	This observation helps identify where a substantial part
	of the cost of our current implementation arises.
	Our algorithm is not given $\ulB_0$ in advance.
	Starting from the original input $B_0$, it must determine
	a sufficiently small subbox $\ulB_0\subseteq B_0$ for which
	the computed end-enclosure satisfies
	\[
		w_{\max}(\olB_1)<\veps.
	\]
	In the current implementation, finding such an initial box
	may require repeated attempts with progressively smaller
	candidate boxes, each involving validated IVP computations.
	The large difference between the running time of \ourAlgo\
	and that of \capdCr\ after $\ulB_0$ is supplied therefore
	suggests that a substantial portion of our computation is
	devoted to determining an appropriate $\ulB_0$, rather than
	to the final propagation of that box.
	This identifies the efficient selection of $\ulB_0$
	as an important direction for future improvement.

	There is a second source of overhead associated with
	the completeness requirement of our algorithm.
	During the computation, \ourAlgo\ periodically constructs
	validated Euler tubes that bound the entire trajectory over
	the corresponding time intervals.
	These trajectory bounds are essential to our termination
	analysis: without such global control, the halting argument
	would no longer apply.

	From a purely numerical point of view, however,
	an Euler tube need not improve the current end-enclosure
	each time it is computed.
	In many stages, its principal role is to provide a certified
	bound on the trajectory, while the actual tightening of the
	end-enclosure is achieved by other refinement operations.
	Consequently, some of the computation required to guarantee
	termination does not directly contribute to reducing the
	width of the final end-enclosure.

	This suggests a second direction for improvement.
	Rather than using the Euler tubes primarily as global
	certificates, we would like to exploit the information
	contained in these tubes more directly in subsequent
	refinement steps.
	Ideally, the computations needed for the termination proof
	would simultaneously contribute to obtaining tighter
	end-enclosures, thereby reducing the practical cost of
	completeness.

	Thus, Table C should not be interpreted simply as a
	head-to-head timing comparison.
	Instead, it exposes two principal sources of overhead in
	our current complete algorithm:
	the search for a suitable $\ulB_0$, and the construction of
	global trajectory bounds needed to guarantee termination.
	Both observations provide concrete directions for improving
	the practical efficiency of \ourAlgo\ without sacrificing
	its completeness guarantee.


%% file: inc/8conclusion.tex
\sect{Conclusion}
	We have presented a complete validated IVP 
	algorithm with the unique ability to pre-specify the
	an $\vareps$-bound on the width of the end-enclosure. 
	Preliminary implementations show promise in comparison
	to current validated software.
	This paper introduces a more structured approach
	to IVP algorithms, opening the way for considerable 
	future development of such algorithms.
	We introduced several novel techniques for
	Step A and Step B, including a new
	exploitation of \lognorm s combined with the radical transform.

	For future work, we plan to do a full scale
	implementation that includes the ability to
	use arbitrary precision arithmetic, in the style
	of \corelib\ \cite{core:home,yu+4:core2}.  
	We will also explore incorporating the Lohner-type
	transform into our radical transform.:w

\savespace{
	It is interesting to contrast our approach to the
	Lohner-type algorithms 
	\cite{nedialkov+2:validated-ode:99}
	where the original differential
	equation $\bfx'=\bff(\bfx)$ is successively transformed into
    $\bfx_1'=\bff_1(\bfx_1)$,
    $\bfx_2'=\bff_2(\bff_2)$, etc.
    The successive computations are
    done in the transformed space of $\bfx_m$ (after $m$ transforms).
    One of the possible issues with Lohner's approach is
	that the iterated transforms can introduce compounded
	errors.  In contrast our radical transform is done
	stepwise, and not iterated.  In the future, we
	plan to investigate generalizations of our radical transform.
    
	Let us give the precise description of the $\bff_i,\bfx_i$
	$\bfx_i'=\bff_i(\bfx_i)$
	for $i=1,2,\ldots$.
	We thus avoid  his iterative sequence of matrix products
	$A_1A_2\cdots A_i$.
}
    
	Nedialkov et al.~\cite[Section 10]{nedialkov+2:validated-ode:99},
	``{\em Some Directions for Future Research}'',
	presented a list of challenges that remain relevant today.
	Our algorithm is one answer to their call for automatic
	step sizes, order control (interpreted as error control)
	and tools for quasi-monotone problems (i.e., contractive systems).

%% file: inc/appendix.tex
\newpage
\appendix
\section{Appendix A: Proofs}
The numberings of lemmas and theorems
in this Appendix are the same as corresponding
results in the text, and are hyperlinked to the text.
Lemmas that do not appear the text are labeled
as Lemma A.1, Lemma A.2, etc.

	\ignore{
	\bthmDIY[Theorem A.1]{{\bf Picard-Lindel\"of Theorem}
		Let $Ball_{\bfp_0}(r)\ib \RR^n$ be the ball
		centered at $\bfp_0$ and $r>0$.
		Then there exists $h>0$ such
			$\ivp(\bfp_0,h,Ball_{\bfp_0}(r))$ is valid.
		\\
		In fact, we can choose any $h\le \min\set{M/r, L}$
		where $M=\sup \set{\|\bff(\bfq)\|: \bfq\inBall_{\bfp_0}(r))}$
		and $L$ is a Lipschitz constant for $\bff$
		 over $Ball_{\bfp_0}(r))$.
	}
	\bpf
		Let $X(h)=C^1([0,h]\to\RR^n)$
		The Picard operator $T=T_\bff: X\to X$ is
		defined as follows: for any $\bfx\in X$,
		$T[\bfx]$ is given by: for all $t\in[0,h]$,
			$$T[\bfx](t)=\bfx(0) + \int_0^t \bff(\bfx(s))ds.$$
		Now let $Y=Y_{\bfp_0}(h,r)$ comprise
		all $\bfy\in C^1([0,h]\to Ball_{\bfp_0}(r))$ satisfying
		$\bfy(0)=\bfp_0$.
		We have two claims:
		\\ (C1) If $h<M/r$ then $T[Y]\ib Y$.
		I.e., $T$ is an endormorphism when its
		range is restricted to $Y$.
		\\ (C2) If $h<\min{L,M/r}$ then $T:Y\to Y$
		is a contraction, i.e., for all $\bfx,\bfy\in Y$,
			$\|T[\bfx]-T[\bfy]\|_{\sup} < \|\bfx-\bfy\|_{\sup}$.
		\\ It follows from (C1) and (C2) that
		there is a unique $\bfy^*\in Y$ such
		that $T[\bfy^*]=\bfy^*$.
		\\
		We see that $\bfy^*:[0,h]\to Ball_{\bfp_0}(r)$ 
		with $\bfy^*(0)=\bfp_0$.  I.e., $\bfy^*$
		is the unique solution to $\ivp(\bfp_0,h, Ball_{\bfp_0})$.
	\epf
	}

\bthmDIY[\refThm{admiss}]{\dt{(Admissible Triple)} \\
		Let $\bff\in C^k(\RR^n\to\RR^n)$, $k\ge 1$ and $h>0$.
		Let $F\ib \RR^n$ be a compact convex set.
		If $E_0$ is contained in the interior of $F_1$ and
		satisfies the inclusion
		\beq
			\sum_{i=0}^{k-1}
		 	[0,h]^i \bff^{[i]}(E_0)+[0,h]^k \bff^{[k]}(F_1)
				\subseteq F_1,
		 \eeq
		then $(E_0,h,F_1)$ is an admissible triple.
	}%
\ccheeX{Bw, I think $k\ge 1$ suffices, why do we need $k\ge 2$?
	If $k\ge 1$, it implies that $\bff$ is Lipshitz.
	Am I missing something?
	}
\bpf
	Note that \refeq{tay} implies that $E_0$ is contained
	in $F_1$ but not necessarily in the interior.
\ccheeX{Bw, I am not sure because we don't know anything
about the remainder term $[0,h]^k \bff^{[k]}(F_1)$
In our computation the remainder term is replace by $Box(\veps)$,
then it is OK.   But here, it is not obvious.
}
	Given $\bfp_0\in E_0$, there exists a solution $\bfx:[0,\tau)\to\RR^n$
	to the IVP given by $\bff'=\bff(\bfx)$
	and $\bfx(0)=\bfp_0$.  Assume that $\tau$ is maximal.
	\\ A1. If $\tau<\infty$ then $\bfx(\tau)=\infty$.
	By way of contradiction, assume
	$\bfx(\tau)\in\RR^n$. By an application of
	Picard-Lindel\"of's Theorem at $\bfx(\tau)$,
	we can extend the solution
	$\bfx(t)$ from $t=\tau$ to $t=\tau+\veps$
	for some $\veps>0$, contradicting
	the maximality of $\tau$.
	\\ A2. For all $t\in [0, \min\set{\tau,h}]$,
		we have $\bfx(t)\in F_1$.
	Assume by way of contradiction that
	the set $S=\set{t\in [0,\min\set{\tau,h}]: \bfx(t)\nin F_1}$
	is non-empty.  Since $S$ is bounded from below, $t^*=\inf S$ exists.
	Since $\bfx(0)\in E_0$ is in the interior of a convex set $F_1$,
	this implies $t^*>0$.
	For all $t\in [0,t^*)$, $\bfx(t)$ has the Taylor expansion
		\beqarryl{tay123}
		\bfx(t) &=& \Big\{
				\sum_{i=0}^{k-1} t^i \bff\supn[i](\bfp_0) \Big\}
					+ t^k \bff\supn[k](\bfx(\xi))
				& \text{(for some $\xi\in [0,t]$;
					see  \refeq{taylorExpand})}\\
			&\in& \Big\{
				\sum_{i=0}^{k-1} h^i \bff\supn[i](\bfp_0) \Big\}
					+ [0,h]^k \bff\supn[k](F_1)
				& \text{($\bfx(\xi)\in F_1$ since $\xi<t^*\le h$)} \\
			&\ib& F_1
				& \text{(by assumption \refeq{tay} )}.
		\eeqarryl
	Thus $\bfx(t^*)$ is a limit point of
	$\set{\bfx(t): 0\le t<t^*}\ib F_1$.
	By the compactness of $F_1$,
	$\bfx(t^*)\in F_1$, contradicting A1.
	\\ A3.  $h<\tau$.  
	Otherwise, we have $\tau=\min\set{\tau,h}$
	and A2 implies that for all $t\le \tau$, $\bfx(t)\in F_1$.
	This means $\bfx(\tau)\in F_1$, contradicting A1.
	\\ A4. $(E_0,h)$ is valid, i.e., for all $\bfp_0\in E_0$,
			$\ivp(\bfp_0,h)$ has a unique solution.
	From A3, we know that $\ivp(\bfp_0,h)$ is non-empty.
	Suppose $\bfx,\bfy\in \ivp(\bfp_0,h)$ with $\bfx\ne\bfy$.
	Let $t^{\ne}\as \inf\set{t\in [0,h]: \bfx(t)\ne\bfy(t)}$.
	Clearly $t^{\ne}$ is well-defined.  Moreover
	$\bfx(t^{\ne})=\bfy(t^{\ne})$ by continuity of solutions.
	Therefore the assumption $\bfx\ne\bfy$ implies $t^{\ne}<h$.
	By Picard-Lindel\"of (\refThm{pl}), there is a $h^{\ne}>0$ such
	$\ivp(\bfx(t^{\ne}),h^{\ne})$ has a unique solution $\bfz$.  But this
	implies that $\bfx(t^{\ne}+t)= \bfy(t^{\ne}+t)=\bfz(t)$
	for all $t\in [0,\min(h^{\ne},h-t^{\ne})]$. 
	This contradicts the minimality $t^{\ne}$.
	\\ A5. $(E_0,h,F_1)$ is an admissible triple.
	We must show that for all $\bfp_0\in E_0$,
	$F_1$ is a full enclosure of $\ivp(\bfp_0,h)$.
	This follows from A2 and A3.
\epf
\ignore{OLD PROOF of Bingwei:
	(1) we must show that $(E_0,h)$ is valid, and
	(2) we must show that $F_1$ is the full-enclosure
	for $\ivp(E_0,h)$.
	
	Proof of (1):
	Since $\bff$ is locally Lipschitz on $\mathbb{R}^n$, repeated application of the Picard--Lindelöf theorem guarantees a unique maximal solution
	\[
	\bfx : [0,\tau) \to \mathbb{R}^n,\qquad \tau>0,
	\]
	to the IVP $\bfx'=\bff(\bfx)$, $\bfx(0)=\bfx_0$. It suffices to show that $\tau>h$.
	
	Assume, for contradiction, that $\tau\le h$. We first claim that $\bfx(t)\in F_1$ for all $t\in[0,\tau)$.
	
	Then, since $\bff$ is continuous on $F_1$ and $\bfx$ is continuous on $[0,\tau)$, the integral representation
	\[
	\bfx(t)=\bfx(0)+\int_0^t \bff(\bfx(s))\,ds
	\]
	implies that the limit
	\[
	\bfy := \lim_{t\to\tau^-} \bfx(t)
	\]
	exists in $\mathbb{R}^n$. Because $F_1$ is closed and $\bfx(t)\in F_1$ for all $t<\tau$, we also have $\bfy\in F_1$.
	
	By applying the Picard--Lindelöf theorem in a 
	neighborhood of $\bfy$, there exists $\delta>0$ and a 
	unique solution $\tilde\bfx$ with initial condition 
	$\tilde\bfx(\tau)=\bfy$, defined on $[\tau,\tau+\delta)$. 
	Concatenating $\bfx$ and $\tilde\bfx$ yields a solution 
	on $[0,\tau+\delta)$, contradicting the maximality 
	of~$\tau$. Hence $\tau>h$, and the unique solution exists 
	on $[0,h]$.

	We now prove the claim. Suppose, for contradiction, that there exists $t'\in[0,\tau)$ such that the solution $\bfx=\ivp(\bfx_0,h)$ satisfies $\bfx(t')\notin F_1$. Define
	\[
	t^{*}:=\inf\{ t\in[0,\tau):\; \bfx(t)\notin F_1\}.
	\]
	
	Applying Taylor's formula to $\bfx$ at $t=0$ and evaluating at $t^{*}$, there exists $\xi\in(0,t^{*})$ such that
	\[
	\bfx(t^{*})
	=
	\sum_{i=0}^{k-1} (t^{*})^{i}\,\bff^{[i]}(\bfx_0)
	+
	(t^{*})^{k}\,\bff^{[k]}(\bfx(\xi)).
	\]
	
	Since $\xi<t^{*}$, we have $\bfx(\xi)\in F_1$, and thus
	$\bff^{[k]}(\bfx(\xi))\in \bff^{[k]}(F_1)$. Consequently,
	\[
	\bfx(t^{*})
	\in
	\sum_{i=0}^{k-1}[0,h]^{i}\,\bff^{[i]}(E_0)
	\;+\;
	[0,h]^{k}\,\bff^{[k]}(F_1)
	\subseteq F_1
	\]
	by~\refeq{tay}.  
	This contradicts the definition of $t^{*}$.
	
	Proof of (2):
	From the above we obtain $\bfx(t)\in F_1$ for all $t\in[0,\tau)$ and $\tau>h$. Thus $\bfx(t)\in F_1$ for all $t\in[0,h]$, i.e., $F_1$ is a full enclosure.
	\epf
}

\bcorDIY[\refCor{cor-1}]{\ \\
		Let $\bfx_i\in \ivp(Ball_{\bfp_0}(r),h,F)$ for $i=1,2$
		and $\olmu\ge \mu_2(J_\bff(F))$.
		\benum[(a)]
		\item
		For all $t\in [0,h]$,
		\beq
		\|\bfx_1(t)-\bfx_2(t)\|_2
		\le \|\bfx_1(0)-\bfx_2(0)\|_2 e^{\olmu t}.\eeq
		\item
		If $\bfx_1(0)=\bfp_0$ then
		an end-enclosure for $\ivp(Ball_{\bfp_0}(r),h,F)$ is
		$$Ball_{\bfx_1(h)}(re^{\olmu h}).$$
		\eenum
	}
	\bpf
	(a) Note that $\bfx_1$ and $\bfx_2$ are solutions of \refeq{bfx'}
	with different initial values.
	We can apply \refThm{ne} to bound $\|\bfx_1(t)-\bfx_2(t)\|$
	by choosing $\delta=\| \bfx_1(0)-\bfx_2(0)\|$ and $\veps=0$.
	Then \refeq{xibfx} implies
	\beqarrys
	\|\bfx_1(t)-\bfx_2(t)\|
	&\le& \clauses{
		\delta e^{\olmu t}+ \frac{\veps}{\olmu}(e^{\olmu t} - 1)
		& \rmif\ \olmu \ne 0, \\
		\delta + \veps t & \rmif\ \olmu = 0}\\
	&=& \clauses{
		\delta e^{\olmu t}
		& \rmif\ \olmu \ne 0, \\
		\delta  & \rmif\ \olmu = 0}
	& \text{(since $\veps=0$)}\\
	&=& \| \bfx_1(0)-\bfx_2(0)\| e^{\olmu t}
	& \text{(for all $\olmu$).}
	\eeqarrys
	\\
	By part (a), for any $\bfq\in Ball_{\bfp_0}(r)$ the 
	corresponding solution 
	$\bfy(t)=\ivp(\bfq,t)$ satisfies the perturbation estimate
	\[
	\|\bfx_1(t)-\bfy(t)\|\le r e^{\olmu t}, \qquad 
	\forall\, t\in[0,h].
	\]
	Evaluating this inequality at $t=h$ gives
	\[
	\|\bfx_1(h)-\bfy(h)\|\le r e^{\olmu h}.
	\]
	Thus $\bfy(h)\in Ball_{\bfx_1(h)}(re^{\olmu h})$.
	\epf
	
	The following is a useful lemma:
	\blemDIY[Lemma A.2]{\ \\
		\label{eulerOneStep}
		Let		$(B_0,H,B_1)$
		be an admissible triple with 
		$\olmu\ge \mu_2(J_{\bff}(B_1))$,
		and
		$ \olM\ge \|\bff^{[2]}(B_1)\|$.
		Denote the Euler step at $\bfq_0\in B_0$
		by the linear function
		$$\ell(t;\bfq_0) = \bfq_0 + t \bff(\bfq_0).$$
		Then for any $\bfp_0\in B_0$ and $t\in [0,H]$,
		$$\|\bfx(t;\bfp_0)-\ell(t;\bfq_0)\|
		\le \|\bfp_0-\bfq_0\| e^{\olmu t}+ \half \olM t^2$$
	}
	\bpf
	By \refCor{cor-1},
	\beql{localcor1}
	\|\bfx(t;\bfp_0)-\bfx(t;\bfq_0)\| \le 
	\|\bfp_0-\bfq_0\|e^{\olmu t}
	\eeql
	We also have
	\beqarrys
	\bfx(t;\bfq_0) &=& \bfq_0 + t\cdot \bff(q_0)
	+ \half t^2 \bfx''(\tau)
	& \text{(for some $\tau\in [0,t]$)}\\
	\lefteqn{\|\bfx(t;\bfq_0) - (\bfq_0 + t\cdot \bff(q_0))\|} \\
	&\le& \| \half t^2 \bfx''(\tau)\| \\
	&=& \| \half t^2 \bff\supn[2](\bfx(\tau;\bfq_0)\| \\
	&\le& \| \half t^2 \olM\|
	& \text{(since $\olM\ge \bff\supn[2](B_1)$)}
	\eeqarrys
	Combined with \refeq{localcor1},
	the triangular inequality shows our desired bound.
	\epf

	\blemDIY[\refLem{eulerStep}]{
		Let $(B_0,H,B_1)$ be an admissible triple,
		$\olmu\ge \mu_2(J_{\bff}(B_1))$ and
		$ \olM\ge \|\bff^{[2]}(B_1)\|$.
		\\	
		Consider the polygonal
		path $\bfQ_{h_1}=(\bfq_0,\bfq_1\dd \bfq_m)$ 
		from the Euler method with $m$ steps of uniform step-size $h_1$
		given by 
		\beql{h1-appendix}
		h_1= h\euler(H,\olM,\olmu,\delta) \as
		\begin{cases}
			\min\set{H, \frac{2\olmu\delta}
				{\olM \cdot (e^{\olmu H}-1)}}
			&\rmif\ \olmu\ge 0\\
			\min\set{H, \frac{2\olmu\delta}
				{\olM \cdot (e^{\olmu H}-1)-\olmu^2\delta}}
			&\rmif\ \olmu<0.
		\end{cases}
		\eeql
		If each $\bfq_i\in B_1$ ($i=0\dd m$) then the path $\bfQ_{h_1}$
		lies inside the $\delta$-tube of $\bfx(t;\bfq_0)$.
		\\ In other words, for all $t\in [0,H]$, we have
		\beql{eulerBd-appendix}
		\|\bfQ_{h_1}(t)-\bfx(t;\bfq_0)\|\le \delta.
		\eeql
	}
	\bpf
	For simplicity, we only prove the lemma
	when $H/h_1$ is an integer.
	We first show that the Euler method with uniform step size $h_1$ 
	has the following error bound:
	\beql{gm}
	\|\bfq-\bfx(H)\|\le 	
	\begin{cases}
		\frac{\olM h_1}{2\olmu}(e^{\olmu H}-1) & \olmu\ge 0,\\
		\frac{\olM h_1}{2\olmu+\olmu^2 h_1}(e^{\olmu H}-1) & 
		\olmu<0.
	\end{cases}
	\eeql
	To show \refeq{gm}, assume the sequence
	$(\bfq_0=\bfx(0),\bfq_1\dd \bfq_m=\bfq)$ from the
	Euler method at times $t_0=0,t_1\dd t_m=H$. 
	Let 
	\[
	g_i := \|\bfq_i - \bfx(t_i)\|_2,\qquad i=0,\dots,m.
	\]
	By Lemma A.2
	we obtain the linear recurrence
	\[
	g_{i+1} \le e^{\olmu h_1}\, g_i + \frac{\olM h_1^2}{2}.
	\]
	
	We solve this recurrence by induction. The base case \(g_0=0\) holds
	by construction.
	Assume that for some \(k\ge0\) we have
	\[
	g_k \le \frac{\olM h_1^2}{2}\sum_{j=0}^{k-1} e^{\olmu h_1 j}.
	\]
	Then
	\begin{align*}
		g_{k+1}
		&\le e^{\olmu h_1} g_k + \frac{\olM h_1^2}{2}\\
		&\le e^{\olmu h_1}\cdot\frac{\olM h_1^2}{2}\sum_{j=0}^{k-1}
		e^{\olmu h_1 j}
		+\frac{\olM h_1^2}{2}\\
		&= \frac{\olM h_1^2}{2}\sum_{j=0}^{k} e^{\olmu h_1 j}.
	\end{align*}
	Hence, by induction we obtain
	\begin{equation}\label{gm-proof}
		g_m \le \frac{\olM h_1^2}{2}\frac{e^{\olmu H}-1}{e^{\olmu h_1}-1}.
	\end{equation}
	
	From \eqref{gm-proof} we deduce the following two useful bounds:
	\[
	g_m \le
	\begin{cases}
		\dfrac{\olM h_1}{2\olmu}\bigl(e^{\olmu H}-1\bigr),
		& \text{if }\ \olmu\ge 0,\\[8pt]
		\dfrac{\olM h_1}{2\olmu+\olmu^2 h_1}\bigl(e^{\olmu H}-1\bigr),
		& \text{if }\ \olmu<0.
	\end{cases}
	\]
	The first inequality follows from
	\(e^{\olmu h_1}-1\ge \olmu h_1\) when \(\olmu\ge0\).
	For the second inequality (the case \(\olmu<0\))
	we use the elementary Taylor-type estimate
	\(e^{\olmu h_1}-1 \le \olmu h_1 + \tfrac{1}{2}\olmu^2 h_1^2\)
	(valid for \(\olmu h_1<0\)),
	which implies
	\(e^{\olmu h_1}-1 \le \olmu h_1(1+\tfrac{1}{2}\olmu h_1)\)
	and yields the stated bound.
	
	Finally, by choosing \(h_1\) as specified in \refeq{h1-appendix},  
	we ensure that \(g_m < \delta\).
	
	\epf

	\blemDIY[\refLem{ad}]{\ \\
	Let $ H > 0 $, $\bfveps = (\veps_1\dd \veps_n)$, and 
	$ E_0 \ib \RR^n $.  
	Also let $\bfM=(M_1\dd M_n)$ with
		$$M_i \as \sup_{\bfp \in \olB}
			\Big|\big(\bff^{[k]}(\bfp)\big)_i\Big|,
			\quad (i=1\dd n)$$
	where $(\bfx)_i$ is the $i$th coordinate of $\bfx\in\RR^n$ and
		$$\olB \as \sum_{i=0}^{k-1} [0, H]^i \bff^{[i]}(E_0)
			+ Box(\bfveps).$$
	If
	\beq
		h = \min \Big\{ H, \min_{i=1}^n
		\Big(\tfrac{\veps_i}{M_i}\Big)^{1/k} \Big\}
		\quad\text{and}\quad
		F_1 \as \sum_{i=0}^{k-1} [0, h]^i \bff^{[i]}(E_0)
			+ Box(\bfveps).
	\eeq 
	then $(E_0,h, F_1)$ is an admissible triple.
	}
	
	\bpf
	Note that $\olB$ and $F_1$ are similar but one
	is based on $H$ while the other is based on $h$.
	To prove the admissibility of $(h,F_1)$ for $E_0$, we invoke 
	\refeq{tay} and this fact:
	\beql{tay1}
		[0,h]^k\bff^{[k]}(F_1)\ib Box(\bfveps)=[\pm \bfveps]. 
		\eeql 
	Here is the proof of \refeq{tay1}:
	\beqarrys
		~[0,h]^k\bff^{[k]}(F_1)
		&=& [0,h]^k \bff^{[k]}\Big(
			\sum_{i=0}^{k-1} [0, h]^i \bff^{[i]}(E_0)
			+ [\pm \bfveps]\Big) 
			&\text{(definition of $F_1$)}\\
		&\ib& [0,h]^k \bff^{[k]}\Big(
			\sum_{i=0}^{k-1} [0, H]^i \bff^{[i]}(E_0)
			+ [\pm \bfveps]\Big) 
			&\text{($h\le H$)}\\
		&=& [0,h]^k \bff^{[k]}(\ol{B})
			&\text{(definition of $\ol{B}$)}\\
		&\ib& [0,h]^k[-\bfM,\bfM]
			&\text{(definition of $\bfM$)}\\
		&\ib& [-\bfveps, \bfveps].
			&\text{(definition of $h$ implies $h^k\bfM\le\bfveps$)}
		\eeqarrys
	\epf

	\blemDIY[\refLem{stepA}]{
		\dt{(Correctness of \stepA)}\\
		$\stepA(E_0,H, \bfveps)\ssa (h,F_1)$ is correct:\\
		I.e., the algorithm halts
		and outputs an $\bfveps$-admissible triple
				$(E_0,h,F_1)$.
	}
	\bpf
	The while-loop of \stepA, we assert this invariant
		$$(E_0,h,\olB) \text{ is $\bfveps$-admissible}$$
	right after $h$ is updated.
	This follows from
	\refLem{ad}.  Let
		$$(E_0,h_i,\olB_i), \quad i=1,2,\ldots$$
	denote the admissible triple at the $i$th iteration.
	Thus, if the algorithm halts after the $i$th iteration,
	the output is admissible. It remains to show termination.
	Observe that 
		$$\olB_1\ibp \olB_2 \ibp \cdots$$
	and therefore the $h_i$'s increases with $i$:ss
	$h_1< h_2 < \cdots$.
	Let $H_i = H/2^i$ be the value of $H$ at the
	end of the $i$th iteration.  
	Clearly, we enter the $1$st iteration
	since $H_0$ is the input $H>0$ and $h_0=0$.
	For $i\ge 1$, we enter the $i$th iteration iff
	$(H_{i-1}>h_{i-1})$ holds and we have:
	$$0=h_0<h_1<\cdots< h_{i-1} < H_{i-1}<H_{i-2}<\cdots<H_0$$
	The while-loop must terminate since
	$H_i\to 0$ as $i\to\infty$ and $h_1>0$.
	\epf
	
	\blemDIY[\refLem{stepB}]{
		\dt{(Correctness of \stepB)} \\
		$\stepB(E_0,h,F_1)\ssa E_2$ is correct,
		i.e., $E_2$ is an end-enclosure for $(E_0,h,F_1)$
	}
	\bpf
	The output $E_2$ is defined as the intersection $E_1\cap B$.  
	Since $E_1$ is already an end-enclosure for $(E_0,h,F_1)$ by \refeq{direct}, 
	it only remains to show that 
	\[
	B = Box_{\bfq_0}\!\big(\tfrac12\|\wmax(E_0)\|\, e^{\mu h}\big)
	+ Box\!\big(h^k\bff^{[k]}(F_1)\big)
	\]
	is also an end-enclosure.
	
	This follows directly from \refCor{cor-1}(b) together with the fact that 
	the solution $\bfx_1=\ivp(m(E_0),h)$ satisfies
	\[
	\bfx_1(h)\in \bfq_0 + Box\!\big(h^k\bff^{[k]}(F_1)\big).
	\]
	
	\epf

	\blemDIY[\refLem{delta-distance}]{
	Consider an admissible triple $(E_0,H,F_1)$ where
		$E_0\as Ball_{\bfp_0}(r_0)$.\\
	Let
			$\olmu = \mu_2(J_\bff(F_1))$,
			$\olM = \|\bff\supn[2](F_1)\|$,
			and $\delta>0$.\\
	If $\bfq_0=\bfp_0+h_1\bff(\bfp_0)$ is obtained from
		$\bfp_0$ by an Euler step of size $h_1$ where
		$h_1\le  h\euler(H,\olM,\olmu,\delta)$ (see~\refeq{h1}),
	then:
	\benum[(a)]
	\item  \dt{($\delta$-Tube)}\\
		The linear function
		$\ell(t)\as (1-t/h_1)\bfp_0+(t/h_1)\bfq_0$
		lies in the $\delta$-tube of $\bfx_0=\ivp(\bfp_0,H)$.
	\item \dt{(End-Enclosure)}\\
		Then
		$Ball_{\bfq_0}(r_0 e^{\olmu h_1}+\delta)$
		is an end-enclosure for $\ivp(E_0,h_1)$.
	\item \dt{(Full-Enclosure)}\\
		The convex hull
			$\chull(Ball_{\bfp_0}(r'), Ball_{\bfq_0}(r'))$
				where
			$r'=\delta+\max(r_0 e^{\olmu h_1},r_0)$
		is a full-enclosure for $\ivp(E_0,h_1)$.
	\eenum
	}
	\bpf
	\benum[(a)]
	\item By \refLem{eulerStep} we have $\ell(t)$
	lies in the $\delta$-tube of $\bfx_0$, since for any
	$t\in[0,h_1]$, $\|\ell(t)-\bfx_0(t)\|\le \delta$.
	\item 	By \refCor{cor-1} we have for any
		$\bfx\in \ivp(Ball_{\bfp_0}(r_0),h_1,F_1)$,
		$\|\bfx(h_1)-\bfx_0(h_1)\|\le r_0e^{\olmu h_1}$.
		Since $\|\bfq_0 - \bfx_0(h_1)\|_2 \leq \delta$, 
		then by the triangular inequality 
		we have 
		$$\|\bfq_0 - \bfx(h_1)\|_2\le \|\bfx(h_1)-\bfx_0(h_1)\|
			+\|\bfq_0 - \bfx_0(h)\|_2 \le r_0 e^{\olmu h}+\delta,
			$$
		so, $\bfx(h_1) \in  Ball_{\bfq_0}(r_0 e^{\olmu h}+\delta)$. 
	\item
	We show that for any $T \in [0, h_1]$,  the end-enclosure of 
	$\ivp(E_0,T)$ is a subset of $Box(Ball_{\bfp_0}(r'),
	Ball_{\bfq_0}(r'))$.
	Note that
	$E_1 =  Ball_{\ell(T)}(r_0e^{\olmu T}+\delta)$ is the
	end-enclosure for $\ivp(E_0, T)$.
	
	Let $\ell(T)_i$ denote the $i$-th component of $\ell(T)$ and
	$r(T)\as r_0e^{\olmu T}+\delta$. Then, we only need to
	prove that for any $i=1\dd n$, the interval $\ell(T)_i \pm r(T)$
	satisfies  
	\[
	\ell(T)_i \pm r(T) \subseteq Box((\bfp_0)_i\pm (r'+\delta),
	(\bfq_0)_i\pm (r'+\delta)),
	\]  
	where $(\bfp_0)_i $ and $(\bfq_0)_i $ are the $i$-th components of
	$\bfp_0$ and $\bfq_0$,
	respectively.  
	
	Since $\ell(T)$ is a line segment, it follows that  
	\[
	\min((\bfq_0)_i, (\bfp_0)_i) \leq \ell(T)_i \leq \max((\bfq_0)_i,
		(\bfp_0)_i).
	\]  
	Additionally, we have $r(T) \leq r' + \delta$.  
	Combining these observations, we conclude that 
	\[
	\ell(T)_i \pm r(T) \subseteq Box((\bfp_0)_i\pm (r'+\delta),
	(\bfq_0)_i\pm (r'+\delta)).
	\] 
	\eenum
	\epf

\blemDIY[Lemma A.3]{\ \\
		\label{lem:Jbfg}
	\benum[(a)]
	\item $\bfg(\bfy)=J_{\whpi}(\whpi\inv(\bfy))
		\Bigcdot \ol\bfg(\whpi\inv(\bfy))$\\
		$=\diag(-d_i y_i^{1+\tfrac{1}{d_i}}: i=1\dd n)
		\Bigcdot \ol\bfg(\whpi\inv(\bfy)) $.
	\item The Jacobian matrix of $\bfg$ with respect to
		$\bfy=(y_1\dd y_n)$ is:
		\beql{Jbfg}
		J_{\bfg}(\bfy)=A(\bfy)
		+ 
		P\inv(\bfy)
		\Bigcdot 
		J_{\ol\bfg}(\whpi\inv(\bfy))\Bigcdot P(\bfy),
		\eeql
		where 
		$$A(\bfy)=\diag\Big(-(d_i+1)y_i^{\frac{1}{d_i}}\cdot
		(\ol\bfg(\olpi\inv(\bfy)))_i): i=1\dd n\Big) $$
		and 
		$$P(\bfy)=\diag\Big(\tfrac{\olpi\inv(\bfy)_i^{d_i+1}}{d_i}:
		i=1\dd n \Big).$$
	\eenum
	}
	\bpf
	\benum[(a)]
	\item 
	For each $ i = 1\dd n $, we have from \refeq{bfy'} that
	$y_i'= g_i(\bfy)$ where $\bfy=(y_1\dd y_n)$,
	$\bfg=(g_1\dd g_n)$, i.e.,
	\beqarrys
	g_i(\bfy) = y_i' 
	&=& \left(\frac{1}{\oly_i^{d_i}}\right)'
	\qquad\text{(by \refeq{bfy'} and $y_i=\oly_i^{-d_i}$)}\\
	&=& -d_i \oly_i^{-(d_i+1)} \oly_i' \\
	&=& -d_i y_i^{1 + \tfrac{1}{d_i}} \Big(\ol\bfg
	\big( y_1^{-\tfrac{1}{d_1}}\dd y_n^{-\tfrac{1}{d_n}} \big)\Big)_i\\
	&=& -d_i y_i^{1+\tfrac{1}{d_i}} (\ol\bfg(\olpi\inv(\bfy)))_i.
	\eeqarrys
	Thus, $$\bfg(\bfy)=
	(g_1(\bfy)\dd g_n(\bfy))
	=\diag(-d_i y_i^{1+\tfrac{1}{d_i}}, i=1\dd n)
	\Bigcdot \ol\bfg(\olpi\inv(\bfy))$$
	\item 
	By plugging 
	$g_i(\bfy) = -d_i y_i^{1+\tfrac{1}{d_i}} (\ol\bfg(\olpi\inv(\bfy)))_i$
	into the Jacobian, we get
	{\small
		\beqarray
		J_\bfg(\bfy)&=& \mmat{\nabla( g_1(\bfy) ) \\ \vdots \\
			\nabla( g_n(\bfy) } 
		=
		\mmat{\nabla( -d_i y_1^{1+\tfrac{1}{d_i}}
			(\ol\bfg(\olpi\inv(\bfy)))_1) \\ \vdots \\
			\nabla (-d_n y_n^{1+\tfrac{1}{d_n}} (\ol\bfg(\olpi\inv(\bfy)))_n)}
		\nonumber\\
		\nonumber\\
		&=& \mmat{\nabla(-d_1 y_1^{1+\tfrac{1}{d_1}})
			(\ol\bfg(\olpi\inv(\bfy)))_1
			\\ \vdots \\
			\nabla (-d_n y_n^{1+\tfrac{1}{d_n}})
			(\ol\bfg(\olpi\inv(\bfy)))_n}
		+ \mmat{ -d_1 y_1^{1+\tfrac{1}{d_1}}
			\nabla((\ol\bfg(\olpi\inv(\bfy)))_1)
			\\ \vdots \\
			-d_n y_n^{1+\tfrac{1}{d_n}} 
			\nabla ((\ol\bfg(\olpi\inv(\bfy)))_n)}
		\label{eq:splitjacobian}
		\eeqarray
	}
	Note that for any $i=1\dd n$, 
	{\small
		$$	\nabla (-d_i y_i^{1+\tfrac{1}{d_i}})	(\ol\bfg(\olpi\inv(\bfy)))_i
		=  
		\big(0\dd 0, -d_i\left(1 + \tfrac{1}{d_i}\right)y_i^{\frac{1}{d_i}}
		(\ol\bfg(\olpi\inv(\bfy)))_i\dd 0 \big)
		$$}
	and
	{\small
		\beqarrys
		-d_i y_i^{1+\tfrac{1}{d_i}}
		\nabla((\ol\bfg(\olpi\inv(\bfy)))_i)
		&=&  \Big( -d_i y_i^{1+\tfrac{1}{d_i}} 
		\frac{\partial (\ol\bfg(\bfx))_i}{\partial x_j}
		(\olpi\inv(\bfy)) \frac{\partial \olpi\inv(\bfy)}{\partial \bfy} : j=1\dd n
		\Big)   \\
		&=& 
		\Big(\tfrac{d_i}{d_j} \left(\tfrac{y_i^{1 + \frac{1}{d_i}} }{y_j^{1 + \frac{1}{d_j}} }\right)
		\frac{\partial (\ol\bfg(\bfx))_i}{\partial x_j}
		(\olpi\inv(\bfy)) : j=1\dd n \Big)\\
		&=&
		\Big( \tfrac{d_i}{d_j}  \olpi\inv(\bfy)_j^{d_j+1}
		\frac{\partial (\ol\bfg(\bfx))_i}{\partial x_j}
		(\olpi\inv(\bfy)) \olpi\inv(\bfy)_i^{-d_i-1} : j=1\dd n \Big).
		\eeqarrys}
	Thus, $$J_{\bfg}(\bfy)=A(\bfy)+P\inv(\bfy)
	\Bigcdot
	J_{\ol\bfg}(\olpi\inv(\bfy))\Bigcdot P(\bfy),$$
	where $$A(\bfy)=\diag(-(d_1+1)y_1^{\frac{1}{d_1}} (\ol\bfg(\olpi\inv(\bfy)))_1)\dd 
	-(d_n+1)y_n^{\frac{1}{d_n}} (\ol\bfg(\olpi\inv(\bfy)))_n) $$
	and 	$$P(\bfy)=\diag(\tfrac{\olpi\inv(\bfy)_1^{d_1+1}}{d_1}\dd \tfrac{\olpi\inv(\bfy)_n^{d_n+1}}{d_n}).$$
	\eenum
	\epf

	\dt{\refThm{keylemma}}
		{\em 
			\benum[(a)]
		\item For any $\bfd=(d_1\dd d_n)$, we have
		\beqarrys
		\mu_2 \big(J_{\bfg} (\pi(F_1))\big)
		&\le&
		\max\set{\tfrac{-(d_i+1)}{\chb_i}:
			i=1\dd n}\\
		&&
		+ \max_{i=1}^n \set{d_i} 
		\cdot
		\|J_{\olbfg}(\olpi(F_1))\|_2
		\cdot
		\max_{i=1}^n \set{\tfrac{(\chb_i)^{d_i+1}}{d_i}}.
		\eeqarrys
		\item If $d_1=\cdots=d_n=d$ then
		$$\mu_2 \left(J_{\bfg} (\pi(F_1))\right)
		\le
		-(d+1)\tfrac{1}{\chb_{\max}}
		+(\chb_{\max})^{d+1}
		\|J_{\olbfg}(\olpi(F_1))\|_2.
		$$
		\eenum
	}
	\bpf
	From {\bf Lemma A.3(b)}
	we have for any $\bfp=(p_1\dd p_n) \in \olpi(F_1)$,
	\beql{splitJacobian1}
	J_{\bfg}(\whpi(\bfp))
	= A(\bfp) + P\inv(\bfp)
	\frac{\partial \olbfg}{\partial \bfx}(\bfp)P(\bfp)
	\eeql
	where
	$P(\bfp)=\diag\big(
	\tfrac{p_i^{d_i+1}}{d_i}: i=1\dd n \big)$
	and
	$A(\bfp) = \diag(a_1\dd a_n)$ with
	\beql{aid1}
	a_i \as -d_i(1 + \tfrac{1}{d_i}) p_i\inv
	\cdot (\olbfg(\bfp))_i. \eeql
	Thus, $A, P$ are diagonal matrices and $p_i\inv$ is well-defined
	since $\bfp\in B_2\ge \1$, \refeq{translation}.
	
	By \refLem{lognorm}(b) and \refeq{aid1}, we conclude that
	the form
	\beql{mudiag}
	\mu_2(A(\bfp))=	\mu_2(\diag(a_1\dd a_n))
	= \max\set{a_i: i=1\dd n}.
	\eeql
	
	From \refeq{olbfg}, we conclude that
	{\small
		\beqarrys
		\mu_2\big(J_{\bfg}(\whpi(\bfp))\big) 
		&=& \mu_2\left(A(\bfp) + P\inv(\bfp)\frac{\partial \olbfg}{\partial
			\bfx}(\bfp)P(\bfp)\right)\\
		&& \text{(by \refeq{splitJacobian1})} \\
		&\le& \mu_2(A(\bfp))
		+ \mu_2\left(P\inv(\bfp)\frac{\partial \olbfg}{\partial
			\bfx}(\bfp)P(\bfp)\right)\\
		&& \text{(by \refLem{lognorm}(a))} \\
		&\le& \mu_2(A(\bfp)) 
		+ \left\|P\inv(\bfp)\frac{\partial \olbfg}{\partial
			\bfx}(\bfp)P(\bfp)\right\|_2 \\
		&& \text{(by \refLem{lognorm}(b))} \\
		&\le& \max\set{\tfrac{-(d_i+1)}{\chb_i}:
			i=1\dd n}\\
		&&
		+\left\|P\inv(\bfp)\right\|\left\|\frac{\partial \olbfg}{\partial
			\bfx}(\bfp)\right\|\left\|P(\bfp)\right\|\\
		&& \text{(by \refeq{matrixnorm})}\\
		&\le& \max\set{\tfrac{-(d_i+1)}{\chb_i}:
			i=1\dd n}\\
		&&
		+ \max_{i=1}^n \set{d_i} 
		\cdot
		\|J_{\olbfg}(\olpi(F_1))\|_2
		\cdot
		\max_{i=1}^n \set{\tfrac{(\chb_i)^{d_i+1}}{d_i}}.
		\eeqarrys
	}
	\epf
	
	\dt{\refLem{Set-d}}
		{\em \
		If $d \ge \old(F_1)$, we have:
		\benum[(a)]
		\item
			$\mu_2 \left(J_{\bfg} (\pi(F_1))\right)
			\le	 (-2+(\chb_{\max})^{d+2})
			\cdot \frac{\|J_{\olbfg}(\olpi(F_1))\|_2}{\chb_{\max}}.$
		\item
			If
			$\log_2(\chb_{\max}) < \tfrac{1}{d+2}$
			then
			$\mu_2 \left(J_{\bfg} (\pi(F_1))\right)< 0$.
		\eenum
	}
	\bpf
	\benum[(a)]
	\item
	By \refThm{keylemma} we have
	{\small
		\beqarrays
		\mu_2\left(J_{\bfg} (\pi(F_1))\right)
		&\le& -(d+1)\frac{1}{\chb_{\max}}
		+(\chb_{\max})^{d+1}\|J_{\olbfg}(\olpi(F_1))\|_2\\
		&=& \Big(\tfrac{-(d+1)}
		{\|J_{\olbfg}(\olpi(F_1))\|_2}+(\chb_{\max})^{d+2} \Big)
		\cdot \tfrac{\|J_{\olbfg}(\olpi(F_1))\|_2}{\chb_{\max}}\\ 
		&&	 \text{(by factoring)} \\
		&\le&  \Big(-2+(\chb_{\max})^{d+2} \Big)
		\cdot \tfrac{\|J_{\olbfg}(\olpi(F_1))\|_2}{\chb_{\max}} \\
		&&
		\text{(By eqn.\refeq{d}, we have $(d+1)
			\ge 2(
			\|J_{\olbfg}(\olpi(F_1))\|_2)$)}. 
		\eeqarrays
	}
	\item
	Since $(\chb_{\max})^{d+2}<2$
	is equivalent to 
	$\log_2(\chb_{\max}) < \tfrac{1}{d+2}$,
	we conclude that
	$\mu_2 \left(J_{\bfg} (\pi(F_1))\right)< 0$.
	\eenum
	\epf
	
\blemDIY[Lemma A.4]{\ \\
	Let $\bfp,\bfq\in B\ib\RR^n$ and $\phi\in C^1(F_1\to \RR^n)$, then 
	$\|\phi(\bfp)-\phi(\bfq)\|_2
	\le \|J_{\phi}(B)\|_2 \cdot \|\bfp-\bfq\|_2$
	}
	\bpf
	\beqarrys
	\|\phi(\bfp)-\phi(\bfq)\|_2
	&\le&
	\|\phi(\bfq)+J_{\phi}(\xi)\Bigcdot(\bfp-\bfq) -\phi(\bfq)\|_2\\
	&& \text{(by Taylor expansion of $\phi(\bfp)$ at $\bfq$)}\\
	&=&\|J_{\phi}(\xi)\Bigcdot(\bfp-\bfq)\|_2\\
	&\le& \|J_{\phi}(\xi)\|_2 \cdot \|(\bfp-\bfq)\|_2\\
	&\le& \|J_{\phi}(B)\|_2   \cdot \|(\bfp-\bfq)\|_2,
	\eeqarrys
	where $\xi\in B$.
	\epf
	
\bcorDIY[\refLem{error-bound-ode}]{
			Let $\bfy=\pi(\bfx)$ and 
			$$
			\mmatx[rcl]{
				\bfx 	&\in&
				\ivp_{\bff}(\bfx_0,h,F_1),\\
				\bfy 	&\in&
				\ivp_{\bfg}(\pi(\bfx_0),h,\pi(F_1)).
			}$$
			For any $\delta>0$ and any point $\bfp\in \RR^n$ satisfying 
			\beq
			\|\pi(\bfp)-\bfy(h)\|_2
			\le \frac{\delta}{\|J_{\pi\inv}(\pi(F_1))\|_2}, \nonumber
			\eeq
			we have 
			$$\|\bfp-\bfx(h)\|_2\le \delta.$$ 
	}
	\bpf
	\beqarrys
	\|\bfp-\bfx(h)\|_2
	&=& \|\pi\inv(\pi(\bfp))-\pi\inv(\pi(\bfx(h)))\|_2\\
	&=& \|\pi\inv(\pi(\bfp))-\pi\inv(\bfy(h))\|_2\\
	&\le&  			  
	\|J_{\pi\inv}(\pi(F_1))\|_2
	\cdot \|\pi(\bfp)-\bfy(h)\|_2\\
	&& \text{(by {\bf Lemma A.4})}\\
	&\le& \delta\qquad
	\text{(by condition \refeq{delta1}.)}
	\eeqarrys
	\epf

	\dt{\refThm{refine}} {\em
		The subroutine $\stage.\Refine(\veps)$ is correct.
		In particular, it halts.}
	\bpf
		The proof is in two parts: (a) partial
		correctness and (b) termination.
		Assume the scaffold $\stage$ has $m$ stages and the
		input for $\Refine$ is $\veps>0$.

	(a) Partial correctness is relatively easy, so we give sketch
		a broad sketch:
		we must show that if the $\Refine$ halts,
		then its output is correct, i.e.,
				$\wmax(E_m(\stage))<\veps$.
		The first line of $\Refine$
		initializes $r_0$ to $\wmax(E_m(\stage))$.
		If $r_0<\veps$, then we terminate without entering
		the while-loop, and the result hold.
		If we enter the while-loop, then we can only exit the while-loop
		if the last line of the while-body assigns to $r_0$
		a value $\wmax(E_m(\stage))$ less than $\veps$.
		Again this is correct.

	(b) The rest of the proof is to show that $\Refine$ halts.
		We will prove termination by way of contradiction.
		If $\Refine$ does not terminate, then it  
		has infinitely many \dt{phases}
		where the $k$th phase ($k=1,2,\ldots$)
		refers to the $k$ iterate of the while-loop.
	\benum[(H1)]
	\item
	\ignore{
		Each phase will refine the stages $1,2\dd m$ (in this order).
		The $i$th stage in phase $k$ has a ``target bound''
			$\delta^k_i>0$ stored as $\delta_i$ in $G_i(\stage)$
			(see \refeq{Gi}).
		For $k=1$, $\delta^1_i$ is $\veps_0$ if $i=m$
		and otherwise inherited from $(m-1)$-stage structure
		before $\Extend(\veps_0,H)$.
		For $k>1$, $\delta^k_i$ is equal to either $\delta^{k-1}_i$
		or half this value (see $\Refine(\veps)$).
		We have this picture of the $\delta$ targets:
			\btable{
				& Stage $1$ & $\cdots$ & Stage $i$
								& $\cdots$ & Stage $m$\\\hline
			Phase $1$: & $\delta^1_1$ & $\cdots$ &	$\delta^1_i$
								& $\cdots$ & $\veps_0$	\\
				$\vdots$ & $\vdots$	& $\ddots$ &$\vdots$
								& $\ddots$&	$\vdots$ \\
			Phase $k$: & $\delta^k_1$	& $\cdots$ & \cored{$\delta^k_i$}
								& $\cdots$ & $\delta^k_m$\\
				$\vdots$ & 	$\vdots$	& $\ddots$ & $\vdots$
								& $\ddots$ & $\vdots$	
			}
		}%
		We will show that $\lim_{k\to\infty}\delta^k_i=0$ for each
		$i=1\dd m$ in
		\refeq{phase-stage}.
		This will yield a contradiction.
	\item
		For a fixed stage $i$, we see the number of times
		that \Refine\ calls \SubrSeven\ is 
			$$d^k_i \as \log_2\Big(\frac{\delta^1_i}{\delta^k_i}\Big).$$
		Similarly, the number of times \Refine\ calls \Bisect\
		is 
			$$\ell^k_i - \ell^1_i$$
		where $\ell^k_i$ is the level of the $(k,i)$ phase-stage.
		So,
			\beql{kdki}
				k = d^k_i + (\ell^k_i - \ell^1_i)\eeql
			since each phase calls either \Bisect\ or \SubrSeven.
			Hence $k\to\infty$.
	\item
		CLAIM: $\lim_{k\to\infty} d^k_i\to \infty$, 
		i.e., \SubrSeven\ is called infinitely often.
		By way of contradiction,
		suppose $d^k_i$ has an upper bound, say $\ol{d}^k_i$.
		Since $\mu_2$, $\Delta t_i$, and $\ol{M}$ are bounded, we have
		$\wh{h} \ge C\cdot 2^{\ol{d}^k_i}$ in \Refine\ (see \refeq{h1}),
		where $C > 0$ is a constant.
	
		Note that each \Bisect($i$) increments the level and thus halves
		$H$.  Therefore, once $H < C \cdot 2^{\ol{d}^k_i}$, \Bisect\ will
		no longer be called.  This implies that
		$\lim_{k\to\infty}(\ell^k_i - \ell^1_i)$ is finite.
		This contradicts the fact that
		$k \to \infty$ since both $d^k_i$
		and $(\ell^k_i - \ell^1_i)$ are bounded.  Thus, our
		CLAIM is proved.
		
	It follows from the CLAIM that
	$\lim_{k\to\infty} \delta^k_i = 0$,
	since \SubrSeven\ is called infinitely often, and after each call,
	$\delta^k_i$ is halved.
	
	\item 
		Consider $(k,i)$ as a \dt{phase-stage}:
	define $r^k_i$ as the radius of the circumball of $E_i(\stage)$ at
	phase $k$.
	For instance, we terminate in phase $k$ if $k \ge 0$ is the first
	phase to satisfy $r^k_m < \half \veps_0$.
	
	Since we call \SubrSeven\ infinitely often, and each call ensures
	that the target $\delta^k_i$ is reached by $r^k_i$
	in the sense of satisfying the inequality
	\beql{subr7}
		r^k_i \le r^k_{i-1} e^{\mu^k_i \Delta t_i} + \delta^k_i,
	\eeql
	
	where $\mu^k_i$ (computed as $\mu_2$ in \Refine) is an upper bound
	for the logarithmic norm over the full enclosure of the $i$th stage.
	
	\item
	A \dt{chain} is a sequence
			$C_1=(1\le k(1)\le k(2)\le \cdots\le k(m))$
		such that \SubrSeven\ is called in phase-stage $(k(i),i)$
		for each $i=1\dd m$.
	The chain contains $m$ inequalities of 
	the form \refeq{subr7}, and we can telescope them into
	a single inequality.  

	\ignore{ Figure of 2 chains C<C', m=3
			\btable{
				& Stage $1$ & $\cdots$ & Stage $i$
								& $\cdots$ & Stage $m$\\\hline
			Phase $1$: & $\delta^1_1$ & $\cdots$ &	$\delta^1_i$
								& $\cdots$ & $\veps_0$	\\
				$\vdots$ & $\vdots$	& $\ddots$ &$\vdots$
								& $\ddots$&	$\vdots$ \\
			Phase $k$: & $\delta^k_1$	& $\cdots$ & \cored{$\delta^k_i$}
								& $\cdots$ & $\delta^k_m$\\
				$\vdots$ & 	$\vdots$	& $\ddots$ & $\vdots$
								& $\ddots$ & $\vdots$	
			}
}%
	But first, to
	simplify these inequalities, let
	$\olmu$ be the largest value of $\mu^1_i$ for $i=1\dd m$,
	$\Delta=\Delta(C_1)$ is 
	$\max_{i=1}^k \delta^{k(1)}_i$,
	and $h_i$ be the step size of the $i$th stage:
		\beqarrys
		r^{k(m)}_m &\le&
			(r^{k(m)}_{m-1}) e^{\mu^{k(m)}_i h_i}
					+ \delta^{k(m)}_i 
						& \text{(by \refeq{subr7} for ($k(m),m$))}\\
				&\le&	 (r^{k(m-1)}_{m-1}) e^{\mu^{k(m)}_i h_i}
							+ \delta^{k(m)}_i 
				 &\text{(since $r^{k(m)}_{m-1}\le r^{k(m-1)}_{m-1})$}\\
				&\le& (
			(r^{k(m-1)}_{m-2}) e^{\mu^{k(m-1)}_{i-1} h_{i-1}}
						+ \delta^{k(m-1)}_{i-1})
				e^{\mu^{k(m)}_i h_i}
							+ \delta^{k(m)}_i 
						& \text{(by \refeq{subr7} for ($k(m-1),m-1$))}\\
				&\le& ((r^{k(m-1)}_{m-2}) e^{\olmu h_{i-1}} + \Delta)
						e^{\olmu h_i} + \Delta 
						& \text{(simplify using $\olmu, h_i, \Delta$)}\\
				&\vdots&\\
				&\le& (r^{k(1)}_0) e^{\olmu \sum_{j=1}^m h_j} +
				\Delta\cdot\Big(
					\sum_{j=0}^{m-1} e^{\olmu \sum_{i=j+1}^m h_i}\Big)\\ 
				&\le&  e^{\olmu} (r^{k(1)}_0 + \Delta \cdot m)
						& \text{(since $1\ge \sum_{i=1}^m h_i$)}\\
		\eeqarrys
		To summarize what we just proved\footnote{
			This proof assumes that $\mu^k_i\le \mu^{k-1}_i$.
			This is true if our interval
			computation of $\mu_2$ is isotonic.  But we
			can avoid assuming isotony by defining $\mu^k_i$ to be
			$\mu^{k-1}_i$ if the computation returns a larger value.
		}
		about a chain $C_1$,
		let $r_m(C_1)$ denote $r^{k(m)}_m$
		and $r_0(C_1)$ denote $r^{k(1)}_0$
		the following \dt{$C_1$-inequality}:
			\beql{chain}
				r_m(C_1) \le e^{\olmu}(r_0(C_1)
								+ \Delta(C_1)\cdot m).
			\eeql
	\item
		If $C=(1\le k(1)\le \cdots \le k(m))$ and
		$C'=(1\le k'(1)\le \cdots \le k'(m))$ are two chains
		where $k(i)<k'(i)$ for $i=1\dd m$, then
		we write $C<C'$. 

		LEMMA H6: If $C<C'$ then
			$$r_m(C')\le\half
				e^{\olmu}(r_0(C)+ \Delta(C)\cdot m)$$
	\item
		It is easy to show that
		there exists an infinite sequence of chains
			$$C_1 < C_2 < C_3 <\cdots.$$
		This comes from the fact that for each $i=1\dd m$,
		there are infinitely many phases that calls $\SubrSeven$.
		It follows by induction using the previous LEMMA H6 that,
		for each $i\ge 2$, 
			$$r_m(C_i)\le (\half)^i
				e^{\olmu}(r_0(C_1)+ \Delta(C_1)\cdot m)$$
		This proves that $\lim_{i\to\infty} r_m(C_i) =0$.
		This contradicts the non-termination of \Refine.
	\eenum
	\epf
	
	\dt{\refThm{correct-main}}
		{\em
			Algorithm $\endEncAlgo(B_0, \veps_0)$ halts,
			provided the interval computation of $\stepA$
			is isotonic.  The output is also correct.
	}
	\bpf
	
	If the algorithm terminates, its correctness is ensured by the
	conclusions in \refSec{stepAB}.
	
	We now proceed to prove the termination of the algorithm.
	Specifically, we need to show that the loop in the algorithm can
	terminate, which means that the time variable $t$ can reach $ 1 $. 
	It suffices to demonstrate that for any inputs
	$B_0$ and $\veps_0 > 0$, there exists a lower bound $\ulh>0$
	such that for the $i$th iteration of the loop has
	step size $\Delta t_i =h_i \ge \ulh$.

	First we define the set
		$\olE\as \image(\ivp(B_0,1))+[-\veps_0,\veps_0]^n$.
		Since $\ivp(B_0,1)$ is valid, $\olE$ is a bounded set.
		Let the pair $(\ulh,\olF)$ be the result of calling
		the subroutine $\stepA(\olE,1,\veps_0)$.
	Note that \stepA\ is implicitly calling box functions
	to compute $\ulh, \olF$ 	(see \refSSec{implicit}), and
	thus $\ulh$ is positive.  Whenever we call \stepA\
	in our algorithm, its arguments are $(E,H,\veps_0)$ for
	some $E\ib\olE$ and $H\le 1$.  If $\stepA(E,H,\veps_0)\ssa (h,F)$,
	then $h\ge \ulh$, provided\footnote{
		If computes $h>\ulh$, we could
		not ``simply'' set $h$ to be $\ulh$ because we do not
		know how to compute a corresponding full enclosure.
		Note that $\olE$ is a full enclosure, but
		we do not know how to compute it.
	}
	$\stepA$ is isotonic.  This proves that the algorithm
	halts in at most $\ceil{1/\ulh}$ steps.
	\epf

\section{Appendix B: The  affine map $\ol\pi$}
	Consider the condition \refeq{0ninolB1}.
	Without loss of generality, assume
	$0 \notin \olI_1$.  To further simplify our
	notations, we assume
	\beql{assume1}  
	\olI_1 >0.
	\eeql  
	In case $\olI_1<0$, we shall indicate the necessary changes
	to the formulas.
	We first describe an invertible linear map
	$\wt\pi:\RR^n\to\RR^n$
	such that
	\beql{pos}
	\wt\pi(\bff(B_1))>\1=(1\dd 1)
	\qquad\text{(Greater-than-One Property of $\wt\pi$)}
	\eeql
	Note that \refeq{pos} means that for each $i=1\dd n$,
	the $i$th component $(\wt\pi(\bff(B_1)))_i$ is greater than one.
	
	To define $\wt\pi$, we first introduce the box $\wtB_1$:
	\beql{olB_1}
	\mmatx{
		\wtB_1
		&\as& Box(\bff(B_1))\\
		&=& \prod_{i=1}^n \olI_i
		& \text{(implicit definition of $\olI_i$)}\\
		&=& \prod_{i=1}^n [\ola_i,\olb_i]
		& \text{(implicit definition of $\ola_i, \olb_i$)}
	} \eeql
	where $Box(S)\in\intbox\RR^n$ is the smallest
	box containing a set $S\ib\RR^n$.
	For instance, $\olI_i=f_i(B_1)$ where $\bff=(f_1\dd f_n)$.
	The assumption \refeq{assume1} says that
	$ \olI_1>0$, i.e., either $\ola_1>0$.
	
	We now define the map $\wt\pi:\RR^n\to\RR^n$ as follows:
	$\wt\pi(x_1\dd x_n) = (\wtx_1\dd \wtx_n)$ where
	\beql{wtpi}
	\wtx_i  \as  \clauses{
		\frac{x_i}{\ola_i}	& \rmif\ \ola_i>0,
		& \text{(i.e., $f_i(B_1)>0$)}\\
		\frac{x_i}{\olb_i}	& \eliF\ \olb_i<0,
		& \text{(i.e., $f_i(B_1)<0$)}\\
		x_i+ x_1\big(
		\tfrac{1+\olb_i-\ola_i}{\ola_1}\big)
		& \elsE
		& \text{(i.e., $0\in f_i(B_1)$).}
	}\\
	\eeql
	Note that if \( \olI_1 < 0 \), we only need to modify the third
	clause in \refeq{wtpi} to \( x_i + x_1 \big( \tfrac{1 + \olb_i -
		\ola_i}{\olb_1} \big) \).

	Observe that $\wt\pi(\wtB_1)$ is generally a parallelopiped,
	not a box.  Even for $n=2$, $\wt\pi(\wtB_1)$ is a parallelogram.
	So we are interested in the box $Box(\wt\pi(\wtB_1))$:
	\beql{B'1}
	\mmatx{
		B'_1 \as Box(\wt\pi(\wtB_1))
		&=& \prod_{i=1}^n I'_i
		& \text{(implicit definition of $I'_i$)}\\
		&=& \prod_{i=1}^n [a'_i,b'_i]
		& \text{(implicit definition of $a'_i, b'_i$)}
	}
	\eeql
	Then we have the following results:

\blemDIY[Lemma B.1]{ \ 
	\benum[(a)]  
	\item  $\wt\pi$ is an invertible linear map given by
	\beql{olpiolA}
	\wt\pi(\bfx) = \olA\Bigcdot\bfx  \eeql
	and the matrix $\olA$ has the structure:
	\[
	\mmat{
		v_1 & & & & \\
		c_2 & v_2 & & & \\
		c_3 & & v_3 & & \\
		\vdots & & & \ddots & \\
		c_n & & & & v_n
	}
	\label{olA}
	\]
	where
	$$v_i=
	\clauses{
		\frac{1}{\ola_i} & \rmif\ \ola_i>0,\\
		\frac{1}{\olb_i}	& \eliF\ \olb_i<0,\\
		1	& \elsE }$$
	$$c_i=
	\clauses{
		0 &\rmif\ 0\nin f_i(B_1),\\
		\tfrac{ 1+ \olb_i-\ola_i}{\ola_1} &\elsE. }$$
	\item The box $Box(\wt\pi(\wtB_1)) = \prod_{i=1}^n I'_i$
	is explicitly given by
	\beql{I'i}
	I'_i = 
	\begin{cases}
		\left[1, \frac{\olb_i}{\ola_i}\right] & \text{if } \ola_i > 0, \\[10pt]
		\left[1, \frac{\ola_i}{\olb_i}\right] & \text{else if } \olb_i < 0, \\[10pt]
		\left[1+\olb_i, \frac{\olb_1}{\ola_1} 
		\big(1 + \olb_i\big(1 + \frac{\ola_1}{\olb_1}\big) - \ola_i\big)\right] 
		& \text{else  }.
	\end{cases}
	\eeql
	\item
	The map $\wt\pi$ has the
	positivity property of \refeq{pos}.
	\eenum
}
	\bpf
	\benum[(a)] 
	\item
	From the definition of $\wtpi$ in \refeq{wtpi},
	we see that the matrix $\olA$ matrix 
	the form described in the lemma.  
	This matrix is clearly invertible.
	\item
	We derive explicit formulas for $I'_i$
	in each of the 3 cases:
	\bitem
	\item If $\ola_i>0$, then it is clear that
	$(\wt\pi(B_1))_i= \left[1, \frac{\olb_i}{\ola_i}\right]$.
	\item Else if $\olb_i<0$, it is also clear that
	$(\wt\pi(B_1))_i= 	\left[1, \frac{\ola_i}{\olb_i}\right]$.
	\item Else, we consider an arbitrary 
	point $\bfx=(x_1\dd x_n)\in \wtB_1$:
	{\small \beqarrys
		(\wt\pi(\bfx))_i &=& x_i+ x_1
		\big(\tfrac{1+\olb_i-\ola_i}{\ola_1}\big)
		&\text{(by definition)}\\
		&\ge& \ola_i +\ola_1\big(
		\tfrac{1+\olb_i-\ola_i}{\ola_1}\big)\\
		&&\text{($x_j\in[\ola_j,\olb_j]$ ($\forall~ j$)
			\& $(1+\olb_i-\ola_i)/\ola_1 > 0$))}\\
		&=& 1+\olb_i.\\
		(\wt\pi(\bfx))_i &=&
		x_i+ x_1\big(
		\tfrac{1+\olb_i-\ola_i}{\ola_1}\big)\\
		&\le& \olb_i +\olb_1\big(
		\tfrac{1+\olb_i-\ola_i}{\ola_1}\big)\\
		&&\text{($x_j\in[\ola_j,\olb_j]$ 
			and $(1+\olb_i-\ola_i)/\ola_1 > 0$)}\\
		&=& \tfrac{\olb_1}{\ola_1}
		\big(1+\olb_i(1+\tfrac{\ola_1}{\olb_1})-\ola_i\big).
		\eeqarrys}
	Since both the upper and lower bounds are
	attainable, they determine the interval $I'_i$
	as claimed.
	\eitem
	\item It is sufficient to show that
	$I'_i\ge1$.
	This is clearly true for the first two clauses of \refeq{I'i}.
	For the last two clauses, we have $I'_i\ge 1+\olb_i$
	by part(b).  The result follows since $0\le \olb_i$.
	\eenum
	\epf

	Let 	\beql{wtB_1}
	\mmatx{
		B^{*}_1
		&\as& Box(\wt\pi(B_1))\\
		&=& \prod_{i=1}^n I^{*}_i
		& \text{(implicit definition of $I*_i$)}\\
		&=& \prod_{i=1}^n [a^{*}_i,b^{*}_i]
		& \text{(implicit definition of $a^{*}_i, b^{*}_i$)}
	} \eeql
	We now define the affine map
	$\ol\pi:\RR^n\to\RR^n$:
	\beql{olpi}
	\mmatx{
		\ol\pi(\bfx) &=& (\olpi_1(x_1),\olpi_2(x_2)\dd\olpi_n(x_n)) \\
		&& \qquad \text{ where } \bfx=(x_1\dd x_n) 
		\text{ and } \\
		\olpi_i(x)
		&\as& \wtpi(x)-a^{*}_i +1.
	}
	\eeql
	Then we have the following results, which is property (Q2):
\blemDIY[Lemma B.2]{ \ 
	$\olpi(B_1)>\bf1.$
	}
	\bpf
	The conclusion follows from the fact that 
	$\wtpi(B_1)\ib \prod_{i=1}^n [a^{*}_i,b^{*}_i]$ and 
	$\olpi(B_1)=\wtpi(B_1)-(a^{*}_1\dd a^{*}_n)+1$.
	
	\epf
	
\ignore{\section{Appendix C: Issues of Validated IVP}
	This is where we will target our sharp criticisms of
	the literature.
	
	Previous assessments of the validated literature has
	been made by Corliss \cite{XXX}, Stetter \cite{XXX}, etc.
	The following remarks brings in other issues as well.
	In numerical computation, the gap between the
	concept of real numbers and their use in computation
	is a foundational question \cite{egctheory}.
	At a more practical level, the rigor of designing specific
	validated algorithms also need to be addressed;
	a 3-Level approach was suggested by \cite{xu-yap}.
	In the following, our remarks will be focused on
	the Validated IVP literature.
	
	\bitem
	\item
		In reviews of the first version of this paper \cite{arxiv-ivp},
		some experts were baffled by our claim that
		previous ``validated IVP algorithms'' were incomplete.
		There are two main issues of incompleteness:
		(1) halting, (2) underspecification of the problem.
	\item
		Inevitably, the halting issue is hidden
		under a common device of ``tolerance''.
		In general, if an algorithm has an escape clause like
		``out of tolerance'', it is important to have a rigorous
		characterization when this clause is invoked (otherwise,
		a trivial solution is to always take the escape clause
		for all inputs.
		This lack of rigor is then traced to the
		usual conflation of ``algorithm'' with ``problem''.
		We must always begin with a ``problem'' with
		a description of its input-output behavior.
		Then an ``algorithm'' to solve that ``problem'' can
		be judged to have correctly solved the problem.
		Without the a priori specification of the problem,
		the term ``algorithm'' is meaningless\footnote{
			Unfortunately, the conflation of
			algorithm with heuristic is a increasing problem
			in the age of AI.
		} or it should 
		be called a ``heuristic''.
	\item
		There are genuinely intractable issues with IVP problem.
		For instance, the standard mathematical proof of existence
		of solution to an IVP gives a local solution: for any
		$\bfp_0$, there exists $h>0$ and solution
		$\bfx\in C^1([0,h]\to\RR^n)$ such that $\bfx(0)=\bfp_0$.
		Since $h$ can be arbitrarily small, it is a local solution.
		This does not guarantee that we can continue this solution
		to any target time, say $h=1$.  
		It is important that our IVP formulation requires $h$ as
		part of the input, and our algorithm
		is conditioned on the promise that $(\bfp_0,h)$ is well-defined.
		Promise problems are common and useful especially when
		the promise holds ``generically'' and
		decision algorithms to check the promise is intractable or
		non-existent.  
	\eitem
}%